\documentclass[times]{aastex61}
\usepackage{amsmath}

\received{January 31, 2018}
\revised{July 19, 2018}
\accepted{July 21, 2018}
\submitjournal{Planetary and Space Science}

\shorttitle{Hot grain dynamics in exozodis}
\shortauthors{Kimura et al.}

\begin{document}

\title{Hot Grain Dynamics by Electric Charging and Magnetic Trapping in Debris Disks}

\correspondingauthor{Hiroshi Kimura}
\email{hiroshi{\_}kimura@perc.it-chiba.ac.jp}

\author{Hiroshi Kimura}
\affiliation{Planetary Exploration Research Center (PERC), Chiba Institute of Technology, Tsudanuma 2-17-1, Narashino, Chiba 275-0016, Japan}
\affiliation{Division of Particle and Astrophysical Science, Graduate School of Science, Nagoya University, Furo-cho, Chikusa-ku, Nagoya 464-8602, Japan}
\affiliation{Graduate School of Science, Kobe University, 1-1 Rokkodai-cho, Nada-ku, Kobe 657-8501, Japan}

\author{Masanobu Kunitomo}
\affiliation{Department of Earth and Planetary Science, The University of Tokyo, 7-3-1, Hongo, Bunkyo-ku, Tokyo 113-0033, Japan}
\affiliation{Division of Particle and Astrophysical Science, Graduate School of Science, Nagoya University, Furo-cho, Chikusa-ku, Nagoya 464-8602, Japan}

\author{Takeru K. Suzuki}
\affiliation{School of Arts \& Sciences, The University of Tokyo, 3-8-1, Komaba, Meguro, Tokyo, 153-8902, Japan}
\affiliation{Division of Particle and Astrophysical Science, Graduate School of Science, Nagoya University, Furo-cho, Chikusa-ku, Nagoya 464-8602, Japan}

\author{Jan Robrade}
\affiliation{Hamburger Sternwarte, Gojenbergsweg 112, 21029 Hamburg, Germany}

\author{Philippe Thebault}
\affiliation{Laboratoire d'\'{E}tudes Spatiales et d'Instrumentation en Astrophysique (LESIA), Observatoire de Paris, Section de Meudon, 92195 Meudon Principal Cedex, France}

\author{Ikuyuki Mitsuishi}
\affiliation{Division of Particle and Astrophysical Science, Graduate School of Science, Nagoya University, Furo-cho, Chikusa-ku, Nagoya 464-8602, Japan}



\begin{abstract}

The recent discovery of hot dust grains in the vicinity of main-sequence stars has become a hot issue among the scientific community of debris disks.
Hot grains must have been enormously accumulated near their sublimation zones, but it is a mystery how such a high concentration of hot grains is sustained.
The most difficult conundrum is that the size of hot dust grains is estimated to lie in the submicrometer range, while submicrometer-sized grains are instantly swept away from near-stellar environments by stellar radiation pressure.
One and only mechanism proposed for prolonging the residence time of hot grains in the near-stellar environments is trapping of charged nanoparticles by stellar magnetic fields.
We revisit the model of magnetic grain trapping around main-sequence stars of various spectral classes by taking into account sublimation and electric charging of the grains.
The model of magnetic grain trapping predicts that hot dust grains are present in the vicinity of main-sequence stars with high rotation velocities and intermediate magnetic-field strengths.
On the contrary, we find that the detection of hot dust grains has no correlation with the rotation velocities of central stars nor the magnetic field strengths of the stars.
Our numerical evaluation of electric grain charging indicates that the surface potential of submicrometer-sized grains in the vicinity of main-sequence stars is typically $4$--$5~\mathrm{V}$, which is one order of magnitude smaller than the value assumed by the model of magnetic grain trapping.
On the basis of our numerical simulation on sublimation of dust grains in the vicinity of a star, it turns out that their lives end due to sublimation in a timescale much shorter than the period of one revolution at the gyroradius.
It is, therefore, infeasible to dynamically extend the dwell time of hot grains inside the sublimation zone by magnetic trapping, while we cannot completely rule out the possibility of magnetic grain trapping outside the sublimation zone where the strength of stellar magnetic field has been underestimated in the previous model.
Nevertheless, the independence of hot dust detection on the stellar rotational velocity and magnetic field strength favors a scenario that some other (yet unnoticed/overlooked) ubiquitous mechanism of grain trapping is at work.

\end{abstract}

\keywords{Hot dust --- 
Debris disks --- Electric grain charging --- Dust sublimation --- Stellar rotational velocity --- Stellar magnetic field}



\section{Introduction} \label{sec:intro}

The central region of debris disks around main-sequence stars is best visible by near-infrared wavelengths at which thermal emission from hot dust grains near the central stars peaks.
Near-infrared interferometric observations of debris disks, in other words, exozodis\footnote{Here we use the term ``exozodis'' as an analog to the zodiacal dust cloud that extends from the Kuiper belt dust ring(s) to the solar F-corona.} have made the discovery of excess emission from hot dust in the vicinity of main-sequence stars \citep{absil-et-al2009,absil-et-al2013,ertel-et-al2014}.
It turned out that the excess emission from hot dust equivalent to approximately 1\% of stelar flux is a serious challenge for classical grain dynamics models.
Currently available parametric models of exozodis attribute the near-infrared excess to an enormous amount of submicrometer-sized grains near the sublimation zone, although they are too small to stay in bound orbits around the central stars against strong stellar radiation pressure \citep{lebreton-et-al2013,vanlieshout-et-al2014,kirchschlager-et-al2017}.
In a classical grain dynamics model, dust grains in orbit around a central star gradually spiral to the star by the Poynting-Robertson effect and accumulate at the outer edge of sublimation zone on account of the increase in the semimajor axis by sublimation \citep{burns-et-al1979,kimura-et-al1997,kobayashi-et-al2009}.
However, the enhancement of hot dust density at the outer edge of sublimation zone is limited to a factor of 10, which is far from sufficient to explain the observed excess emission from hot dust \citep{kobayashi-et-al2009,kobayashi-et-al2011,vanlieshout-et-al2014}.
There are proposed mechanisms to supply dust grains in situ by electrostatic ejection of the grains from their parent bodies or through cometary activity of planetesimals in inner mean-motion resonances with eccentric planets \citep{kimura-et-al2014,faramaz-et-al2016}.
Unfortunately, a vast supply of submicrometer-sized grains into the near-stellar environments would result in a considerable amount of dust outflows, which has not been observed to date.
Moreover, these mechanisms are unable to prolong the resident time of the grains in the near-stellar environments, while trapping of dust grains in the vicinity of stars by a yet another mechanism, if exists, is relatively straightforward.

\citet{rieke-et-al2016} have proposed a model of magnetic trapping in which charged nanoparticles are dynamically confined in the stellar magnetic field inside the sublimation zone.
Nonetheless, we find it difficult to justify the underlying assumptions of the magnetic grain trapping;
First, it is not obvious whether the production of nanoparticles in the vicinity of a star by whatever mechanisms results in magnetic trapping instead of radiative blowout.
Second, the model explicitly assumes that dust grains around A-type stars obtain the same electrical charges as around young stellar objects (YSOs), irrespective of a significant difference in their plasma temperatures, on which the electrostatic potential may be dependent \citep[cf.][]{telleschi-et-al2005,draine-salpeter1979}.
Third, the model implicitly assumes that the radiation spectrum of A-type stars follows Planck's law with an effective temperature of 10,000~K, while the deviation of stellar spectra from Planck's law is significant in the ultraviolet, which might affect the photoelectric charging.
Fourth, the model ignores the effect of sublimation on the dynamics of hot dust grains, while sublimation may prevent the grains from being trapped for a considerable period of time.
Fifth, the model suggests that the duration of magnetic grain trapping is on the order of months or a few years at most, but the resident time of hot nanoparticles expected from observations is 5--50 years \citep{su-et-al2016}.
Sixth, the model predicts an association between the detection of near-infrared excess emission and the rotation velocity of stars, but rapid rotators do not always have a near-infrared excess \citep{absil-et-al2013}.
Last but not least, since \citet{rieke-et-al2016} ignored collisional destruction and stellar-wind velocity in their simulations, the magnetic trapping scenario needs to be further investigated with detailed numerical simulations, as emphasized by \citet{kral-et-al2017}.

Inevitably, dust particles in debris disks are electrically charged by virtue of exposure to ultraviolet radiation and plasma flow from central stars, irrespective of their physical and chemical properties.
As a result, the magnetic field of the central star might influence the dynamics of dust particles by exerting a force, called the Lorentz force, on the particles, the smallest ones, in particular \citep{parker1964,belton1966}.
This is because the Lorentz force is proportional to the radius of dust particles and the gravitational force is proportional to the volume of the particles.
Numerical simulations on the dynamics of charged dust particles in the close vicinity of the Sun have shown that the Lorentz force plays an important role in the dynamics of the particles near the sublimation zone \citep{rusk1988,krivov-et-al1998,czechowski-mann2010,czechowski-kleimann2017}.
The magnitude of the Lorentz force acting on a charged grain is proportional to the electrical charges on the surface of the grain or the electrostatic surface potential in other words.
Therefore, it is clear that the electrical charge on the grain surface is the key quantity to elucidate the dynamics of dust particles near the sublimation zone in a stellar magnetic field.
The electrostatic potential of dust particles, which is mainly determined by the current balance between plasma electrons and photoelectrons, has been estimated for the zodiacal dust disk around the Sun \citep{mukai1981,kimura-mann1998a}.
In the vicinity of stars, thermionic emission and secondary electron emission might also make major contributions to the electric charging of grains \citep{belton1966,lefevre1975,mukai1981}.
Unfortunately, it is not obvious how the electrostatic potential varies in the sublimation zone nor how the spectral class of main-sequence stars affects the potential.
Accordingly, a proper estimate of electrical charges on the surface of dust grains could help us understand the dynamics of dust grains in the near-stellar environments.  

In this paper, we revisit the magnetic trapping of hot dust grains around main-sequence stars of various spectral classes by taking into account sublimation and electric charging of the grains.
We first model electrostatic grain charging in the vicinity of main-sequence stars with various spectral types and stellar wind parameters.
Subsequently we present our numerical results on the equilibrium temperatures, electrostatic potentials, and charge-to-mass ratios of dust particles near the sublimation zone, as well as the ratio of forces acting on the particles in the stellar magnetic and gravitational fields.
The results are utilized to compare characteristic timescales for sublimation to periods of gyrating motion, and the detections of near-infrared excess emission to stellar rotation velocities and magnetic field strengths.
Finally we discuss the plausibility of magnetic grain trapping based on our results of grain dynamics and our findings about the ubiquitous nature of hot grains.

\section{Grain charging}

Dust grains are electrically charged by a variety of electric currents $J$ such as photoelectron emission, impinging of plasma particles, secondary electron emission, and thermionic emission \citep{belton1966}.
The derivative of the electric charge $Q$ on the surface of a dust grain with respect to time $t$ equals to the sum of electric currents, namely, 
	\begin{eqnarray}
		\frac{dQ}{dt} = \sum_{k} J_k ,
		\label{dqdt}
	\end{eqnarray}
where $J_k$ is the electric current of the $k$-th charging process.
It is clear that the condition $dQ/dt = 0$ provides an equilibrium charge $Q$ on the surface of the grain and an equilibrium surface potential $U_\mathrm{g}$ in turn by the following relation:
	\begin{eqnarray}
		U_\mathrm{g} = \frac{Q}{4 \pi \varepsilon_0 a} ,
		\label{surface-potential}
	\end{eqnarray}
where $a$ is the grain radius and $\varepsilon_0$ is the permittivity of vacuum.
Here we shall first describe the details of our modeling on each electric current\footnote{\citet{mukai1981} provides figures to show contributions of individual electric currents to grain charging, which manifests photoelectron emission being the major charging process. In Appendix~\ref{appendix-b}, we provide a crude estimate of electric currents due to photoelectron emission and electron impinging to demonstrate the predominance of photoelectron emission.}, which is used to determine the equilibrium surface potentials of dust grains in a variety of stellar environments.

\subsection{Photoelectron emission}

The electric current due to photoelectron emission, ${J}_\mathrm{ph}$, is given by \citep{mukai1981,kimura-mann1998a}
	\begin{eqnarray}
		{J}_\mathrm{ph} = e\int_{W}^{\infty }d(h\nu ) 
		{C}_\mathrm{abs}(h\nu )F(h\nu )Y({h\nu })
		\int_{{\epsilon}_{\min}}^{{\epsilon}_{\max}}d\epsilon 
		{\rho}_\mathrm{ph} (\epsilon) ,
		\label{Jph}
	\end{eqnarray}
where $e$, $h$, $\nu$, ${C}_\mathrm{abs}(h\nu )$, $F(h\nu )$, $Y({h\nu })$, and ${\rho}_\mathrm{ph} (\epsilon)$ are the elementary charge, Planck's constant, the frequency of incident photon, the absorption cross section of a photon having energy $h\nu$, the stellar photon flux in the range of frequency from $\nu$ to $\nu + d\nu$, the photoelectric quantum yield at $h\nu$, and the energy distribution of photoelectrons in the range of energy from $\epsilon$ to $\epsilon + d\epsilon$, respectively.
We compute ${C}_\mathrm{abs}(h\nu )$ and $Y({h\nu })$ in the framework of Mie theory as functions of photon energy, grain radius, and grain composition \citep{kimura2016}.
The refractive indices of silicate and carbon\footnote{\citet{rouleau-martin1991} provide two sets of refractive indices for amorphous carbon, named AC1 and BE1, while we use the data for BE1 throughout the paper.}, which have been popular grain compositions for modeling of debris disks, are used for the computations \citep{laor-draine1993,rouleau-martin1991}.
The work function $W$ decreases with grain radius $a$ as \citep{brus1983,makov-et-al1988,wong-et-al2003}
\begin{eqnarray}
W = W_\infty + \frac{1}{4 \pi \varepsilon_0} \frac{3}{8} \frac{e^2}{a} \frac{\varepsilon-1}{\varepsilon} ,
\end{eqnarray}
where $W_\infty$ is the work function for a bulk and $\varepsilon$ is the static dielectric constant relative to vacuum.
Table~\ref{tab:grain-parameters} gives the grain parameters such as $W_\infty$ and $\varepsilon$ used for computations of grain charging and sublimation.
The energy distribution of photoelectrons may be approximated by \citep{grard1973}
\begin{eqnarray}
{\rho}_\mathrm{ph} (\epsilon)  = \frac{\epsilon}{\phi_\mathrm{ph}^2} \exp{\left({-\frac{\epsilon}{\phi_\mathrm{ph}}}\right)},
\end{eqnarray}
with $\phi_\mathrm{ph} = 1~\mathrm{eV}$ in the range of energy between ${\epsilon}_{\min} = \max(0,eU_\mathrm{g})$ with $U_\mathrm{g}$ being the surface potential of the grain and ${\epsilon}_{\max} = h\nu -W$.
The stellar photon flux $F(h\nu )$ at a distance $r$ from the central star is described as
\begin{eqnarray}
F(h\nu ) = 2\pi\left\{{1-\left[{1-\left({\frac{R_\star}{r}}\right)^{2}}\right]^{1/2}}\right\} P_\star (h\nu) ,
\end{eqnarray}
where $R_\star$ is the stellar radius and $P_\star (h\nu)$ is the stellar radiance at photon energy $h\nu$ (see Appendix~\ref{appendix-a}).

\subsection{Plasma impingement}

The electric current due to impingements of plasma particles, ${J}_\mathrm{imp}$, is given by \citep{draine-salpeter1979,mukai1981,kimura-mann1998a}
	\begin{eqnarray}
		{J}_\mathrm{imp} = \sum\limits_{j}^{} {2\pi }{Z}_{j}e 
		\int_{{v}_{0}}^{\infty }dv\int_{0}^{\pi }d\theta 
		{\sigma }_{j}\left({v}\right)
		{f}_{j}\left({v,\theta }\right){v}^{3}\sin\theta ,
		\label{Jimp}
	\end{eqnarray}
where ${Z}_{j}e$, ${\sigma }_{j}\left({v}\right)$, and ${f}_{j}\left({v,\theta }\right)$ are the electric charge of the $j$-th plasma component, the collisional cross section of the $j$-th plasma component having radial velocity $v$ at infinity, and the velocity distribution of the $j$-th plasma component at infinity in the ranges of radial velocity from $v$ to $v+dv$ and polar angle from $\theta$ to $\theta + d\theta$.
The minimum velocity ${v}_{0}$ of the $j$-th plasma component to collide a grain is given by
\begin{eqnarray}
{v}_{0}  = 
  \begin{cases}
    0  & {Z}_{j} \phi_{j} \le 0 , \\
    \left({2{Z}_{j}eU_\mathrm{g}/{m}_{j}}\right)^{1/2} & {Z}_{j} \phi_{j}> 0 ,
  \end{cases}
\end{eqnarray}
where ${m}_{j}$ is the mass of the $j$-th plasma component.
The dimension-less variable $\phi_{j}$ is defined by 
\begin{eqnarray}
\phi_{j} \equiv \frac{eU_\mathrm{g}}{k_\mathrm{B} T_{j}} ,
\end{eqnarray}
where $k_\mathrm{B}$ is Boltzmann's constant and $T_{j}$ is the temperature of the $j$-th plasma component.
The collisional cross section of the $j$-th plasma particles is given by
\begin{eqnarray}
{\sigma }_{j}\left({v}\right) = \pi a^2 \left({1-\frac{2 Z_J e U_\mathrm{g}}{m_j v^2}}\right) .
\end{eqnarray}
We assume thermodynamic equilibrium of plasma at a great distance from the surface of a grain so that the velocity distribution of the $j$-th plasma component is described by the Maxwell-Boltzmann distribution:
\begin{eqnarray}
{f}_{j}\left({v,\theta }\right) = n_j \left({\frac{m_j}{2 \pi k_\mathrm{B} T_j}}\right)^{3/2} \exp\left[{-\frac{m_j}{2 k_\mathrm{B} T_j} \left({v^2 + w^2 - 2 v w \cos{\theta}}\right)}\right] ,
\end{eqnarray}
where $\theta$ is measured from the bulk velocity vector $\mathbf{w}$ of the plasma in the reference frame of the grain.
If the bulk velocity of the plasma and the orbital velocity of the grain with respect to the central star are denoted as $\mathbf{v_\mathrm{sw}}$ and $\mathbf{v_\mathrm{g}}$, respectively, we have $\mathbf{w}=\mathbf{v_\mathrm{sw}}-\mathbf{v_\mathrm{g}}$.

\subsection{Plasma penetration}

The electric current due to penetration of plasma particles, ${J}_\mathrm{pen}$, is given by \citep{draine-salpeter1979,kimura-mann1998a}

	\begin{eqnarray}
		{J}_\mathrm{pen} = -\sum\limits_{j}^{} {2\pi }{Z}_{j}e 
		{\int_{{v}_{1,j}}^{\infty }dv\int_{0}^{\pi }d\theta {\sigma }_{j}
		\left({v}\right)f_j\left({v,\theta }\right){v}^{3}\sin\theta } ,
 		\label{jpen}
	\end{eqnarray}
The minimum velocity ${v}_{1,j}$ of the $j$-th plasma particle to penetrate a grain without neutralization is given by \citep{draine-salpeter1979,kimura-mann1998a}
\begin{eqnarray}
{v}_{1,j}  = 
  \begin{cases}
    \left({2 k_\mathrm{B} T_{j} \psi_{j} / m_{j}}\right)^{1/2}  & {Z}_{j} \phi_{j} \le 0 , \\
    \left[{2 k_\mathrm{B} T_{j} (\psi_{j} + {Z}_{j} \phi_{j}) / m_{j}}\right]^{1/2} & {Z}_{j} \phi_{j}> 0 ,
  \end{cases}
\end{eqnarray}
where $\psi_{j}$ is related to the threshold energy $\Delta_{j}$ of penetration as
\begin{eqnarray}
\psi_{j}  = 
  \begin{cases}
    \Delta_{j}/k_\mathrm{B} T_{j}  & {Z}_{j} < 0 , \\
    (\Delta_{j} + \Gamma_{j})/k_\mathrm{B} T_{j} & {Z}_{j}  > 0 ,
  \end{cases}
\end{eqnarray}
under the assumption that positive ions emerge neutral unless their exit energies exceed $\Gamma_{j} = (1/2) m_{j} c^2 \alpha^2$ with $c$ and $\alpha$ being the speed of light and the fine structure constant.
The threshold energy $\Delta_{j}$ of penetration may be determined by the condition of $R_{j}(\Delta_{j}) = 4a/3$ where the projected range $R_{j}(E_0)$ for the $j$-th plasma particle having incident energy $E_0$ may be approximated by \citep{draine-salpeter1979}
\begin{eqnarray}
R_{j}(E_0) = \hat{R}_{j} \rho^{p_{j}} E_0^{q_{j}} ,
\end{eqnarray}
with fit parameters $\hat{R}_{j}$ $p_{j}$, and $q_{j}$ for the $j$-th plasma particle and the volumetric mass density $\rho$ of dust grains \citep[see][]{kimura-mann1998b}.

\subsection{Secondary electron emission}

The electric current due to secondary electron emission, ${J}_\mathrm{see}$, is given by \citep{draine-salpeter1979,mukai1981,kimura-mann1998a}
	\begin{eqnarray}
		{J}_\mathrm{see} = \sum\limits_{j}^{} {2\pi }e 
		{\int_{{v}_{0}}^{\infty }dv\int_{0}^{\pi }d\theta 
		{\delta }_{j}\left({{E}_{0}}\right){\sigma }_{j}
		\left({v}\right)f_j\left({v,\theta }\right){v}^{3}\sin\theta } \int_{{\epsilon}_{\min}}^{\infty }d\epsilon 
		{\rho }_{j}\left({\epsilon}\right) , 
		\label{Jsee}
	\end{eqnarray}
where ${\delta }_{j}({E}_{0})$ is the secondary electron yield at ${E}_{0}$ and ${\rho }_{j}(\epsilon)$ is the energy distribution of secondary electrons in the range of energy from $\epsilon$ to $\epsilon+ d\epsilon$.
We compute ${\delta }_{j}({E}_{0})$ as a function of incident energy in the framework of an elementary theory by taking into account the curvature of grain surface \citep{kimura-mann1998b}.
The secondary electron yield ${\delta }_{j}({E}_{0})$ depends on the grain composition\footnote{The dependence of the secondary electron yields for slabs on incident energy shows a material-independent universal curve when the yield and the energy are normalized to $\delta_{\max}$ and $E_{\max}$, respectively.} through the maximum yield $\delta_{\max}$ for a slab and the incident energy $E_{\max}$ for $\delta_{\max}$.
The energy distribution of secondary electrons induced by bombardments of electrons may be approximated by \citep{draine-salpeter1979}
\begin{eqnarray}
{\rho }_\mathrm{e}(\epsilon) = \frac{\epsilon}{2 \phi_\mathrm{e}^2}\left[{1 + \frac{1}{2}\left({\frac{\epsilon}{\phi_\mathrm{e}}}\right)^2}\right]^{-3/2} ,
\end{eqnarray}
with $\phi_\mathrm{e}= 2~\mathrm{eV}$.
The energy distribution of secondary electrons induced by bombardments of positive ions may be approximated by \citep{draine-salpeter1979}
\begin{eqnarray}
{\rho }_\mathrm{i}(\epsilon) = \frac{1}{\phi_\mathrm{i}}\left[{1 + \frac{1}{2}\left({\frac{\epsilon}{\phi_\mathrm{i}}}\right)^2}\right]^{-2} ,
\end{eqnarray}
with $\phi_\mathrm{i}= 1~\mathrm{eV}$.

\subsection{Thermionic emission}

The electric current due to thermionic emission, $J_\mathrm{th}$, is given by \citep{sodha1961}
\begin{eqnarray}
J_\mathrm{th} = 4 \pi e a^2 \kappa \left({\frac{4 \pi m_e k_\mathrm{B}^2}{h^3}}\right)
 \left({1+\frac{{\epsilon}_{\min}}{k_\mathrm{B} T_\mathrm{g}}}\right) T_\mathrm{g}^2 \,
\exp\left({-\frac{W+{\epsilon}_{\min}}{k_\mathrm{B} T_\mathrm{g}}}\right) ,
\end{eqnarray}
where $m_e$ is the electron mass, $\kappa$ is the material-dependent correction factor ($0 < \kappa \le 1$) and $T_\mathrm{g}$ is the equilibrium temperature of a grain.
It is common practice to compute the equilibrium temperature of a grain by solving the following equation of energy balance among the absorption of stellar radiation, thermal emission, and sublimation \citep{mukai-yamamoto1979,kimura-et-al1997}:
\begin{eqnarray}
\int_{0}^{\infty} C_\mathrm{abs}(\lambda, a) F'(\lambda) d\lambda  = 4 \pi \int_{0}^{\infty} C_\mathrm{abs}(\lambda, a) P(\lambda, T_\mathrm{g}) d\lambda - \frac{dm}{dt}L ,
\label{equilibrium-temperature}
\end{eqnarray}
where $C_\mathrm{abs}(\lambda, a)$, $F'(\lambda)$, $P(\lambda, T_\mathrm{g})$, ${dm}/{dt}$, and $L$ are the absorption cross section of the grain having radius $a $ at a wavelength of $\lambda$, the stellar radiance at a wavelength of $\lambda$, the Planck function at a wavelength of $\lambda$ and a grain temperature of $T_\mathrm{g}$, the mass loss rate due to sublimation, and the latent heat of sublimation, respectively.
The stellar photon flux $F'(\lambda)$ in the range of wavelength from $\lambda$ to $\lambda + d\lambda$ is related to $F(h\nu)$ by
\begin{eqnarray}
F'(\lambda) = \frac{\nu^2}{c} F(h\nu) .
\end{eqnarray}
The mass loss rate ${dm}/{dt}$ of a grain is given by \citep{mukai-yamamoto1979,kimura-et-al1997}
\begin{eqnarray}
\frac{dm}{dt} = -4 \pi a^2 \sqrt{\frac{M u}{2 \pi k_\mathrm{B} T_\mathrm{g}}} p_\infty \exp \left({-\frac{M u}{k_\mathrm{B} T_\mathrm{g}}L}\right),
\end{eqnarray}
where $u$, $M$, $p_\infty$ are the unified atomic mass unit, the molecular weight of grain substance, and the vapor pressure at the utmost limit of grain temperatures.

\section{Grain dynamics}

Since classical grain dynamics models fail to account for the enormous accumulation of hot dust in the vicinity of central stars, we focus on the magnetic grain trapping scenario, regardless of dynamical mechanism for the injection of nanoparticles into near-stellar environments.
The interaction between moving grain charges and stellar magnetic fields might play a vital role in the dynamics of dust grains near the sublimation zone, in particular, if the charge-to-mass ratio $|Q/m|$ of a dust grain increases with decreasing grain size due to sublimation.
Note that the value of $|Q/m|$ is a decisive parameter to describe the importance of stellar magnetic fields compared with stellar gravitational fields.
Because the grain charge and mass are given by $Q = 4 \pi \varepsilon_0 a U_\mathrm{g}$ and $m = (4/3) \pi a^3 \rho$, respectively, we may express the charge-to-mass ratio in terms of the surface potential per unit area:
\begin{eqnarray}
\frac{Q}{m} = \frac{3 \varepsilon_0}{\rho} \left({\frac{U_\mathrm{g}}{a^{2}}}\right) .
\end{eqnarray}

\subsection{The influence of a magnetic field}

If the stellar magnetic field is strong enough to trap a dust grain with the surface potential $U_\mathrm{g}$, then the grain gyrates with a period $\tau_\mathrm{g}$ given by 
\begin{eqnarray}
\tau_\mathrm{g} = \left|{\frac{2 \pi a^2 \rho}{3 \varepsilon_0 U_\mathrm{g} B \sin\Theta}}\right| ,
\label{gyration-timescale}
\end{eqnarray}
where $B$ is the strength of the magnetic field and $\Theta$ is an angle between the velocity vector and the magnetic field vector.
To evaluate the maximum effect of stellar magnetic field on the dynamics of hot grains in the vicinity of central stars, we hereafter adopt $\Theta = 90\deg$, indicating that the magnetic field is perpendicular to the orbital plane of the grains.
Eq.~(\ref{gyration-timescale}) implies that the gyration period $\tau_\mathrm{g}$ decreases with decreasing the surface area of the grain due to sublimation, if the surface potential $U_\mathrm{g}$ is independent of grain size. 
 
\subsection{The influence of sublimation}

In case that the mass losses of dust grains due to sublimation are crucial to their survival in the sublimation zone, magnetic trapping might not help prolonging the life of the grains in the vicinity of a star.
Therefore, it is worth comparing the period of gyration with the timescale for sublimation in order to clarify whether the grain dynamics is dominated by magnetic trapping or sublimation.
The characteristic timescale $\tau_\mathrm{s}$ for sublimation is given by \citep{kimura-et-al2002}
\begin{eqnarray}
\tau_\mathrm{s} = \left|{\frac{m}{dm/dt}}\right| ,
\label{sublimation-timescale}
\end{eqnarray}
which might impose restrictions on the dwell time for magnetic trapping.
Note that Eq.~(\ref{sublimation-timescale}) underestimates the characteristic timescale for sublimation if there is no influence of stellar magnetic field upon the dynamics of dust grains.
Therefore, we focus our discussion on the influence of sublimation under the assumption that the magnetic trapping of dust grains takes place in the near-stellar environments.
When the sublimation timescale $\tau_\mathrm{s}$ reaches $10^2$--$10^3$ times the orbital period of the particles, sublimation starts to influence the dynamics of dust particles significantly \citep{mukai-yamamoto1979}.
Accordingly, we do not consider the influence of stellar radiation pressure on the dynamics of dust particles inside the sublimation zone nor the influence of the Poynting-Robertson effect, the timescale of which is much longer than the orbital period.

\section{Grain environments}

To scrutinize the plausibility of magnetic grain trapping, we shall first compile data for stellar and stellar-wind parameters by restricting the data to main-sequence stars, for which the detection or non-detection of hot grains was reported.

\subsection{Stellar parameters}
\label{stellar}

The rotation velocities $V_{\star}$ of main-sequence stars are often available in the literature, otherwise, in the form of $V_{\star} \sin i$ where $i$ denotes the inclination of the stellar rotation axis with respect to the line of sight.
Accordingly, only if there is a lack of $V_{\star}$ in the literature and the inclination $i_\mathrm{d}$ of a debris disk has been estimated, then we assume $i = i_\mathrm{d}$ and derive $V_{\star}$ from $V_{\star} \sin i / \sin i_\mathrm{d}$ \citep[see][]{greaves-et-al2014}.
In case of no available data for $V_{\star}$ nor $i_\mathrm{d}$ in the literature, we are forced to evaluate the rotation velocities $V_{\star}$ of main-sequence stars using the following relation \citep{noyes-et-al1984}:
\begin{eqnarray}
V_{\star} = \frac{2 \pi R_\star}{\tau_\mathrm{c} Ro} ,
\label{rotation-velocity}
\end{eqnarray}
where $\tau_\mathrm{c}$ and $Ro$ are the characteristic convection turnover time and the Rossby number, respectively.
Except for A-type stars, we may utilize an empirical function for $\tau_\mathrm{c}$ \citep{noyes-et-al1984}:
\begin{eqnarray}
\log\left({\tau_\mathrm{c}/\tau_0}\right) = 
\begin{cases}
    1.362-0.166x+0.025x^2-5.323x^3 & x \ge 0 \\
    1.362-0.14x & x < 0
  \end{cases} ,
  \label{turnover-time}
\end{eqnarray}
where $\tau_0 = 1~\mathrm{day}$ and $x=1-(B-V)$ with $B-V$ being the B--V color index\footnote{For the B--V color indices of all main-sequence stars used in this study, we simply take values given in the SIMBAD database.}.
Note that this formula is in accord with an empirical function proposed for low-mass main-sequence stars by \citet{wright-et-al2011}, as long as the mass range of stars in Tables~\ref{tab:hot-grain-stars} and \ref{tab:no-hot-grain-stars} is concerned.
If the X-ray luminosity\footnote{The X-ray luminosity is typically measured in the range of photon energy from $\sim 0.2$ to $\sim 2~\mathrm{keV}$, but the exact range is slightly mission dependent.} $L_\mathrm{X}$ and the stellar luminosity $L_\star$ fulfills the condition $L_\mathrm{X} / L_\star \la {10}^{-3}$, then the Rossby number can be derived from the following empirical formula \citep{wright-drake2016}.
\begin{eqnarray}
Ro = c_0 \left({\frac{L_\mathrm{X}}{L_\star}}\right)^{-\eta} ,
\label{rossby-number}
\end{eqnarray}
with $c_0 = 1.1\times{10}^{-2}$ and $\eta = 0.37$. 

According to the magnetic trapping of dust grains proposed by \citet{rieke-et-al2016}, we may assume a dipole magnetic field \begin{eqnarray}
B=B_\star \left({\frac{r}{R_\star}}\right)^{-3} ,
\end{eqnarray}
where $B_\star$ is the strength of magnetic field at stellar surface.
The strength of magnetic field at stellar surface, $B_\star$, is often not available in the literature, but we may estimate the value of $B_\star$ using the following empirical formula \citep{babel-montmerle1997,gallet-et-al2017}:
\begin{eqnarray}
B_\star = \hat{B}_\star\left({\frac{Ro}{\hat{R}o}}\right)^{-\xi} ,
\label{magnetic-field-strength}
\end{eqnarray}
with $\hat{B}_\star = 3 \times {10}^{-2}~\mathrm{T}$, $\hat{R}o = 2.19$, and $\xi = 2$ for stars with a surface radiative zone and an effective temperature above $6600~\mathrm{K}$ and $\hat{B}_\star = 2 \times {10}^{-4}~\mathrm{T}$, $\hat{R}o = 1.96$, and $\xi = 1$ for cooler stars with a surface convection layer.
It is worthwhile to note that Eqs.~(\ref{rossby-number}) and (\ref{magnetic-field-strength}) enable us to estimate the X-ray luminosity $L_\mathrm{X}$ and also, excluding A-type stars, the rotation velocity through Eq.~(\ref{rotation-velocity}) with the help of Eq.~(\ref{turnover-time}) and the same with the converse.
For the sake of simplicity, we assume that the magnetic dipole of a central star is not tilted with respect to the rotation axis of the star and thus the orbital plane of dust particles is perpendicular to the rotation and dipole axes.

Tables~\ref{tab:hot-grain-stars} and \ref{tab:no-hot-grain-stars} list the radius $R_{\star}$, mass $M_{\star}$, luminosity $L_{\star}$, rotation velocity $V_{\star}$, X-ray luminosity $L_\mathrm{X}$, and magnetic field strength $B_\star$ of main-sequence stars with the detection and non-detection of hot grains, respectively \citep{absil-et-al2009,absil-et-al2013,ertel-et-al2014,ertel-et-al2016,nunez-et-al2017}.
The values are in principle taken from the literature in the last column of Tables~\ref{tab:hot-grain-stars} and \ref{tab:no-hot-grain-stars}, otherwise, the values of $V_{\star}$ and $B_\star$ are in part estimated as described above.
Note that an underlined value and an overlined one indicate that the values are the lower limit and the upper limit of the quantity, respectively.

\subsection{Stellar wind parameters}
\label{stellar-wind}

The bulk velocity, the spatial density, and the temperature of a stellar wind are requisites for numerical calculations of the electric current due to the exposure of dust grains in the stellar wind.
If observational data on the wind parameters are available in the literature, then we prefer to use the data, otherwise we shall derive them from our theoretical consideration.
Concerning the velocity law, which is the dependence of stellar-wind velocity on the distance from the central star, we shall assume the so-called $\beta$-law \citep[][Eq.~2.3]{lamers-cassinelli1999}. 
While stellar radiation pressure accelerates winds from hot stars rapidly, which results in $\beta\approx 0.5-0.8$, the acceleration of winds by cool stars is weaker, yielding a larger $\beta$ value \citep[e.g.,][]{castor-et-al1975,pauldrach-et-al1986,krticka-kubat2011}.
By adopting $\beta=1$, the bulk velocity $v_\mathrm{sw}$ of the stellar wind can be given by
\begin{eqnarray}
v_\mathrm{sw} \approx v_\mathrm{sw}^\infty \left[{1 - \left({\frac{r}{R_\star}}\right)^{-1}}\right],
\end{eqnarray}
where $v_\mathrm{sw}^\infty$ is the wind velocity at a great distance from the central star.
As a crude estimate, the wind velocity $v_\mathrm{sw}^\infty$ at a great distance from the central star is equal to the escape velocity:
\begin{eqnarray}
v_\mathrm{sw}^\infty \approx \sqrt{\frac{2 G M_\star}{R_\star}},
\end{eqnarray}
where $M_\star$ and $R_\star$ denote the mass and radius of the central star and $G$ is the gravitational constant.
As far as a steady state is concerned, the spatial density $n_\mathrm{sw}$ and the bulk velocity $v_\mathrm{sw}$ of the stellar wind are related to the stellar mass-loss rate, $\dot{M}_\star$, as 
$\dot{M}_\star = 4 \pi r^2 \mu m_\mathrm{u} n_\mathrm{sw} v_\mathrm{sw}$,
where $\mu$, $m_\mathrm{u}$, and $r$ are the mean molecular weight of stellar wind particles, the atomic mass unit, and the distance from the center of the star, respectively.
Therefore, if the stellar mass-loss rate has been derived from measurements, then we may estimate the spatial density $n_\mathrm{sw}$ of the stellar wind as
\begin{eqnarray}
n_\mathrm{sw} = \frac{\dot{M}_\star}{4 \pi r^2 \mu m_\mathrm{u} v_\mathrm{sw}} .
\end{eqnarray}
In case of no observational data on $\dot{M}_\star$, we approximate the stellar mass-loss rate using the following formula proposed by \citet{andriesse1979,andriesse2000} who states its typical accuracy within a factor of two for main-sequence stars:
\begin{eqnarray}
\dot{M}_\star = \left({\frac{L_\star^6 R_\star^9}{G^7 M_\star^9}}\right)^{1/4},
\label{mass-loss-rate}
\end{eqnarray}
where $L_\star$ is the stellar luminosity.
Note that the mass-loss rate given by Eq.~(\ref{mass-loss-rate}) does not deviate from the one given by the commonly-used Reimers' law for late-type giants and supergiants and its extension to cooler stars \citep[cf.][]{schroeder-cuntz2005,suzuki2007}. 
However, we are aware that Eq.~(\ref{mass-loss-rate}) does not account for the evolution of stellar mass-loss rates observed for solar-type stars and thus might significantly overestimate or underestimate the mass-loss rates \citep[cf.][]{wood-et-al2002,wood-et-al2005}.
If the coronal temperature $T_\mathrm{cor}$ is available in the literature, we estimate the stellar-wind temperature $T_\mathrm{sw}$ as
\begin{eqnarray}
T_\mathrm{sw} \approx T_\mathrm{cor} \left({\frac{r}{R_\star}}\right)^{-0.5} ,
\end{eqnarray}
although the radial dependence of $T_\mathrm{sw}$ may slightly differ from star to star depending on the density. 
In the absence of observational data on $T_\mathrm{cor}$, the coronal temperature $T_\mathrm{cor}$ can be determined from the following empirical relation with the X-ray surface flux $L_\mathrm{X}/(4 \pi R_\star^2)$ for low-mass main-sequence stars \citep{telleschi-et-al2005,johnstone-guedel2015}
\begin{eqnarray}
T_\mathrm{cor} \approx \hat{T}_\mathrm{cor} \left({\frac{L_\mathrm{X}}{\hat{L}_\mathrm{X}}}\right)^{0.26} \left({\frac{R_\star}{R_\sun}}\right)^{-0.52},
\end{eqnarray}
where $\hat{T}_\mathrm{cor} = 1.7\times{10}^{6}~\mathrm{K}$ and $\hat{L}_\mathrm{X} = {10}^{20.37}~\mathrm{J~s^{-1}}$.

Tables~\ref{tab:hot-grain-stars} and \ref{tab:no-hot-grain-stars} include the mass-loss rate $\dot{M}_\star$, coronal temperature $T_\mathrm{cor}$, and stellar-wind velocity $v_\mathrm{sw}^\infty$ at a great distance from central stars with the detection and non-detection of hot grains, respectively.
Note that the X-ray luminosity $L_\mathrm{X}$ for A0--A5 stars might be caused by low-mass companions, typically K- or M-dwarfs, and thus the values of $T_\mathrm{cor}$ for A0--A5 stars should be used with caution.
Nevertheless, the weak dependence of $T_\mathrm{cor}$ on $L_\mathrm{X}$ assures that the values of $T_\mathrm{cor}$ are accurate within a factor of two, even if the values of $L_\mathrm{X}$ deviate from the true values by one order of magnitude.

\section{Results}

\subsection{Equilibrium temperature}

\begin{figure}
\plottwo{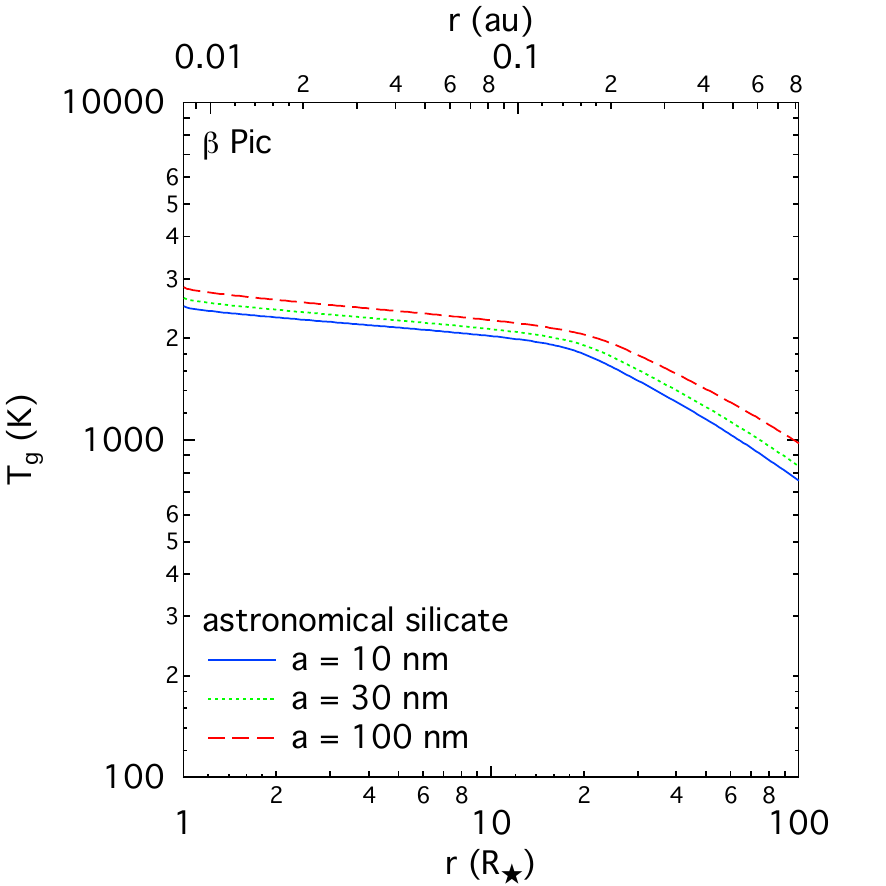}{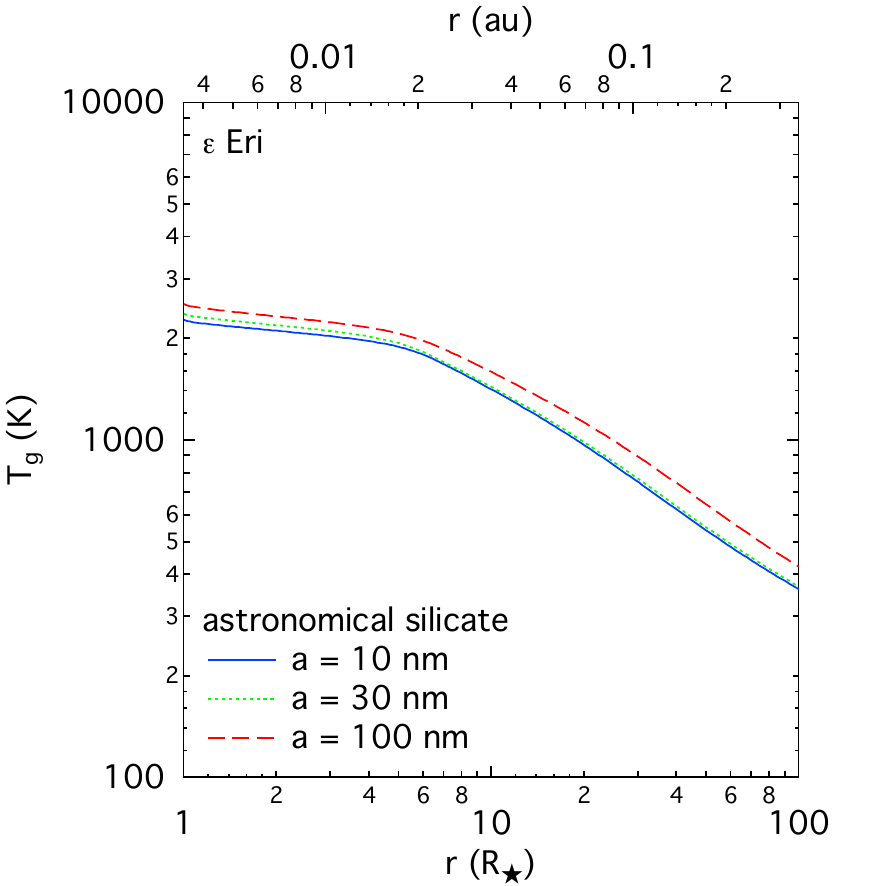}\\
\plottwo{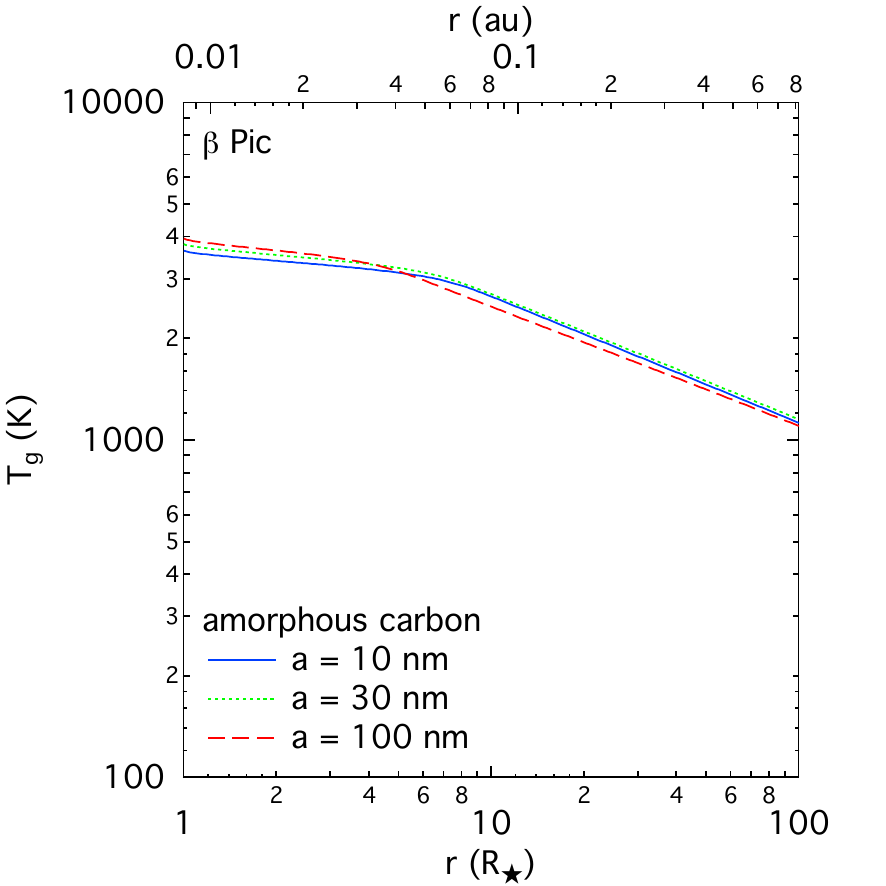}{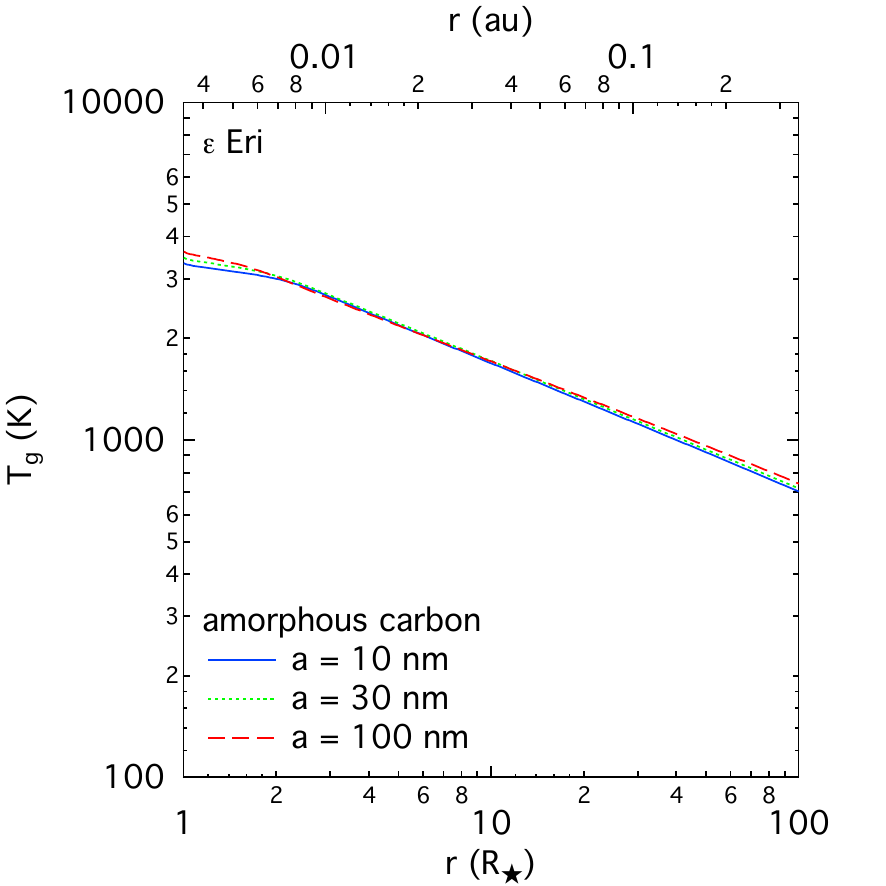}
\caption{Equilibrium temperature of dust grains consisting of the so-called ``astronomical silicate'' (upper panel) and amorphous carbon ``BE1'' (lower panel) within $100~R_\star$ from the A6V star $\beta$ Pictoris (left) and the K2V star $\epsilon$ Eridani (right). Solid line: grain radius $a = 10~\mathrm{nm}$; dotted line: $a = 30~\mathrm{nm}$; dashed line: $a = 100~\mathrm{nm}$.
\label{fig:1}}
\end{figure}

The temperature $T_\mathrm{g}$ of dust grains is an essential physical quantity to correctly understand grain charging and dynamics in the sublimation zone through thermionic emission and mass loss.
Figure~\ref{fig:1} shows the equilibrium temperatures of dust grains around $\beta$~Pic (A6V) and $\epsilon$~Eri (K2V) within $r = 100~R_\star$ from the stars (see Appendix~\ref{appendix-a} for their spectra).
Carbon grains attain higher temperatures than silicate grains when compared at the same distance from the central star, owing to the high absorptivity of carbon grains.
A kink in the radial distribution of temperatures determines the outer edge of sublimation zone, inside which hot dust grains suffer from intense sublimation.
Dust grains around $\epsilon$~Eri sublimate at smaller distances from the central star than those around $\beta$~Pic, because the former have lower temperatures compared to the latter.

\subsection{Equilibrium surface potential}

\begin{figure}
\plottwo{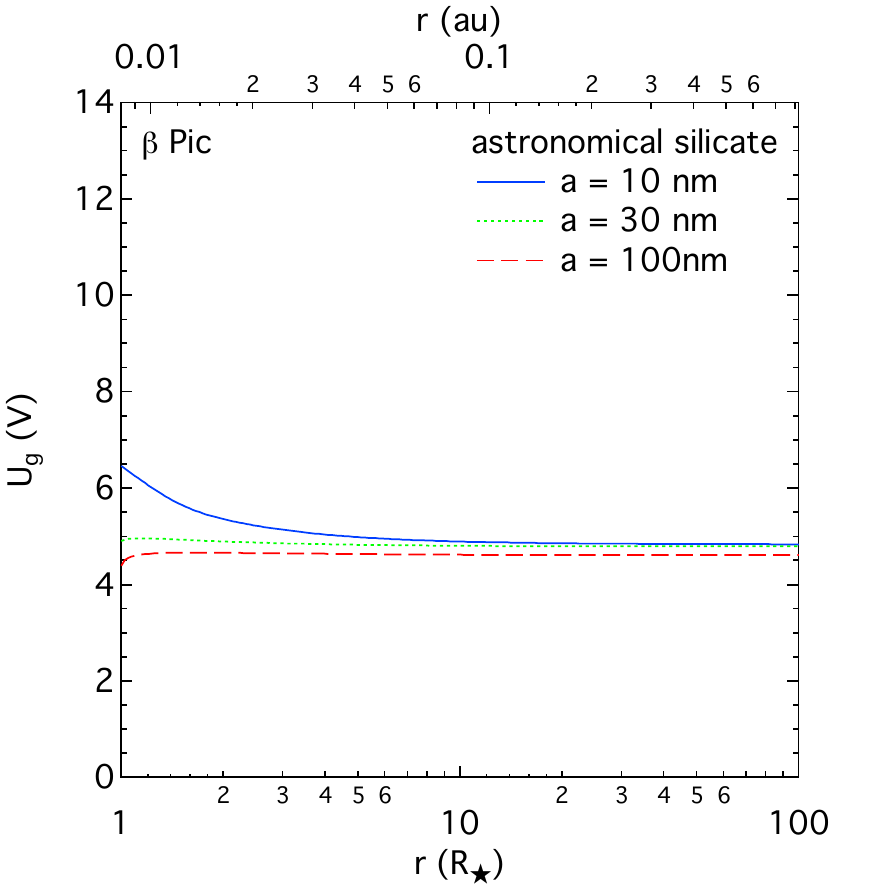}{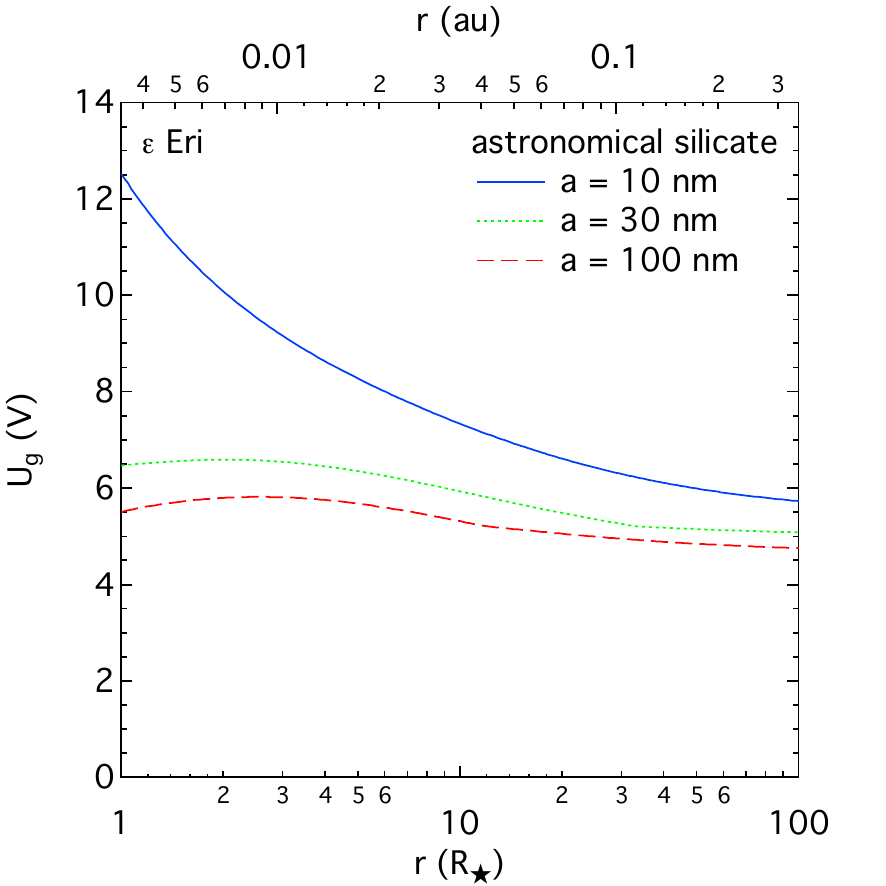}\\
\plottwo{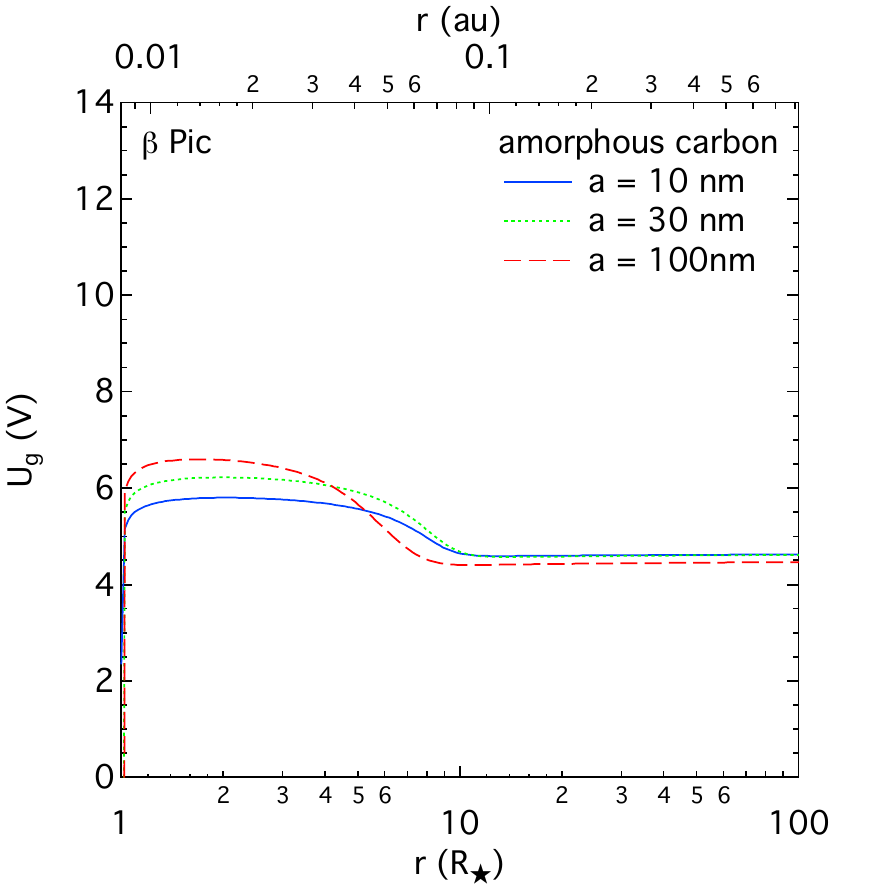}{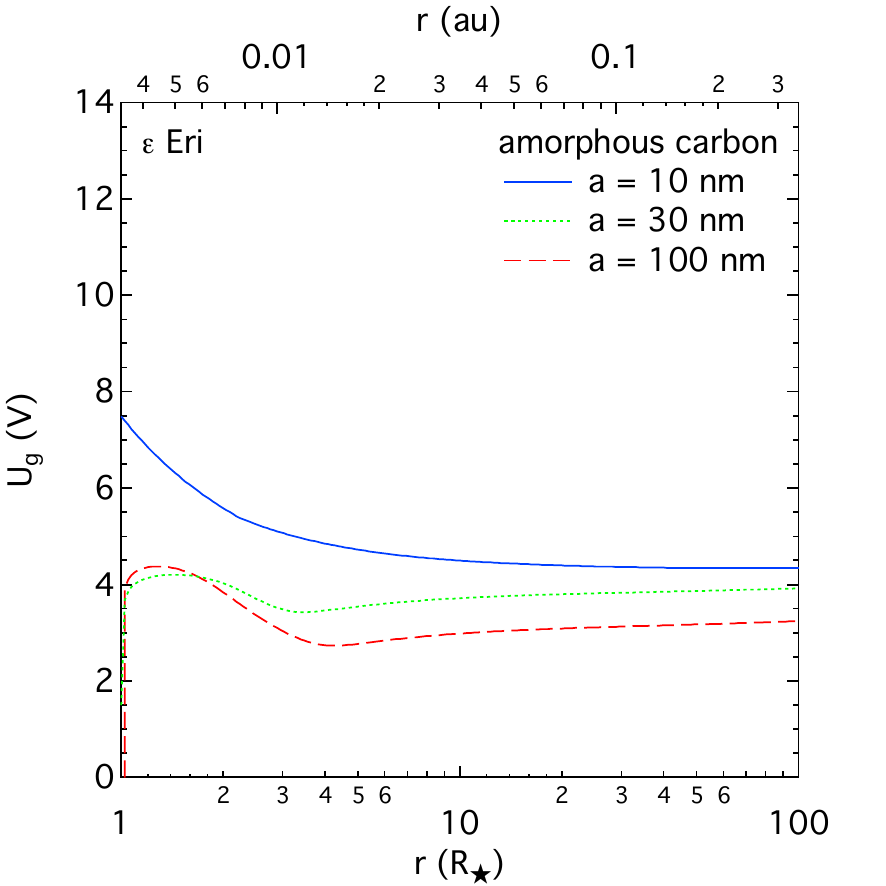}
\caption{Equilibrium surface potential of dust grains consisting of astronomical silicate (upper panel) and amorphous carbon BE1 (lower panel) in the vicinity of $\beta$ Pictoris (left) and $\epsilon$ Eridani (right). Solid line: grain radius $a = 10~\mathrm{nm}$; dotted line: $a = 30~\mathrm{nm}$; dashed line: $a = 100~\mathrm{nm}$. 
\label{fig:2}}
\end{figure}

Following \citet{rieke-et-al2016}, we consider a dust grain launched at a periapsis (i.e., $\mathbf{v_\mathrm{sw}}\perp\mathbf{v_\mathrm{g}}$) and a velocity $v_\mathrm{g} = \sqrt{G M_\star / r}$ with respect to the central star.
Figure~\ref{fig:2} depicts the equilibrium surface potential $U_\mathrm{g}$ of dust grains as a function of radial distance from the central star.
Our results reveal that a typical value of the equilibrium surface potential is $U_\mathrm{g} \approx 4$--$5~\mathrm{V}$, irrespective of stellar parameters and stellar-wind parameters.
This can be ascribed to the condition that the grain charging is dominated by the electric current due to photoelectron emission, in which the highest energy of photoelectrons limits the surface potential of the grains.
The potentials for small grains of $a = 10~\mathrm{nm}$, in particular, near $\epsilon$~Eri tend to rise in the vicinity of the stars where the secondary electron emission dominates over the photoelectron emission.
It turns out that thermionic emission is negligible for electric charging of silicate grains, while it gives a noticeable contribution to electric charging of carbon grains at the outer edge of sublimation zone.
The positive values of surface potentials imply that the electric currents due to plasma impingement and penetration are lower than the current of photoelectrons (see Appendix~\ref{appendix-b}).

\subsection{Charge-to-mass ratio}

\begin{figure}
\plottwo{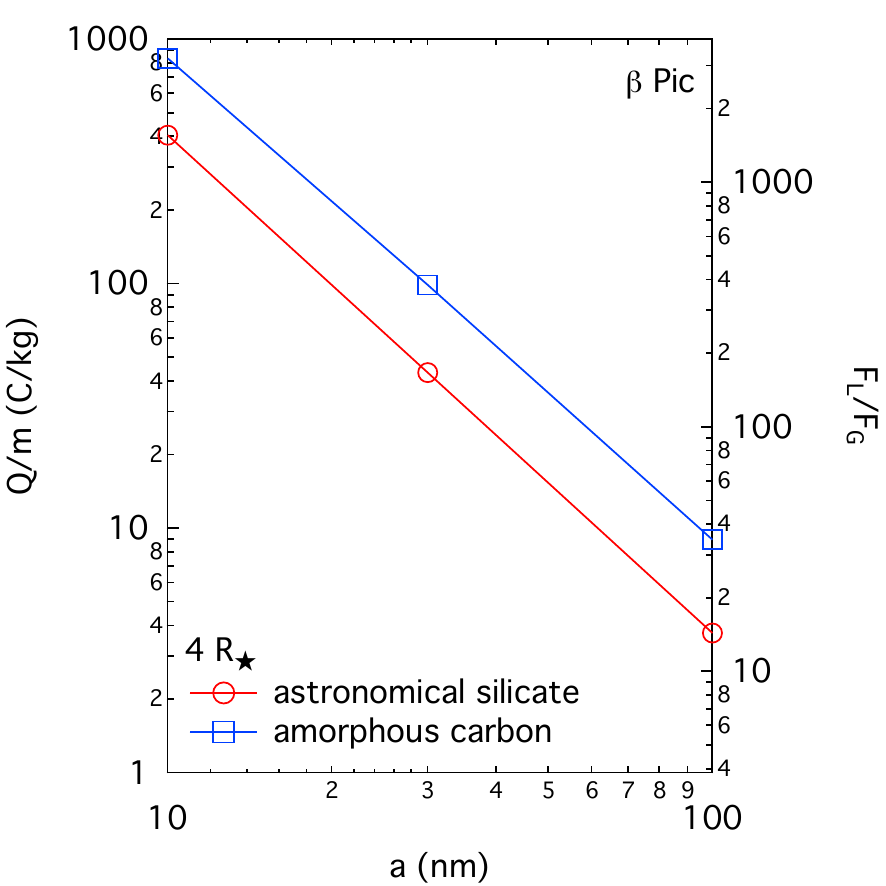}{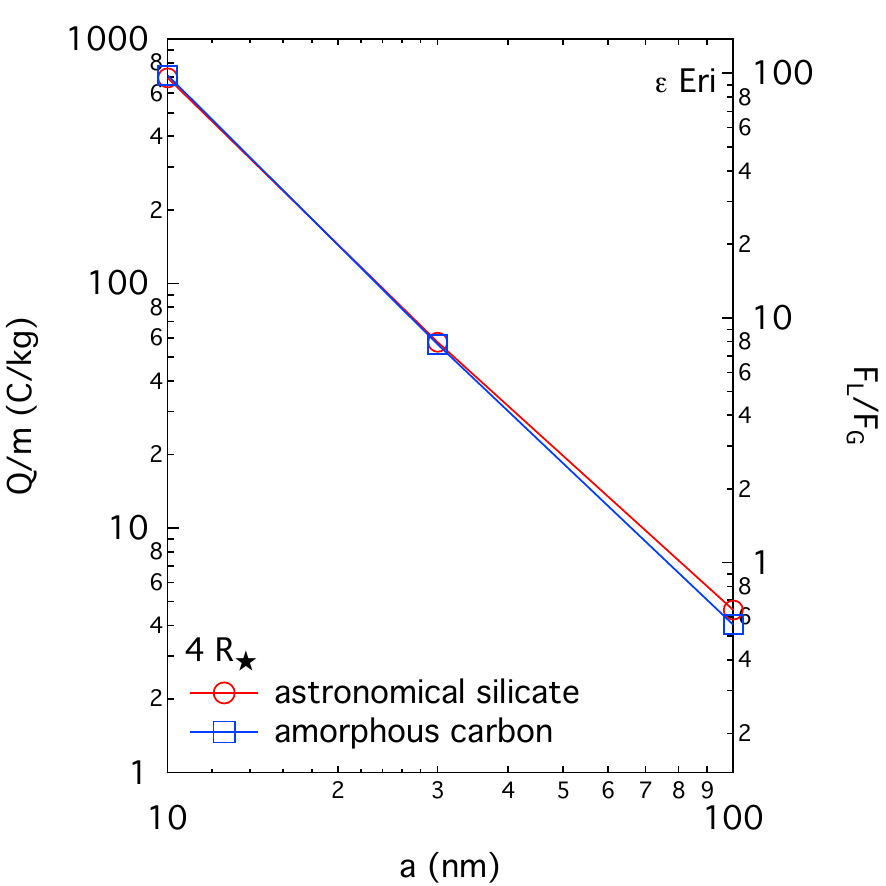}\\
\plottwo{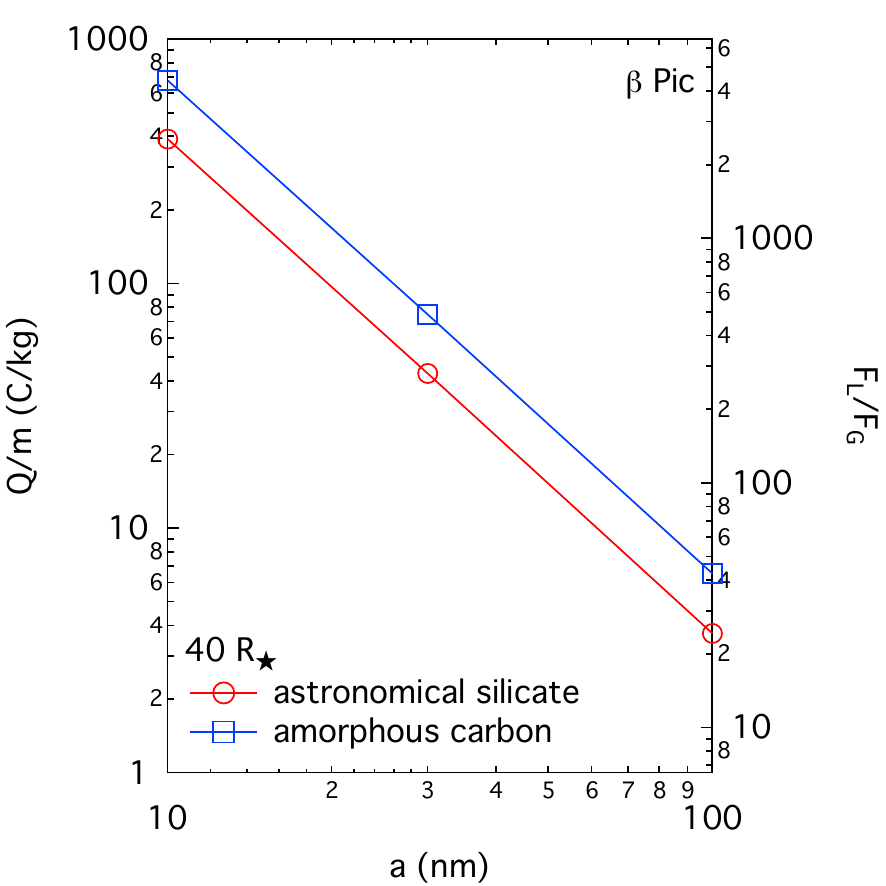}{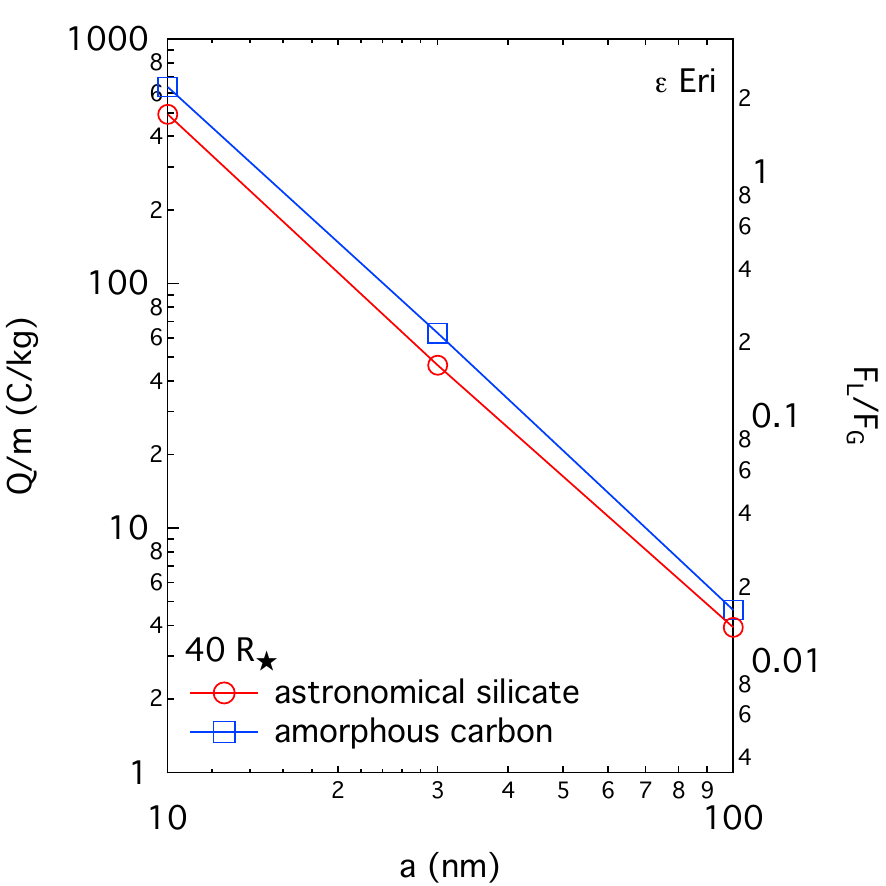}
\caption{The charge-to-mass ratio $Q/m$ of dust grains inside the sublimation zone at $4~R_\star$ (upper panel) and outside the sublimation zone at $40~R_\star$ (lower panel) from $\beta$ Pictoris (left) and $\epsilon$ Eridani (right). Open circles: dust grains consisting of astronomical silicate; open squares: dust grains of amorphous carbon BE1. The ratio of Lorentz force $F_\mathrm{L}$ to gravitational force $F_\mathrm{G}$ acting on the grains is given in the right vertical axis.
\label{fig:3}}
\end{figure}

Figure~\ref{fig:3} shows the charge-to-mass ratio $Q/m$ along the left vertical axis and the ratio of Lorentz force $F_\mathrm{L}$ to gravitational force $F_\mathrm{G}$ along the right axis as a function of grain radius $a$.
The upper and lower panels depicts the results inside the sublimation zone at $4~R_\star$ and outside the sublimation zone $40~R_\star$, respectively, from the center of $\beta$~Pic (left panel) and $\epsilon$~Eri (right panel).
The values for $Q/m$ and $F_\mathrm{L}/F_\mathrm{G}$ of dust grains consisting of astronomical silicate and amorphous carbon BE1 are given by open circles and open squares, respectively.
It is clear that the charge-to-mass ratio $Q/m$ and the ratio $F_\mathrm{L}/F_\mathrm{G}$ of Lorentz force to gravitational force are approximately proportional to $a^{-2}$, irrespective of grain material and stellar parameters.
Dust grains composed of amorphous carbon BE1 have slightly larger values of $Q/m$ and $F_\mathrm{L}/F_\mathrm{G}$ than those composed of astronomical silicate, but the differences are only a factor of two at most.
Because the $F_\mathrm{L}/F_\mathrm{G}$ values exceed 10 both inside and outside the sublimation zone around $\beta$~Pic, the Lorentz force plays a significant role in the dynamics of dust grains having radius $a = 10$--$100~\mathrm{nm}$ around $\beta$~Pic.
In contrast, the ratio of $F_\mathrm{L}/F_\mathrm{G}$ for dust grains inside the sublimation zone at $4~R_\star$ around $\epsilon~\mathrm{Eri}$ exceeds unity for $a \la 80~\mathrm{nm}$, but that for dust grains outside the sublimation zone at $40~R_\star$ around $\epsilon~\mathrm{Eri}$ is smaller than unity for $a \ga 20~\mathrm{nm}$.

\subsection{Period of gyration}

\begin{figure}
\plottwo{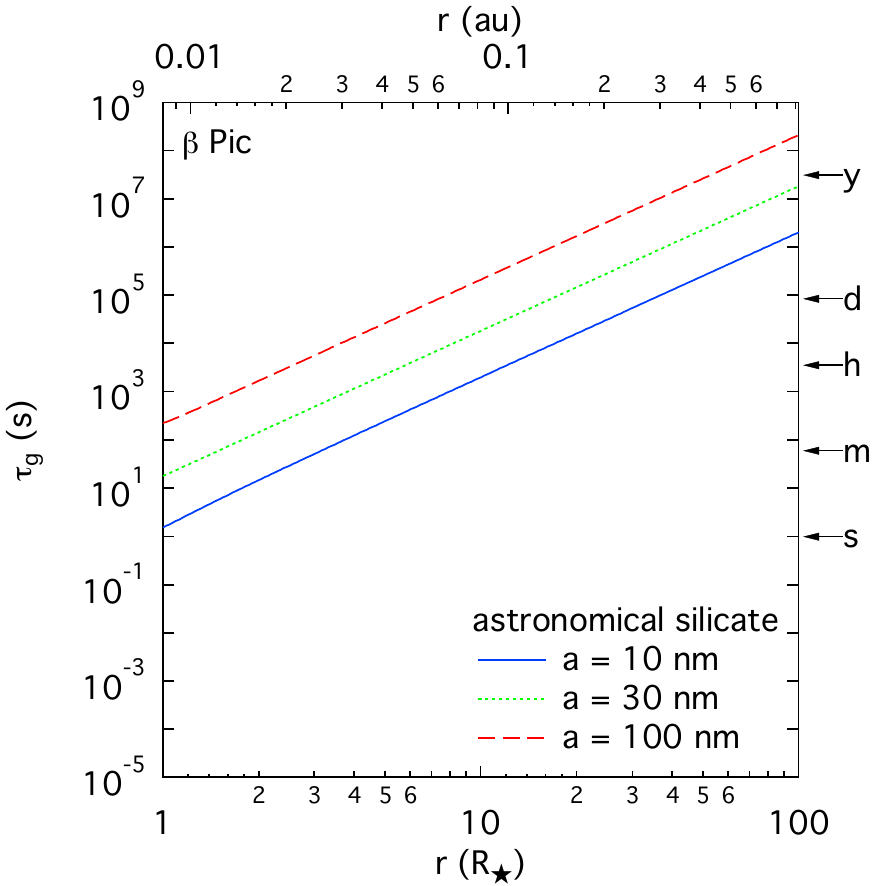}{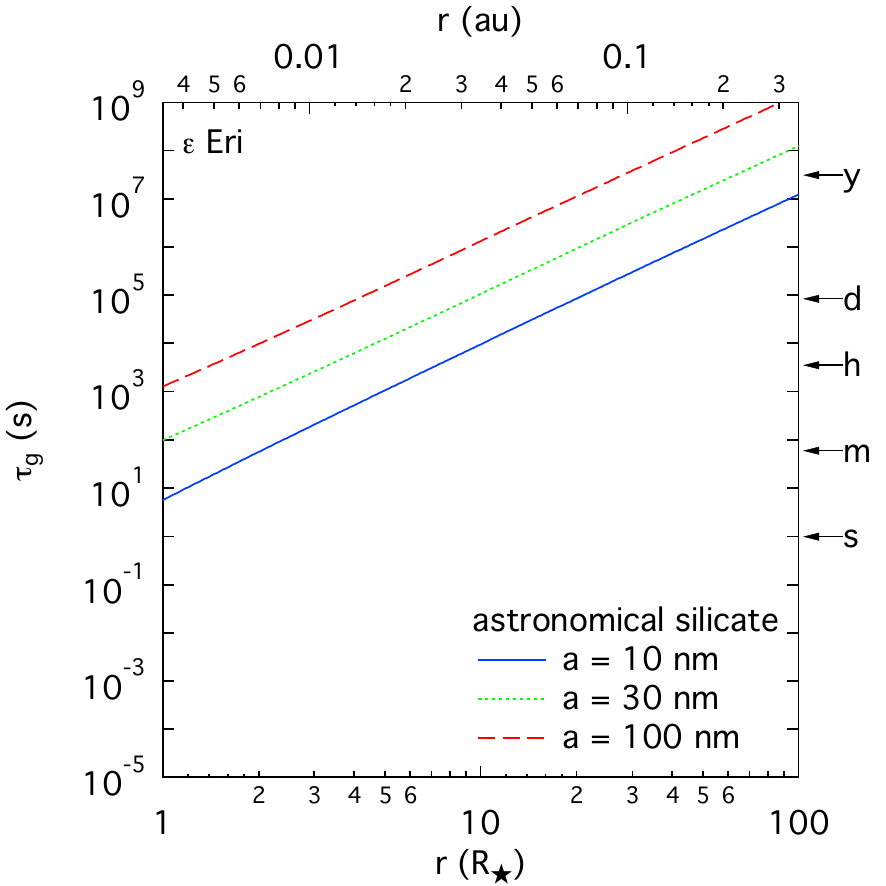}\\
\plottwo{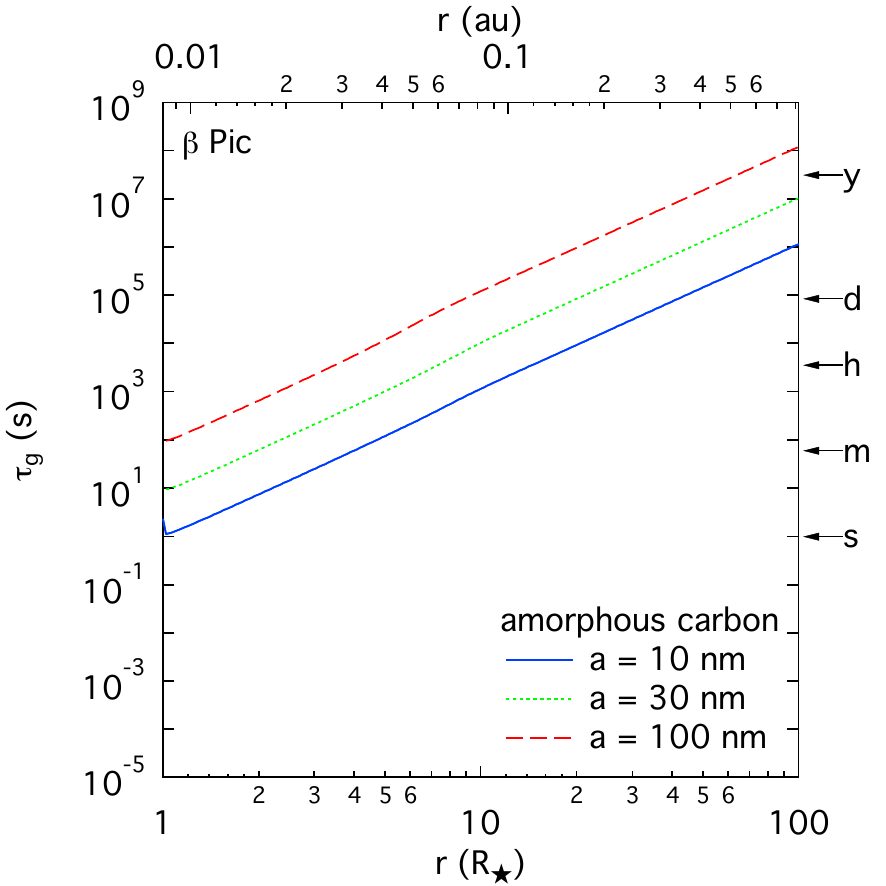}{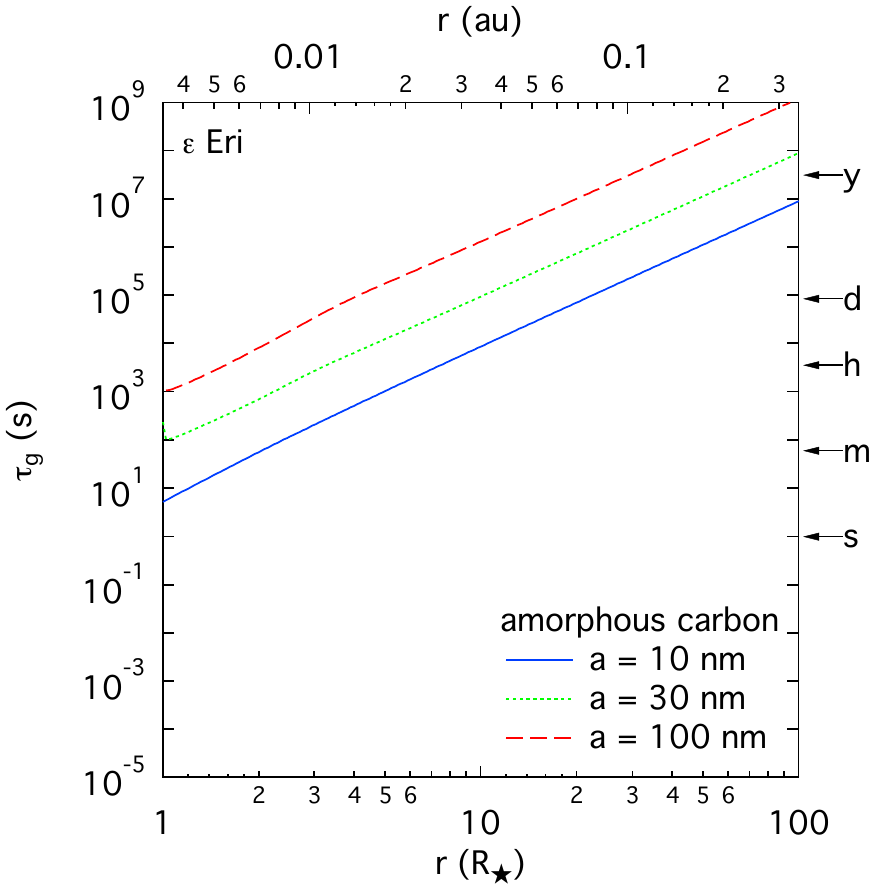}
\caption{Periods for gyration of grains consisting of astronomical silicate (upper panel) and amorphous carbon BE1 (lower panel) in magnetic fields of $\beta$ Pictoris (left) and $\epsilon$ Eridani (right). Solid line: grain radius $a = 10~\mathrm{nm}$; dotted line: $a = 30~\mathrm{nm}$; dashed line: $a = 100~\mathrm{nm}$.
\label{fig:4}}
\end{figure}

In order for magnetic trapping to have considerable influence on grain dynamics, the period $\tau_\mathrm{g}$ of circular motion must be significantly shorter than the timescale $\tau_\mathrm{s}$ for sublimation.
Figure~\ref{fig:4} shows the period $\tau_\mathrm{g}$ for the gyrating motion of dust grains in a magnetic field of $\beta$~Pic (left) and $\epsilon$~Eri (right) as a function of radial distance from the central star.
The period for gyration motion increases with distance $r$ from the central star and radius $a$ of hot grains, irrespective of grain material and stellar parameters.
As a result, the results are consistent with the gyration period that follows $\tau_\mathrm{g} \propto r^3 a^2$, which comes from $B \propto r^{-3}$ and $U_\mathrm{g} \sim \mathrm{const.}$ in Eq.~(\ref{gyration-timescale}).
The gyration period is on the order of days for grains with radius $a = 100~\mathrm{nm}$ and on the order of hours for $a = 10~\mathrm{nm}$ at $r \sim 10~R_\star$, while the period is shorter in $\beta$~Pic than in $\epsilon$~Eri, because of its inverse proportionality to the magnetic field strength.
Because sublimation reduces the radius of grains, the period for the gyrating motion, $\tau_\mathrm{g}$, drastically decreases as sublimation proceeds.
Nevertheless, as shown below, the period $\tau_\mathrm{g}$ of gyrating motion is still significantly longer than the timescale $\tau_\mathrm{s}$ for sublimation.

\subsection{Timescale for sublimation}

\begin{figure}
\plottwo{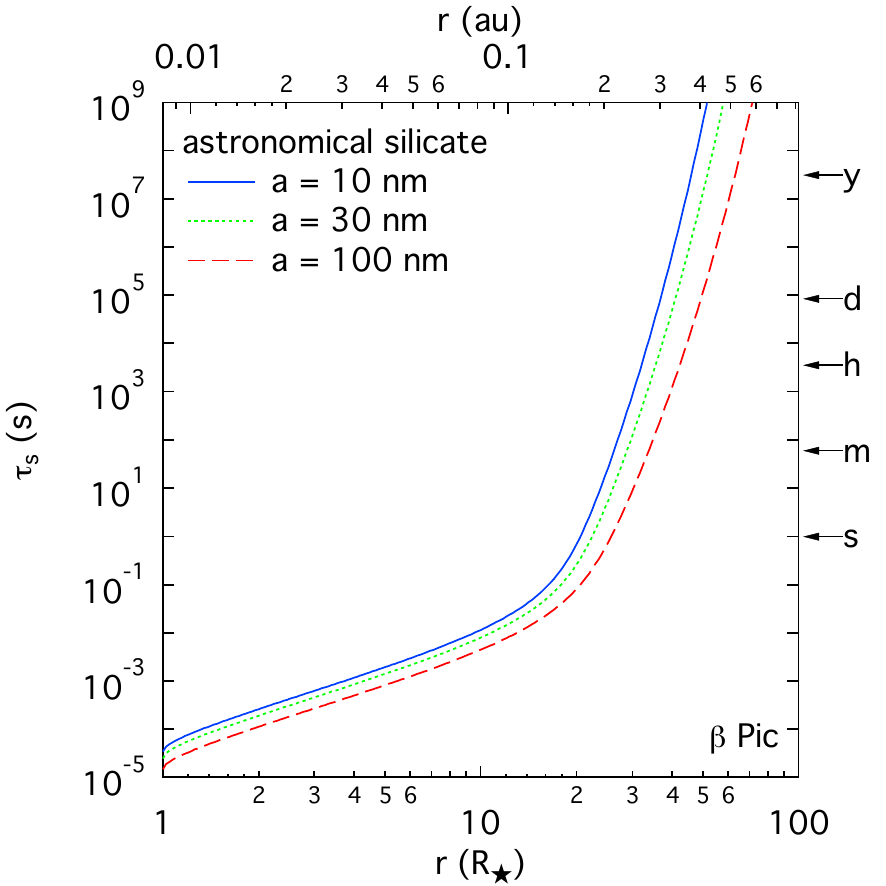}{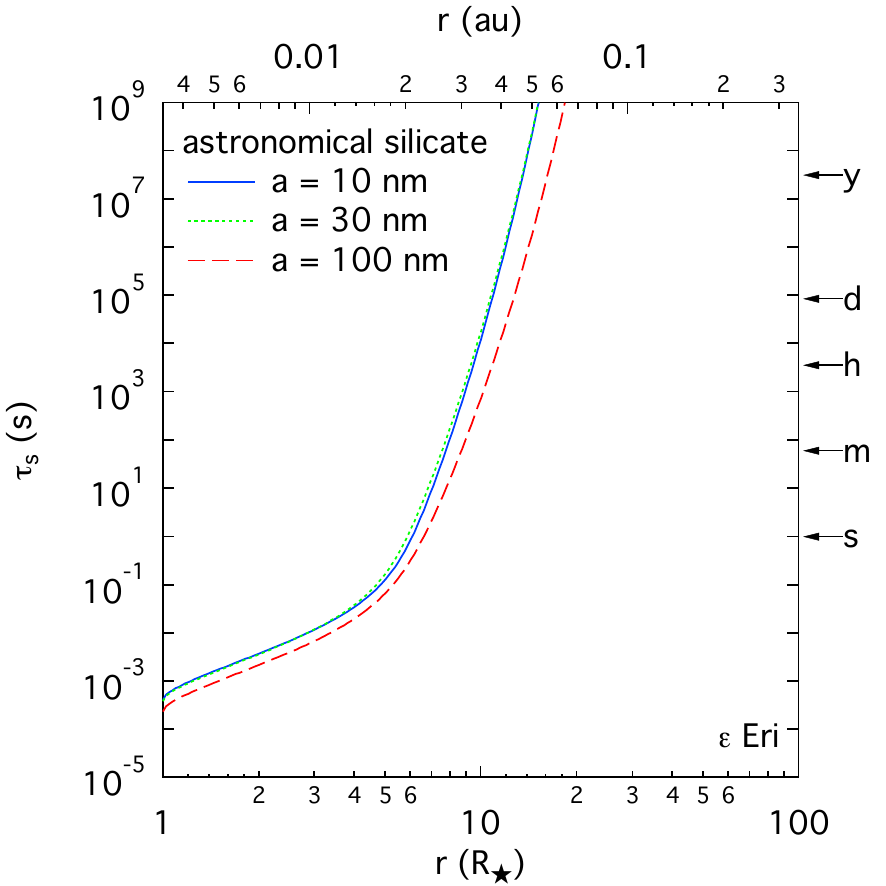}\\
\plottwo{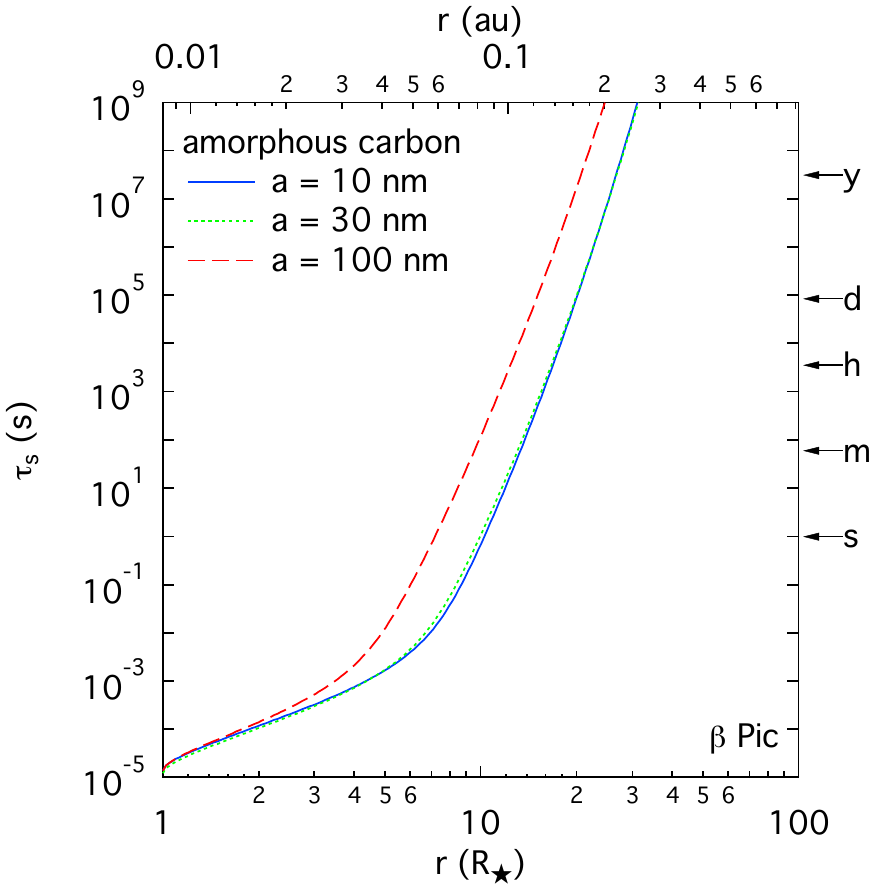}{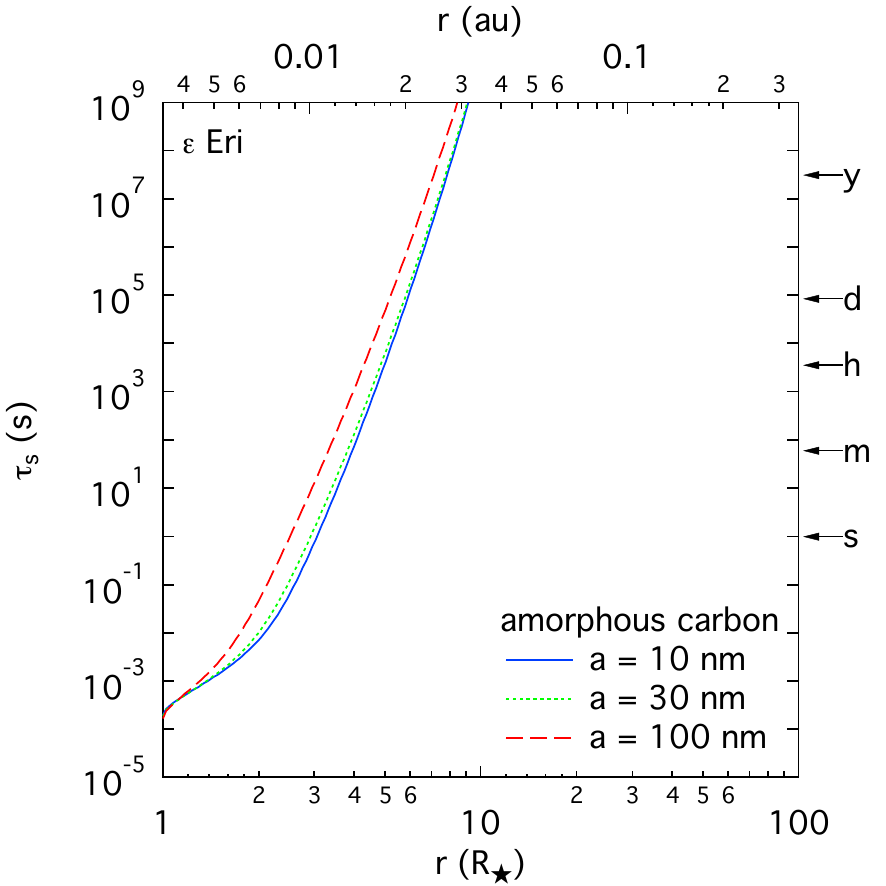}
\caption{Characteristic timescales for sublimation of grains consisting of astronomical silicate (upper panel) and amorphous carbon BE1 (lower panel) in the vicinity of $\beta$ Pictoris (left) and $\epsilon$ Eridani (right). Solid line: grain radius $a = 10~\mathrm{nm}$; dotted line: $a = 30~\mathrm{nm}$; dashed line: $a = 100~\mathrm{nm}$.
\label{fig:5}}
\end{figure}

The timescale for sublimation varies widely with the temperature of dust grains through the dependence of $dm/dt$ on the grain temperature.
Figure~\ref{fig:5} shows the characteristic timescales $\tau_\mathrm{s}$ for sublimation as a function of radial distance from the central star.
There is a kink at the outer edge of sublimation zone where the second term in the right-hand side of Eq.~(\ref{equilibrium-temperature}) becomes comparable to the first term.
Sublimation zone around $\beta$~Pic is larger than that around $\epsilon$~Eri and silicate grains sublimate at larger distances compared to carbon grains. 
Because sublimation reduces the radius of grains, the characteristic timescales $\tau_\mathrm{s}$ for sublimation tend to increase or decrease if they are composed of astronomical silicate or amorphous carbon, respectively.
The timescale for sublimation is so short inside the sublimation zone that hot dust grains cannot survive even for a second inside their sublimation zones, while it rises dramatically beyond the outer edge of sublimation zone.

\subsection{The effect of stellar rotational velocity}

Interferometric observations of debris-disk stars in the near-infrared wavelength range have revealed that the excess emission from hot dust grains appears in approximately 10--30\% of the stars \citep{absil-et-al2009,absil-et-al2013,ertel-et-al2014,nunez-et-al2017}.
\citet{rieke-et-al2016} claim that the near-infrared excesses depend on the rotational period of the central star, if the excesses are associated with the magnetic trapping of nanometer-sized dust particles. 
Therefore, we shall investigate whether the observed interferometric near-infrared excess of a debris disk around a main-sequence star correlates with the rotational velocity of the star.

\begin{figure}
\plottwo{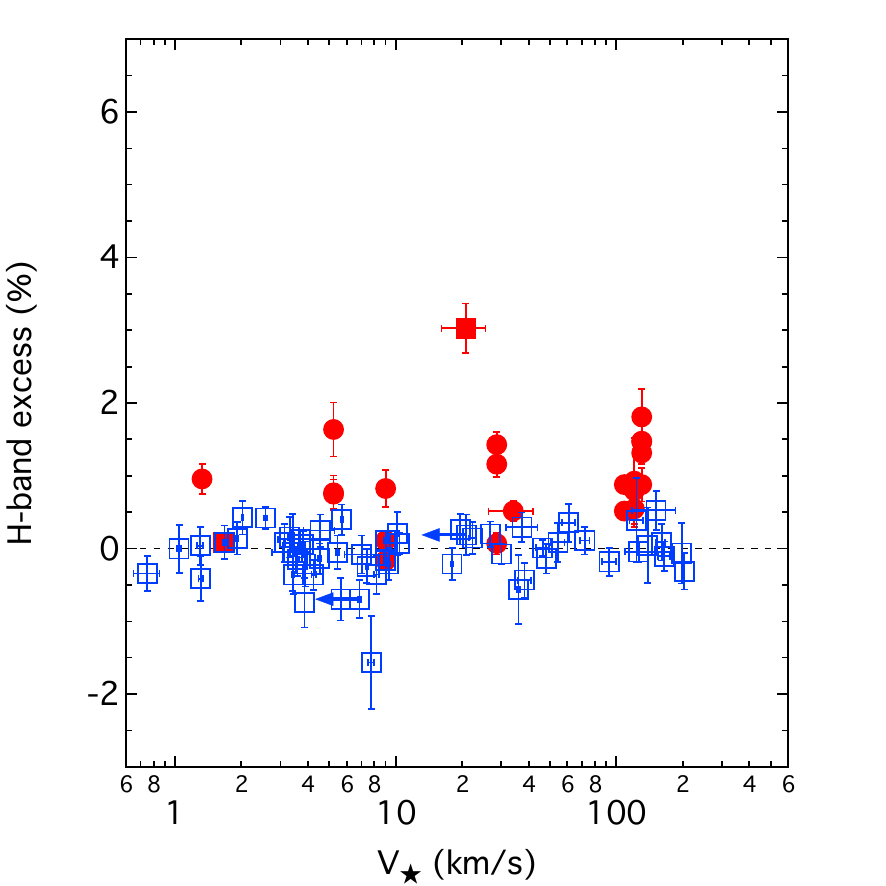}{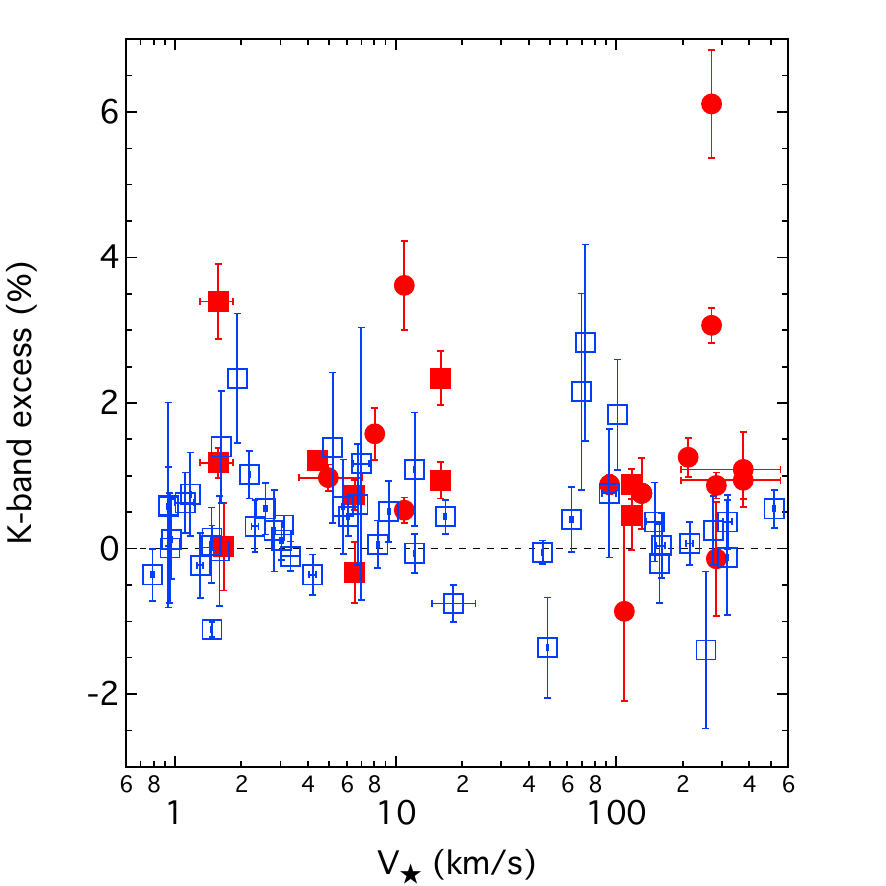}
\caption{Interferometric near-infrared excess of debris disks in the H-band (left) and the K-band (right) as a function of stellar equatorial rotation velocity. 
Red closed circles: clear detections of interferometric near-infrared excess; red closed squares: tentative detections of interferometric near-infrared excess; blue open squares: no detections of interferometric near-infrared excess.
\label{fig:6}}
\end{figure}

Figure~\ref{fig:6} depicts interferometric near-infrared excesses in the H-band (left) and the K-band (right) versus the rotation velocity of central stars.
Red closed circles, red closed squares, and blue open squares indicate stars with clear detections, with tentative detections, and without detections of near-infrared excesses in observed interferometric data, respectively.
As opposed to the hypothesis of magnetic grain trapping, we find no evidence of the relationship between the detection of hot dust grains in the vicinity of central stars and the fast rotation of the stars.

\subsection{The effect of stellar magnetic field}

\begin{figure}
\plottwo{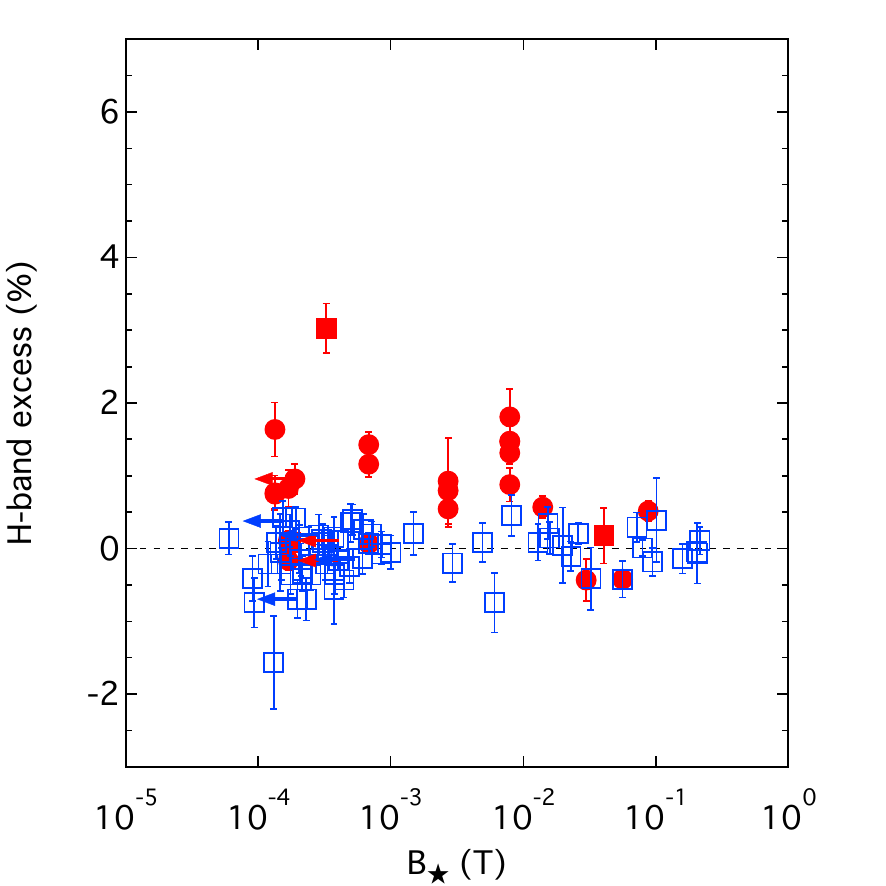}{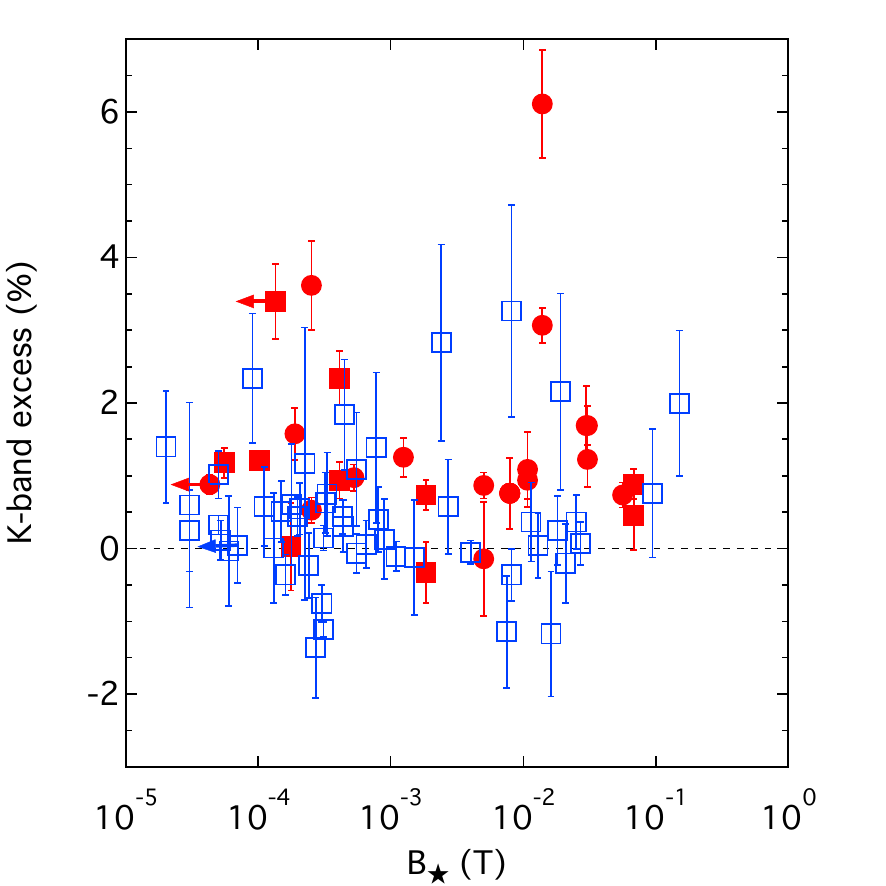}
\caption{Interferometric near-infrared excess of debris disks in the H-band (left) and the K-band (right) as a function of stellar magnetic field strength. 
Red closed circles: clear detections of interferometric near-infrared excess; red closed squares: tentative detections of interferometric near-infrared excess; blue open squares: no detections of interferometric near-infrared excess.
\label{fig:7}}
\end{figure}

According to \citet{rieke-et-al2016}, strong stellar magnetic fields do not help magnetic grain trapping, because a small gyroradius of grains due to strong magnetic fields enhances a collisional probability of the grains.
Weak stellar magnetic fields are ineffective in magnetic grain trapping either and thus magnetic grain trapping is limited to a certain range of magnetic field strength.
Figure~\ref{fig:7} presents how interferometric near-infrared excesses in the H-band (left) and the K-band (right) depend on the magnetic field strength of central stars.
As in Fig.~\ref{fig:6}, stars with clear detections, with tentative detections, and without detections of near-infrared excesses in observed interferometric data are denoted by red closed circles, red closed squares, and blue open squares, respectively.
There seems to be no correlation between the excess emission and the strength of stellar magnetic field, $B_\star$, in the H-band nor in the K-band.
Even though the product of the magnetic field strength and the grain charge is the key quantity, the weak dependence of grain charge on stellar parameters does not change the unfavorable situation to the magnetic grain-trapping scenario.

\section{Discussion}

We have revisited the magnetic grain trapping by modeling the electric charging of dust grains in the vicinity of main-sequence stars as the electric charge on the grain surface is the key parameter for the strength of magnetic grain trapping.
The model utilizes our estimates of stellar parameters and stellar-wind parameters, which are requisites for computing electric currents of various charging processes.
The most advantage of our model is that the computation of grain charges is not restricted to spectral classification of the stars nor physical properties of the grains.
Therefore, one can apply our model of grain charging to any stellar environments and grain properties, as far as main-sequence stars are concerned.
It has turned out that dust grains in exozodis are electrically charged up to $U \approx 4$--$5~\mathrm{V}$ at least for stellar environments and grain properties considered in this paper.
Although nanometer-sized grains may attain higher surface potentials in the sublimation zone, we expect that the intense electrical charging has only a limited effect on the dynamics of the grains compared to sublimation.
A typical value of $U \approx 4$--$5~\mathrm{V}$ in exozodis agrees with the surface potential of dust grains in the zodiacal cloud around the Sun \citep{mukai1981,kimura-mann1998a}.
Consequently, if only an order of magnitude estimate is required for the surface potential of grains in exozodis, it is safe to assume the surface potential of $U \approx 4$--$5~\mathrm{V}$.

Contrary to the electrostatic surface potential of $U = 40~\mathrm{V}$ assumed in the model of magnetic grain trapping by \citet{rieke-et-al2016}, we have revealed in Fig.~\ref{fig:2} that a typical value of surface potential be $U \approx 4$--$5~\mathrm{V}$ in exozodis.
This stems from the fact that the major charging process is the photoelectron emission where the maximum energy of photoelectrons determines the surface potential.
As a result, the model of magnetic grain trapping clearly overestimates the surface potential of nanoparticles in the sublimation zone by one order of magnitude.
In contrast, they assumed the strength of magnetic field at stellar surface to be $B_\star = {10}^{-4}~\mathrm{T}$, while realistic strengths of stellar magnetic fields might be one to two orders of magnitude higher than the assumed value.
Because the product of grain electric charge and magnetic field strength is the key quantity for the dynamics of hot grains in a stellar magnetic field, the motion of hot grains might be reasonably well simulated by the model of \citet{rieke-et-al2016}.
Therefore, we cannot dismiss the model of magnetic grain trapping only by our numerical estimates of electric grain charging.

We have shown that the sublimation of grains takes place in a much shorter timescale than the magnetic trapping of the grains inside the sublimation zone at least in the cases of $\beta$~Pic and $\epsilon$~Eri.
The strength of magnetic field at the surface of $\beta$~Pic has been estimated to be $B_\star \approx 8 \times {10}^{-3}~\mathrm{T}$, more than one order of magnitude larger than the value assumed by \citet{rieke-et-al2016}.
Therefore, the correction of the stellar magnetic field strength to the model of \citet{rieke-et-al2016} does not significantly reduce the period of gyration short enough to prolong the lifetime of grains trapped in the stellar magnetic field.
In consequence, the magnetic trapping of grains in the sublimation zone does not seem to be a potential mechanism to prolong the lifetimes of the grains in the near-stellar environments.

While we have assumed $B \propto r^{-3}$ according to \citet{rieke-et-al2016}, we are aware that the radial dependence of the magnetic field strength becomes weaker as the radial distance $r$ increases.
Such a variation in the radial dependence of the magnetic field strength is associated with the nature of magnetic field lines frozen in the stellar wind.
As a result, the azimuthal component of magnetic fields could predominate over the radial component in the far distance from the central star, which leads to $B \propto r^{-1}$.
This might imply that the periods for gyration motion, $\tau_\mathrm{g}$, shown in Fig.~\ref{fig:4} be significantly overestimated in the far distance.
In particular, the radial distance beyond which the azimuthal magnetic field becomes the dominant component is closer to the central star for fast rotating stars in comparison with slow rotators \citep{weber-davis1967,belcher-macgregor1976,sakurai1985}.
It should be, however, noted that the assumption of $\Theta = 90\deg$ (i.e., $\mathbf{v_\mathrm{g}} \perp \mathbf{B}$) is out of harmony with the predominance of the azimuthal component.
According to \citet{parker1958}, we may assume $\sin\Theta \approx (v_\mathrm{sw}^\infty/V_{\star}) (R_{\star}/r)$ in the far distance from the central star.
This results in $\tau_\mathrm{g} \propto r^2$ at large $r$, instead of $\tau_\mathrm{g} \propto r^3$, indicating that we might have overestimated $\tau_\mathrm{g}$ by an order of magnitude, but not more than two orders of magnitude within $r = 1~\mathrm{au}$.
Therefore, we still obtain $\tau_\mathrm{s} < \tau_\mathrm{g}$ inside the sublimation zone and thus rapid sublimation prevents the magnetic trapping of charged nanoparticles from taking place in this region.
This does not exclude the possibility of magnetic grain trapping {\it outside} the sublimation zone where the strength of stellar magnetic fields has been underestimated in the model of \citet{rieke-et-al2016}.

Even if hot dust grains accumulate outside the sublimation zone, the near-infrared excess emission from hot grains implies that the location of the grains should not be very far from the outer edge of the sublimation zone.
In case that the Lorentz force acting on dust grains is smaller the gravitational force on the particles at distances very far from the outer edge of the sublimation zone, they must first be transported to near-stellar environments by the Poynting-Robertson effect.
However, small dust grains cannot stay in bound orbits after released from their parent bodies, provided that the radiation pressure force exceeds half the gravitational force.
By studying the ratio of forces acting on nanoparticles in the magnetic and gravitational fields of a central star, we confirm that the Lorentz force dominates the dynamics of nanoparticles smaller than $a=100~\mathrm{nm}$ near the outer edge of the sublimation zone (see Fig.~\ref{fig:3}).
Therefore, the dynamics of nanoparticles, if they exist, is controlled by the magnetic field and only sublimation poses a threat to the scenario for the magnetic grain trapping.
If the mechanism of trapping hot grains in the vicinity of a central star is not associated with stellar magnetic fields, then we must anticipate that the size of hot dust grains lies in the range of $a \ga 100~\mathrm{nm}$.

One of our important results is that the observed interferometric data do not show any correlation between the detection of hot grains and the magnitude of stellar rotations nor the strength of stellar magnetic fields (Figs.~\ref{fig:6} and \ref{fig:7}).
This is clearly at odds with the proposed model of magnetic grain trapping by \citet{rieke-et-al2016} that predicts the presence of hot grains around rapidly rotating stars with intermediate magnetic fields.
It is worth noting that \citet{absil-et-al2013} suggested a possible correlation between the near-infrared excess emission and the rotation velocities of A stars, while \citet{ertel-et-al2014} claimed no correlation.
The arguments in the latter are based on the presence of fast rotators without hot grains, but a temporal variation in the detection of hot grains may weaken their arguments.
In contrast, our results reveal the presence of slow rotators with hot grains providing strong evidence for no correlation between the near-infrared excess emission and fast stellar rotation.
Our results imply that the proposed model of magnetic grain trapping needs to be rejected or at least largely modified to get rid of the dependence on the stellar rotation and magnetic field strength.
Since the magnetic grain trapping is so far the one and only mechanism proposed for prolonging the lifetime of hot grains in the vicinity of main-sequence stars, the refutal of the magnetic grain trapping leads us to the conclusion that the mechanism of trapping hot grains in the vicinity of a central star remains a mystery.

The presence of hot dust grains has been identified among approximately 10--30\% of main-sequence stars by near-infrared interferometric observations \citep{absil-et-al2009,absil-et-al2013,ertel-et-al2014,nunez-et-al2017}.
On the one hand, the detection of hot grains in the vicinity of main-sequence stars does not show clear dependences on stellar parameters nor spectral classes.
On the other hand, a temporal variation in the presence of hot grains implies that an unsuccessful detection of hot grains in a specific observation does not guarantee the absence of hot grains in other periods of time \citep{ertel-et-al2016,nunez-et-al2017}.
This does not conflict with a temporal variation in the detection of near-infrared excess around $4~R_\sun$ fron the G-type Sun, which has been attributed to the solar dust ring \citep[see][]{kimura-mann1998b}.
These could be interpreted as the fact that the presence of hot grains in the vicinity of main-sequence stars is more ubiquitous among main-sequence stars than previously thought.

\acknowledgments

We are grateful to JSPS's Grants-in-Aid for Scientific Research (KAKENHI \#23244027, \#26400230, \#15K05273, \#17H01105, \#16H02160, and \#17H01153).
We would like to thank Hiroshi Kobayashi for fruitful discussions on the dynamical behavior of hot dust grains while preparing this paper and anonymous reviewers for their useful comments.
This research has made use of the SIMBAD data base, operated at CDS, Strasbourg, France.

%






\appendix

\section{Stellar spectra}
\label{appendix-a}

\begin{figure}
\plotone{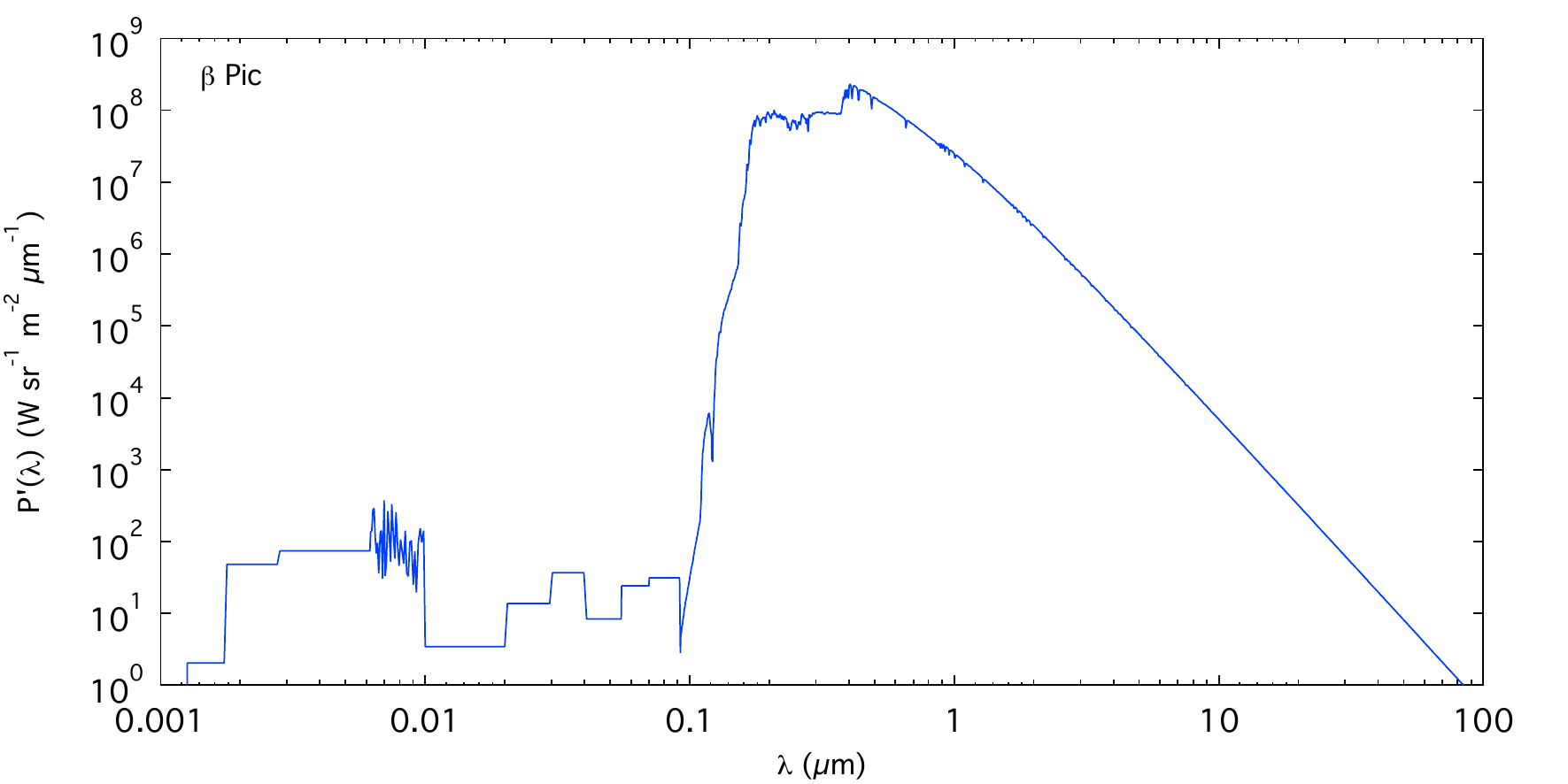}\\
\plotone{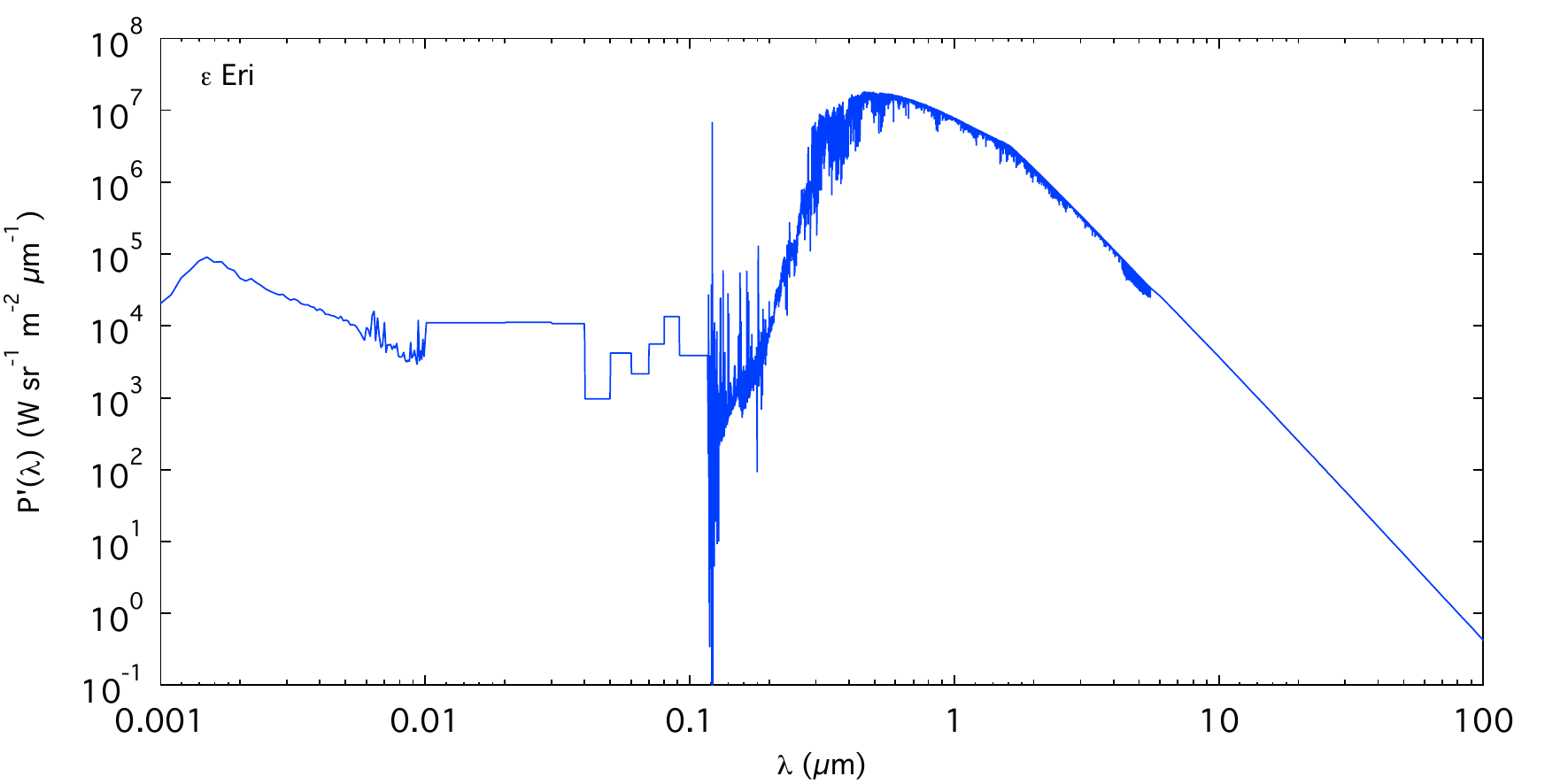}
\caption{Stellar spectra of the A6V star $\beta$ Pictoris (top) and the K2V star $\epsilon$ Eridani (bottom).
\label{fig:a1}}
\end{figure}

The stellar spectrum $P_\star (h\nu)$ from ultraviolet to infrared wavelengths is a requisite for computations of the electric currents due to photoelectric emission and thermionic emission.
We choose the K2V star $\epsilon$~Eri and the A6V star $\beta$~Pic to represent Vega-like main-sequence stars without and with the presence of hot grains, respectively.
\citet{loyd-et-al2016} constructed the stellar spectrum for M and K-type stars by combining observed data in the X-ray, empirical estimates in the extreme ultraviolet (EUV), and a model spectrum in the visible to infrared (IR) wavelength range.
We adopt the stellar spectrum for $\epsilon$~Eri given in \citet{loyd-et-al2016}, while we construct the stellar spectrum for $\beta$~Pic in a similar way as \citet{loyd-et-al2016}.
For the spectrum of $\beta$~Pic, we combine the X-ray data given in \citet{hempel-et-al2005} and given by the APEC coronal model of \citet{smith-et-al2001}, adopt a model of \citet{sanzforcada-et-al2011} in the EUV, and utilize the plasma simulation code {\sc Cloudy} (ver.~13.04), last described by \citet{ferland-et-al2013}, and ATLAS atmosphere models to describe the model spectra in the FUV to infrared (IR) wavelength range \citep{castelli-kurucz2003}.
Figure~\ref{fig:a1} shows the stellar spectra $P'_\star (\lambda)$ for $\beta$~Pic and $\epsilon$~Eri where $P'_\star (\lambda)$ is linked to $P_\star (h\nu)$ as follows:
\begin{eqnarray}
P'_\star(\lambda) = \frac{\nu^2}{c} P_\star(h\nu) .
\end{eqnarray}
The stellar spectra from visible to infrared wavelengths show less spectral lines in $\beta$~Pic compared with $\epsilon$~Eri, since the former has a higher effective temperature and thus a higher ionization degree than the latter.

\section{Electric current ratio of electron impingement to photoelectron emission}
\label{appendix-b}

To demonstrate the dominance of photoelectric emission over grain charging, we compare the electric current due to electron impingement, $J_\mathrm{e}$, and the electric current due to photoelectron emission, $J_\mathrm{ph}$.
The ratio of the currents, $J_\mathrm{e}/J_\mathrm{ph}$, for neutral (i.e., $Q=0$) dust grains at rest ($w=0$) in the reference frame of stellar wind can be written as 
\begin{eqnarray}
\frac{J_\mathrm{e}}{J_\mathrm{ph}} = {\left({\frac{R_\star}{r}}\right)}^{1/4} \left[{\frac{v_\mathrm{cor}/v_\mathrm{sw}^{\infty}}{1-\left({R_\star/r}\right)}}\right] \left({\frac{\dot{M}_\star}{4\pi {r_0}^2 \mu m_\mathrm{u}}}\right) {I_0}^{-1} ,
\label{current-ratio}
\end{eqnarray}
where $v_\mathrm{cor}$ is the mean thermal velocity of electrons in the corona:
\begin{eqnarray}
v_\mathrm{cor} = \sqrt{\frac{8 k_\mathrm{B} T_\mathrm{cor}}{\pi m_\mathrm{e}}} ,
\end{eqnarray}
with $m_\mathrm{e}$ being the mass of an electron, and $I_0$ is the photoelectron flux at $r_0 = 1~\mathrm{au}$ from the central star:
\begin{eqnarray}
I_0 = \int_{W}^{\infty }d(h\nu ) 
		{Q}_\mathrm{abs}(h\nu )\pi {\left({\frac{R_\star}{r_0}}\right)}^{2} P_\star(h\nu )Y({h\nu }) \left[{1-\left({1+\frac{h\nu - W}{\phi_\mathrm{ph}}}\right) \exp\left({-\frac{h\nu - W}{\phi_\mathrm{ph}}}\right)}\right] ,
		\label{photoelectron-flux}
\end{eqnarray}
with ${Q}_\mathrm{abs}(h\nu ) = {C}_\mathrm{abs}(h\nu )/({\pi a^2})$ being the absorption efficiency of the grains at photon energy $h\nu$.
To obtain the upper limit to the current ratio, we shall take $I_0 = 5.9 \times {10}^{13}~\mathrm{electrons~m^{-2}~s^{-1}}$, which corresponds to the photoelectron flux from lunar fine grains under solar irradiation \citep{feuerbacher-et-al1972,senshu-et-al2015}.
\citet{feuerbacher-et-al1972} noted that photoelectrons emitted from a fine grain are most likely reabsorbed by the other grains, because the photoelectron flux from lunar fine grains is one order of magnitude lower than the fluxes from solid surfaces, regardless of the surface materials \citep[cf.][]{feuerbacher-fitton1972}. 
In addition, the photoelectric quantum yield $Y(h\nu)$ for dust grains with $a=100~\mathrm{nm}$ or smaller is higher than the yield for a bulk surface of the same material \citep{kimura2016}.
Therefore, we may expect that the photoelectron flux of single small grains are significantly higher than $I_0 = 5.9 \times {10}^{13}~\mathrm{electrons~m^{-2}~s^{-1}}$.
If the solar values are inserted into Eq.~(\ref{current-ratio}), we obtain 
\begin{eqnarray}
\frac{J_\mathrm{e}}{J_\mathrm{ph}} &\approx& 0.23\,{\left({\frac{r}{1~\mathrm{au}}}\right)}^{-1/4} {\left[{1-4.65\times{10}^{-3}{\left({\frac{r}{1~\mathrm{au}}}\right)}^{-1}}\right]}^{-1} {\left({\frac{T_\mathrm{cor}}{1.6~\mathrm{MK}}}\right)}^{1/2} \left({\frac{v_\mathrm{sw}^{\infty}}{448~\mathrm{km~s^{-1}}}}\right)^{-1} \left({\frac{\dot{M}_\star}{2.5\times{10}^{-14}~M_\sun~\mathrm{yr^{-1}}}}\right) \nonumber \\ 
&& \times {\left(\frac{I_0}{5.9 \times {10}^{13}~\mathrm{electrons~m^{-2}~s^{-1}}}\right)}^{-1} .
\end{eqnarray}
Namely, the condition of $J_\mathrm{e}/J_\mathrm{ph} < 1$ is fulfilled except for the sublimation zone at $r < 3~R_\sun$, indicating that photoelectron emission is indeed the dominant charging process.
We confirm $J_\mathrm{e}/J_\mathrm{ph} < 1$ by the current ratios $J_\mathrm{e}/J_\mathrm{ph}$ numerically computed from Eqs.~(\ref{Jph}) and (\ref{Jimp}) not only around the Sun but also around $\beta$~Pic and $\epsilon$~Eri, except for carbon grains in the immediate vicinity of $\epsilon$~Eri at $r < 6~R_\star$.




\clearpage

\clearpage

\begin{deluxetable}{lLLLLLLRLLl}
\tablecaption{Physical Parameters of Grain Material \label{tab:grain-parameters}}
\tablecolumns{11}
\tablewidth{0pt}
\tablehead{
\colhead{Material} & \colhead{$W_\infty$} & \colhead{$\varepsilon$} & \colhead{$\rho$} & \colhead{$\delta_{\max}$} & \colhead{$E_{\max}$} & \colhead{$\kappa$} & \colhead{$M$} & \colhead{$L$} & \colhead{$p_\infty$} &
\colhead{Reference}\\
\colhead{} & \colhead{(eV)} & \colhead{} & \colhead{(kg~m$^{-3}$)} & \colhead{} & \colhead{(eV)} & \colhead{} & \colhead{} & \colhead{(J~kg$^{-1}$)} & \colhead{(N~m$^{-2}$)} & \colhead{}
}
\startdata
Silicate & 4.97 & 7.27 & 3.30 \times 10^{3} & 2.4 & 400 & 0.102 & 169 & 3.21 \times {10}^{6} & 6.72 \times 10^{13} & 1, 2, 3, 4, 5\\
Carbon & 4.75 & 7.00 & 1.81 \times 10^{3} & 1.0 & 300 & 0.383 & 24 & 3.74 \times{10}^{7} & 9.23 \times 10^{15} & 6, 7, 8, 9, 10, 11 \\
\enddata
\tablerefs{(1) \citet{feuerbacher-et-al1972}; (2) \citet{shannon-et-al1991}; (3) \citet{laor-draine1993}; (4) \citet{hwang-daily1992}; (5) \citet{nagahara-et-al1994}; (6) \citet{feuerbacher-fitton1972}; (7) \citet{louh-et-al2005}; (8) \citet{rouleau-martin1991}; (9) \citet{bruining1954}; (10) \citet{ivey1949}; (11) \citet{clarke-fox1969}.
}
\end{deluxetable}

\clearpage

\startlongtable
\begin{deluxetable*}{lcCCCCCCCCCl}
\tablecaption{Stellar and Wind Parameters for Stars With the Presence of Hot Grains \label{tab:hot-grain-stars}}
\tablecolumns{12}
\tablewidth{0pt}
\tablehead{
\colhead{Name} & \colhead{Type} & \colhead{$R_{\star}$} & \colhead{$M_{\star}$} & \colhead{$L_{\star}$} & \colhead{$V_{\star}$} & \colhead{$L_\mathrm{X}$} & \colhead{$B_{\star}$} & \colhead{$\dot{M}_\star$} & \colhead{$T_\mathrm{cor}$} & \colhead{$v_\mathrm{sw}^\infty$} & 
\colhead{Reference}\\
\colhead{} & \colhead{} & \colhead{($R_{\sun}$)} & \colhead{($M_{\sun}$)} & \colhead{($L_{\sun}$)} & \colhead{($\mathrm{km~s^{-1}}$)} & \colhead{($\mathrm{J~s^{-1}}$)} & \colhead{($\mathrm{T}$)} & \colhead{($M_{\sun}~\mathrm{yr^{-1}}$)} & \colhead{($\mathrm{MK}$)} &\colhead{($\mathrm{km~s^{-1}}$)} &  \colhead{} 
}
\decimals
\startdata
$\kappa$ Phe & A5IV & 0.98 & 1.7 & 3.8 & ... & ... & ... & 8.6\times{10}^{-14} & ... & 815 & 1 \\
$\kappa$ Tuc\tablenotemark{$\ddagger$} & F6V & 1.1 & 1.3 & 1.8 & 28.6 & 1.6\times{10}^{22} & 6.8\times 10^{-4} & 6.9\times{10}^{-14} & 4.9 & 671 & 2, 3, 4 \\
$\upsilon$ And\tablenotemark{$\ddagger$} & F9V & 1.6 & 1.6 & 3.3 & 10.9  & 1.3\times{10}^{21} & 2.5\times{10}^{-4} & 2.3\times{10}^{-13} &  2.1 & 627 & 5, 6, 7, 8\\
$\tau$ Cet & G8.5V & 0.82 & 0.85 & 0.51 & 4.95 & 5.1\times{10}^{19} & 5.2\times{10}^{-4} & 1.3\times{10}^{-14} & 1.3 & 630 & 9, 10, 11, 12, 13 \\
HD 14412 & G8V & 0.76 & 1.0 & 0.43 & 1.32 & \overline{1.2}\times{10}^{20} & \overline{1.9}\times 10^{-4} & 5.8\times{10}^{-15} &  \overline{1.7}  & 720 & 5, 12 \\
$\sigma$ Cet\tablenotemark{$\dagger$} & F5V & 1.2 & 1.4 & 2.2 & 20.7 & 2.7\times{10}^{21} & 3.2\times 10^{-4} & 1.0\times{10}^{-13} & 2.9 & 656 & 2, 3, 4 \\
e Eri & G8III & 0.93 & 1.3 & 0.66 & 5.22 & 7.2\times{10}^{19} & 1.3\times 10^{-4} & 1.0\times{10}^{-14} & 1.3 & 731 & 5, 10, 12, 14 \\
10 Tau\tablenotemark{$\dagger$} & F9V & 1.6 & 1.5 & 3.0 & 4.42 &  1.5\times{10}^{20}  & 1.0\times 10^{-4} & 2.4\times{10}^{-13} & 1.2 & 595 & 5, 15, 16\\
b Tau\tablenotemark{$\S$} & A7V & 2.5 & 2.0 & 22 & 108 & ... & ... & 7.1\times{10}^{-12} & ... & 546 & 17, 18, 19\\
$\beta$ Pic & A6V & 1.8 & 1.8 & 8.8 & 130 & 3.2\times{10}^{19} & 7.9\times{10}^{-3} & 1.1\times{10}^{-14}  & 1.1 & 60 & 9, 20, 21, 22, 23\\
$\eta$ Lep\tablenotemark{$\dagger \P$} & F2V & 1.5 & 1.4 & 5.4 & 117 & 4.0\times{10}^{21} & 6.8\times 10^{-2} & 1.2\times{10}^{-12} & 2.9 & 508 & 2, 3, 24 \\
$\lambda$ Gem\tablenotemark{$\flat$} & A3V & 2.8 & 2.1 & 27 & ... & 1.6\times{10}^{22} & 5.6\times{10}^{-2} & 1.1\times{10}^{-11} & 3.0 & 537 & 25, 26 \\
$\beta$ Leo\tablenotemark{$\P$} & A3V & 1.7 & 1.8 & 15 & 374 & \overline{1.3}\times{10}^{19} & 1.1\times{10}^{-2} & 2.2\times{10}^{-12} & \overline{0.61} & 627 & 9, 22, 18, 27, 28\\
HD 104731 & F5V & 1.3 & 1.3 & 3.0 & 34.0 & 3.1\times{10}^{21} & 8.7\times 10^{-2} & 2.0\times{10}^{-13} & 2.9 & 626 & 2, 3, 4\\
$\delta$ Crv & A0IV & 2.4 & 2.6 & 59 & ... & 5.1\times{10}^{21} & 1.4\times{10}^{-2} & 1.6\times{10}^{-11} & 2.4 & 636 & 17, 29\\
$\xi$ Boo\tablenotemark{$\dagger\P$} & G7V & 0.77 & 0.98 & 0.52 & 6.52 & 4.8\times{10}^{21} & 1.8\times{10}^{-3} & 5.8\times{10}^{-14} & 4.4 & 695 & 5, 7, 30, 31, 32\\
$\kappa$ CrB\tablenotemark{$\dagger \amalg \sharp$} & K1IV & 5.1 & 1.5 & 13 & 1.57 & 1.3\times{10}^{20} & 5.5\times{10}^{-5} & 2.9\times{10}^{-11} & 0.63 & 339 & 8, 33, 34 \\
$\alpha$ Lyr & A0V & 2.8 & 2.4 & 44 & 211 & \overline{1.8}\times{10}^{18} & 1.2\times{10}^{-3} & 1.6\times{10}^{-11} & \overline{0.29} & 577 & 35, 36, 37\\
HD 172555 & A7V & 1.5 & 2.0 & 9.5 & 121 & 7.1\times{10}^{21} & 2.7\times{10}^{-3} & 6.4\times{10}^{-13} & 3.3 & 709 & 4, 38, 39, 40\\
110 Her\tablenotemark{$\dagger$} & F6V & 1.9 & 1.2 & 5.5 & 16.0 & 1.3\times{10}^{22} & 4.1\times 10^{-4} & 1.3\times{10}^{-12} & 3.5 & 500 & 5, 15, 41\\
$\zeta$ Aql & A0IV & 2.1 & 2.4 & 35 & ... & 8.5\times{10}^{21} & 3.0\times{10}^{-2} & 6.0\times{10}^{-12} & 3.0 & 662 & 1, 35, 42\\
b Aql\tablenotemark{$\dagger$} & G7IV & 1.3 & 1.4 & 1.7 & 1.66 & 3.9\times{10}^{20} & 1.8\times 10^{-4} & 8.3\times{10}^{-14} & 1.7 & 625 & 4, 5, 30\\
$\alpha$ Aql & A7V & 2.0 & 1.8 & 11 & 269 & 1.4\times{10}^{20} & 1.4\times{10}^{-2} & 1.8\times{10}^{-12} & 2.5  & 585 & 35, 43, 44 \\
$\alpha$ Cep\tablenotemark{$\P$} & A7IV & 2.7 & 1.9 & 18 & 283 & 7.1\times{10}^{20} & 5.0\times{10}^{-3} & 6.9\times{10}^{-12} & 1.4 & 517 & 35, 42, 45 \\
$\mu$ PsA\tablenotemark{$\dagger$} & A1.5IV & 2.0 & 2.1 & 25 & ... & 9.0\times{10}^{21} & 4.0\times 10^{-2} & 4.7\times{10}^{-12} & 3.1 & 624 & 42, 46, 47\\
$\tau$ PsA\tablenotemark{$\ddagger \sharp$} & F6V & 1.4 & 1.1 & 3.0 & 9.01 & 6.0\times{10}^{20} & 1.7\times 10^{-4} & 4.0\times{10}^{-13} & 1.8 & 533 & 5, 15, 48 \\
$\theta$ Peg\tablenotemark{$\S \flat$} & A1V & 2.5 & 2.3 & 38 & ... & 9.1\times{10}^{21} & 3.0\times{10}^{-2} & 1.2\times{10}^{-11} & 2.7 & 586 & 17, 35\\
$\epsilon$ Cep\tablenotemark{$\dagger$} & F0V & 2.6 & 1.8 & 11 & ... & ... & ... & 3.4\times{10}^{-12} & ... & 507 & 2, 49 \\
$\alpha$ PsA & A4V & 1.8 & 2.0 & 18 & 93 & \overline{6.3}\times{10}^{17} & \overline{4.3}\times 10^{-5} & 2.5\times{10}^{-12} & \overline{0.27} & 644 & 9, 43, 50 \\
$\iota$ Psc & F7V & 1.6 & 1.2 & 3.4 & 8.04 & 9.3\times{10}^{20} & 1.9\times{10}^{-4} & 4.3\times{10}^{-13} & 1.9 & 545 & 4, 51, 52 \\
Sun\tablenotemark{$\ddagger$} & G2V & 1.0 & 1.0 & 1.0 & 2.02 & 2.3\times{10}^{20} & 1.0\times{10}^{-4} & 2.5\times{10}^{-14}  & 1.6 & 448 & 30, 53, 54, 55 \\
\enddata
\tablecomments{Underlined values and overlined ones indicate that the values are the lower limit and the upper limit of the quantity, respectively.}
\tablenotetext{\dagger}{The near-infrared excess has been designated as a tentative detection \citep{absil-et-al2013,ertel-et-al2014}.}
\tablenotetext{\ddagger}{A temporal variation in the near-infrared excess including non-detection of the excess has been reported \citep{kimura-mann1998b,ohgaito-et-al2002,ertel-et-al2016,nunez-et-al2017}.}
\tablenotetext{\S}{The near-infrared excess has been detected in one of the two (H and K) bands, but not in the other band \citep{ertel-et-al2014,nunez-et-al2017}.}
\tablenotetext{\P}{The follow-up observations did not show a statistically significant excess, while uncertainties in the follow-up observations were larger than the previous observations \citep{nunez-et-al2017}.}
\tablenotetext{\amalg}{The follow-up observations showed a statistically significant excess, although it is not clear whether or not the near-infrared excess shows a temporal variation \citep{nunez-et-al2017}.}
\tablenotetext{\flat}{The magnetic field strength $B_\star$ is used to derive the Rossby number $Ro$ from Eq.~(\ref{magnetic-field-strength}) and then the X-ray luminosity $L_\mathrm{X}$ from Eq.~(\ref{rossby-number}).}
\tablenotetext{\sharp}{The rotation velocity $V_\star$ is used to derive the Rossby number $Ro$ from Eqs.~(\ref{rotation-velocity}) and (\ref{turnover-time}), and then the X-ray luminosity $L_\mathrm{X}$ from Eq.~(\ref{rossby-number}).}
\tablerefs{(1) \citet{malagnini-morossi1990}; (2) \citet{pasinettifracassini-et-al2001}; (3) \citet{nordstroem-et-al2004}; (4) \citet{huensch-et-al1998b}; (5) \citet{valenti-fischer2005}; (6) \citet{simpson-et-al2010}; (7) \citet{marsden-et-al2014}; (8) \citet{kashyap-et-al2008}; (9) \citet{difolco-et-al2004}; (10) \citet{schroeder-et-al2009}, (11) \citet{watson-et-al2011}; (12) \citet{schmitt1997}; (13) \citet{borra-et-al1984}; (14) \citet{kennedy-et-al2015}; (15) \citet{wright-et-al2004}; (16) \citet{schmitt-liefke2004}; (17) \citet{zorec-royer2012}; (18) \citet{royer-et-al2007}; (19) \citet{morales-et-al2016}; (20) \citet{royer-et-al2002}; (21) \citet{guenther-et-al2012}; (22) \citet{hubrig-et-al2006}; (23) \citet{bruhweiler-et-al1991}; (24) \citet{panzera-et-al1999}; (25) \citet{boyajian-et-al2013}; (26) \citet{bychkov-et-al2003}; (27) \citet{stock-et-al2010}; (28) \citet{schmitt1997}; (29) \citet{berghoefer-et-al1996}; (30) \citet{baliunas-et-al1996}; (31) \citet{wood-et-al2005}; (32) \citet{johnstone-guedel2015}; (33) \citet{jofre-et-al2015}; (34) \citet{bonsor-et-al2013}; (35) \citet{monin-et-al2002}; (36) \citet{pease-et-al2016}; (37) \citet{hill-et-al2010}; (38) \citet{diaz-et-al2011}; (39) \citet{smith-et-al2012}; (40) \citet{schroeder-et-al2008}; (41) \citet{marshall-et-al2013}; (42) \citet{schroeder-schmitt2007}; (43) \citet{hadjara-et-al2014}; (44) \citet{robrade-schmitt2009}; (45) \citet{vanbelle-et-al2006}; (46) \citet{david-hillenbrand2015}; (47) \citet{anderson-francis2012}; (48) \citet{piters-et-al1998}; (49) \citet{bruntt-et-al2007}; (50) \citet{poppenhaeger-et-al2017}; (51) \citet{ghezzi-et-al2010}; (52) \citet{james-jeffries1997}; (53) \citet{pizzolato-et-al2003}; (54) \citet{allen2000}; (55) \citet{koehnlein1996}.}
\end{deluxetable*}

\clearpage

\startlongtable
\begin{deluxetable*}{lcCCCCCCCCCl}
\tablecaption{Stellar and Wind Parameters for Stars Without the Presence of Hot Grains \label{tab:no-hot-grain-stars}}
\tablecolumns{12}
\tablewidth{0pt}
\tablehead{
\colhead{Name} & \colhead{Type} & \colhead{$R_{\star}$} & \colhead{$M_{\star}$} & \colhead{$L_{\star}$} & \colhead{$V_{\star}$} & \colhead{$L_\mathrm{X}$} & \colhead{$B_{\star}$} & \colhead{$\dot{M}_\star$} &  \colhead{$T_\mathrm{cor}$} & \colhead{$v_\mathrm{sw}^\infty$} & 
\colhead{Reference}\\
\colhead{} & \colhead{} & \colhead{($R_{\sun}$)} & \colhead{($M_{\sun}$)} & \colhead{($L_{\sun}$)} & \colhead{($\mathrm{km~s^{-1}}$)} & \colhead{($\mathrm{J~s^{-1}}$)} & \colhead{($\mathrm{T}$)} & \colhead{($M_{\sun}~\mathrm{yr^{-1}}$)} & \colhead{($\mathrm{MK}$)} &\colhead{($\mathrm{km~s^{-1}}$)} &  \colhead{} 
}
\startdata
HD 142 & F7V & 1.4 & 1.0 & 2.5 & \overline{6.86} & \overline{7.9}\times{10}^{20} & \overline{2.0}\times{10}^{-4} & 3.0\times{10}^{-13} & \overline{2.0} & 537 & 1, 2\\
$\beta$ Cas & F2III & 3.8 & 1.9 & 21 & 215 & 4.1\times{10}^{21} & 2.7\times{10}^{-2} & 1.9\times{10}^{-11} & 1.8 & 438 & 3, 4, 5 \\
$\zeta$ Tuc & F9.5V & 1.1 & 1.2 & 1.3 & 3.90 & 1.0\times{10}^{20} & 1.2\times {10}^{-4} & 4.5\times{10}^{-14} & 1.3 & 652 & 1, 6, 7 \\
$\lambda^2$ Phe & F5V & 2.0 & 1.3 & 6.6 & 60.8 & 2.5\times{10}^{22} & 4.9\times {10}^{-4} & 1.9\times{10}^{-12} & 4.0 & 496 & 8, 9, 10\\
54 Psc & K0V & 0.88 & 0.80 & 0.52 & 1.11 & 4.0\times{10}^{20} & 3.2\times{10}^{-4} & 1.9\times{10}^{-14} & 2.1 & 589 & 1, 11, 12, 13 \\
HD 3823\tablenotemark{$\sharp$} & G0V & 1.4 & 1.2 & 2.4 & 4.55 & 5.2\times{10}^{20} & 1.7\times {10}^{-4} & 2.0\times{10}^{-13} & 1.8 & 580 & 1, 14, 15\\
$\eta$ Cas A & F9V & 1.0 & 1.1 & 1.2 & 3.11 & 1.2\times{10}^{20} & 5.0\times{10}^{-5} & 4.9\times{10}^{-14} & 1.4  & 633 & 1, 7, 13, 16 \\
HR 244 & F9V & 1.7 & 1.2 & 3.5 & 6.72 & 8.2\times{10}^{20} & 1.8\times{10}^{-4} & 6.1\times{10}^{-13} & 1.8  & 511 & 17, 18 \\
$\theta$ Cas & A7V & 2.8 & 2.1 & 29 & ... & ... & ... & 1.2\times{10}^{-11} & ...  & 535 & 19 \\
$\nu$ Phe & F9V & 1.2 & 1.2 & 1.9 & 3.43 & 4.3\times{10}^{20} & 1.7\times {10}^{-4} & 1.2\times{10}^{-13} & 1.8 & 603 & 1, 7\\
107 Psc & K1V & 0.81 & 0.81 & 0.43 & 1.17 & 4.4\times{10}^{19} & 3.3\times{10}^{-4} & 1.1\times{10}^{-14} & 1.2 & 618 & 1, 12, 13, 20\\
$q^1$ Eri & F9V & 1.1 & 0.98 & 1.5 & 7.02 & 1.9\times{10}^{21} & 3.3\times {10}^{-4} & 9.9\times{10}^{-14} & 2.8 & 583 & 1, 21, 22, 23\\
V987 Cas & K0V & 0.87 & 0.96 & 0.58 & 1.91 & 2.2\times{10}^{21} & 9.0\times{10}^{-5} & 1.4\times{10}^{-14} & 3.3 & 649 & 13, 17, 20, 24\\
$\chi$ Cet & F0V & 1.3 & 1.5 & 3.3 & 197 & 1.1\times{10}^{22} & 2.0\times{10}^{-1} & 1.7\times{10}^{-13} & 4.1 & 667 & 25, 26, 27\\
$\gamma$ Tri & A1V & 2.1 & 2.3 & 32 & 254 & ... & ... & 6.2\times{10}^{-12} & ... & 640 & 19, 28, 29\\
$\delta$ Hyi & A2V & 2.1 & 2.2 & 40 & ... & ... & ... & 9.1\times{10}^{-12} &  ... & 631 & 8, 9\\
$13$ Tri\tablenotemark{$\flat$} & G0V & 1.9 & 1.1 & 3.7 & 0.93 & 7.3\times{10}^{18} & 3.0\times{10}^{-5} & 9.6\times{10}^{-13} & 0.5 & 473 & 13, 30\\
$\theta$ Per & F8V & 1.2 & 1.4 & 2.1 & 6.08 & 6.6\times{10}^{20} & 2.0\times {10}^{-4} & 9.2\times{10}^{-14} & 2.0  & 660 & 1, 7, 16\\
$\iota$ Hor & F8V & 1.2 & 1.1 & 1.7 & 7.40 & 6.0\times{10}^{21} & 4.9\times {10}^{-4} & 1.0\times{10}^{-13} & 3.7 & 595 & 1, 12, 31 \\
HR 857 & K1V & 0.75 & 0.81 & 0.38 & 5.44 & 5.0\times{10}^{21} & 1.0\times {10}^{-3} & 8.2\times{10}^{-15} & 4.4 & 641 & 1, 12, 20, 32\\
$\rho^3$ Eri & A5V & 1.9 & 1.8 & 12 & ... & ... & ... & 2.0\times{10}^{-12} & ...  & 595 & 19\\
$\iota$ Per & G0V & 1.5 & 1.2 & 2.2 & 4.20 & 3.9\times{10}^{20} & 1.6\times{10}^{-4} & 2.3\times{10}^{-13} & 1.6 & 544 & 17, 33, 34, 35\\
$\kappa^1$ Cet & G5V & 0.92 & 0.71 & 0.82 & 5.17 & 7.7\times{10}^{21} & 7.7\times{10}^{-4} & 9.7\times{10}^{-13} & 4.6 & 543 & 13, 17, 20, 36, 37, 38\\
$\zeta^1$ Ret & G4V & 0.91 & 1.1 & 0.78 & 3.14 & 8.1\times{10}^{20} & 3.1\times {10}^{-4} & 1.7\times{10}^{-14} & 2.5 & 689 & 1, 7\\
$\zeta^2$ Ret & G0V & 0.97 & 1.2 & 1.0 & 3.42 & 1.4\times{10}^{20} & 1.5\times {10}^{-4} & 2.5\times{10}^{-14} &  1.5 & 684 & 1, 7, 39\\
$\kappa$ Ret & F3IV & 1.2 & 1.5 & 3.0 & 37.3 & 2.3\times{10}^{21} & 7.2\times {10}^{-2} & 1.2\times{10}^{-13} & 2.8 & 690 & 23, 40 \\
$\epsilon$ Eri & K2V & 0.74 & 0.90 & 0.32 & 3.33 & 2.1\times{10}^{21} & 1.1\times{10}^{-3} & 3.5\times{10}^{-13}  & 3.5 & 680 & 10, 11, 13, 41, 42, 43\\
$\delta$ Eri & K0+IV & 2.3 & 1.2 & 3.3 & 1.62 & 9.0\times{10}^{19} & 2.0\times{10}^{-5} & 1.0\times{10}^{-12}  & 0.87 & 447 & 13, 17, 20, 44\\
HD 25457 & F7V & 1.2 & 1.2 & 2.1 & 30.1 & 3.4\times{10}^{22} & 8.5\times {10}^{-4} & 1.2\times{10}^{-13} & 5.7 & 623 & 4, 16, 45, 46, 47\\
$o^2$ Eri & K0V & 0.81 & 0.82 & 0.41 & 0.948 & 4.1\times{10}^{20} & 1.3\times {10}^{-4} & 1.0\times{10}^{-14} & 2.2 & 621 & 13, 17, 20, 48\\
58 Eri & G1.5V & 0.98 & 0.86 & 0.96 & 4.52 & 6.5\times{10}^{21} & 6.1\times{10}^{-4} & 5.1\times{10}^{-14} &  4.1 & 578 & 1, 7, 20\\
$\pi^3$ Ori & F6V & 1.3 & 1.3 & 2.7 & 16.8 & 9.8\times{10}^{21} & 4.3\times{10}^{-4} & 2.0\times{10}^{-13} & 4.3 & 604 & 1, 7, 13, 16, 43 \\
$\pi^1$ Ori\tablenotemark{$\flat$} & A3V & 1.7 & 2.1 & 20 & ... & 4.0\times{10}^{21} & 2.6\times{10}^{-2} & 2.4\times{10}^{-12} & 2.7 & 674 & 19, 49\\
HR 1604 & F5V & 2.2 & 1.2 & 7.6 & ... & ... & ... & 3.3\times{10}^{-12} &  ... & 456 & 8, 9\\
$\zeta$ Dor & F9V & 1.1 & 1.1 & 1.5 & 19.6 & 1.0\times{10}^{22} & 6.2\times {10}^{-4} & 7.3\times{10}^{-14} & 4.4 & 618 & 23, 25 \\
$\lambda$ Aur\tablenotemark{$\flat$} & G1.5IV & 1.3 & 1.0 & 1.8 & 2.81 & 3.5\times{10}^{18} & 3.0\times {10}^{-5} & 1.6\times{10}^{-13} & 0.49 & 548 & 13, 18, 35 \\
HD 34721 & G0V & 1.3 & 1.2 & 2.1 & 4.25 & 1.3\times{10}^{21} & 2.5\times {10}^{-4} & 1.7\times{10}^{-13} & 2.3 & 576& 1, 16, 23\\
$\zeta$ Lep & A2IV & 1.5 & 2.0 & 14 & 517 & ... & ... & 1.0\times{10}^{-12} & ... & 718 & 40, 50, 51\\
HD 38858 & G2V & 0.94 & 1.1 & 0.86 & 5.65 & 4.1\times{10}^{20} & 2.3\times {10}^{-4} & 2.5\times{10}^{-14} & 2.0 & 653 & 1, 23, 52, 53\\
HD 40307 & K2.5V & 0.71 & 0.78 & 0.25 & 0.748 & 9.8\times{10}^{19} & 2.1\times {10}^{-4} & 4.0\times{10}^{-15} & 1.6 & 648 & 1, 23, 54\\
HD 43162 & G6.5V & 0.90 & 0.99 & 0.66 & 5.70 & 1.4\times{10}^{22} & 5.1\times{10}^{-4} & 1.7\times{10}^{-14} & 5.2 & 647 & 10, 13, 16, 55, 56\\
HD 45184\tablenotemark{$\sharp$} & G1.5V & 1.1 & 0.97 & 1.2 & 2.56 & 3.4\times{10}^{20} & 1.9\times{10}^{-4} & 6.2\times{10}^{-14} & 1.8  & 591 & 1, 16\\
$\xi$ Gem & F5IV & 2.7 & 1.7 & 12 & 48.7 & 8.7\times{10}^{21} & 2.7\times {10}^{-4} & 4.3\times{10}^{-12} & 2.6 & 496 & 4, 18\\
38 Gem & A8V & 1.8 & 1.6 & 8.1 & ... & 1.8\times{10}^{22} & 1.5\times{10}^{-1} & 1.3\times{10}^{-12} & 3.8 & 575 & 17, 19\\
HD 53705 & G0V & 1.1 & 1.3 & 1.3 & ... & ... & ... & 4.5\times{10}^{-14} & ... & 664 & 1\\
HD 69830 & G8V & 0.90 & 0.86 & 0.6 & 1.30 & 3.2\times{10}^{20} & 2.4\times {10}^{-4} & 2.1\times{10}^{-14} & 2.0 & 604 & 1, 7, 11\\
30 Mon & A0V & 2.0 & 2.8 & 34 & 160 &  2.7\times{10}^{21} & 1.3\times {10}^{-2} & 3.9\times{10}^{-12} & 2.2 & 727 & 28, 29, 45, 57\\
HD 72673\tablenotemark{$\sharp$} & G9V & 0.76 & 0.98 & 0.39 & 1.04 & 1.3\times{10}^{20} & 2.0\times {10}^{-4} & 5.5\times{10}^{-15} & 1.7 & 703 & 1, 7, 16\\
HD 76151 & G3V & 0.98 & 1.2 & 0.97 & 3.30 & 2.2\times{10}^{21} & 3.7\times{10}^{-4} & 2.3\times{10}^{-14} & 3.1 & 695 & 1, 12, 13, 20\\
HD 76932 & G2V & 1.4 & 0.78 & 2.0 & ... & ... & ... & 4.4\times{10}^{-13} & ...  & 460 & 30\\
$\psi$ Vel & F3V & 2.2 & 1.4 & 11 & 123 & 1.3\times{10}^{22}  & 1.0\times {10}^{-1} & 3.5\times{10}^{-12} & 3.3 & 505 & 8, 9, 23\\
$\alpha$ Leo & B8IV & 4.2 & 3.4 & 350 & 317 & 3.8\times{10}^{21}  & 1.5\times {10}^{-3} & 4.1\times{10}^{-10} & 1.7 & 558 & 23, 34, 58\\
q Vel & A2V & 1.8 & 2.2 & 22 & ... & ... & ... & 2.6\times{10}^{-12} & ... & 687 & 40\\
HD 90132 & A8V & 1.8 & 1.7 & 11 & ... & 3.0\times{10}^{20} & 6.0\times{10}^{-3} & 1.7\times{10}^{-12} & 1.3 & 590 & 8, 9\\
HD 91324 & F9V & 1.1 & 1.3 & 2.3 & ... & ... & ... & 9.3\times{10}^{-14} & ... & 673 & 26, 59, 60\\
$\beta$ UMa\tablenotemark{$\flat$} & A1IV & 2.3 & 2.5 & 44 & 46.2 & 7.0\times{10}^{20} & 4.0\times{10}^{-3} & 1.0\times{10}^{-11} & 1.5 & 638 & 28, 29, 40, 61 \\
$\delta$ Leo & A5IV & 2.8 & 2.2 & 33 & ... & \overline{4.5}\times{10}^{19} & 7.5\times{10}^{-3} & 1.3\times{10}^{-11} & \overline{0.66} & 552 & 40, 58, 62 \\
$\gamma$ Crt & A7V & 2.0 & 1.8 & 13 & ... & 1.3\times{10}^{21} & 1.5\times {10}^{-2} & 2.5\times{10}^{-12} & 1.9 & 583 & 19, 57\\
HD 102365 & G2V & 0.96 & 1.2 & 0.83 & 2.02 & \overline{1.3}\times{10}^{20} & \overline{1.5}\times {10}^{-4} & 1.7\times{10}^{-14} & \overline{1.5} & 700 & 1, 7, 31\\
$\beta$ Vir & F9V & 1.6 & 1.6 & 3.5 & 8.29 & 1.9\times{10}^{21} & 6.5\times{10}^{-4}  & 2.7\times{10}^{-13} & 2.3 & 612 & 1, 7, 16, 61\\
$\delta$ UMa\tablenotemark{$\flat$} & A2V & 2.5 & 2.0 & 23 & 319 & 4.4\times{10}^{21} & 2.5\times{10}^{-2} & 7.0\times{10}^{-12} & 2.3  & 558 & 61, 63\\
$\eta$ Crv & F2V & 1.6 & 1.4 & 5.1 & 149 & 5.0\times{10}^{21} &  1.1\times{10}^{-2} & 6.1\times{10}^{-13} & 3.0 & 580 & 23, 29, 33, 45, 64, 65\\
$\tau$ Cen & A1IV & 2.8 & 2.4 & 51 & ... & ... & ... & 2.0\times{10}^{-11} & ... & 577 & 19\\
61 Vir & G7V & 0.96 & 1.0 & 0.81 & 1.68 & 9.3\times{10}^{19} & 1.4\times {10}^{-4} & 3.8\times{10}^{-15} & 1.4 & 629 & 1, 7, 20, 66\\
70 Vir & G4V & 1.9 & 1.5 & 3.0 & 3.04 & 2.5\times {10}^{19} & 5.2\times {10}^{-5} & 3.4\times{10}^{-13} & 0.69 & 551 & 1, 7, 67\\
$\tau$ Boo & F6IV & 1.4 & 1.3 & 3.0 & 17.9 & 8.9\times{10}^{21} & 3.2\times{10}^{-4} & 2.7\times{10}^{-12} & 2.0 & 300 & 1, 13, 20, 24, 68\\
$\iota$ Vir & F7III & 2.5 & 1.5 & 8.7 & 18.2 & 2.1\times{10}^{22} & 3.0\times{10}^{-4} & 3.3\times{10}^{-12} & 3.4 & 478 & 4, 20, 40, 69 \\
$\sigma$ Boo & F4V & 1.4 & 1.1 & 3.2 & 62.4 & 3.2\times{10}^{21} & 8.0\times{10}^{-4} & 4.0\times{10}^{-13} & 2.9 & 545 & 1, 17, 69 \\
$\alpha$ Cir\tablenotemark{$\flat$} & A7V & 2.0 & 1.7 & 11 & 22.2 & 1.1\times{10}^{21} & 1.6\times{10}^{-2} & 1.9\times{10}^{-12} & 1.8 & 574 & 70, 71, 72\\
$\mu$ Vir & F2V & 2.1 & 1.7 & 9.8 & 126 & 3.2\times{10}^{22} & 2.0\times {10}^{-1} & 2.0\times{10}^{-12} & 4.2 & 557 & 23, 40 \\
109 Vir & A0III & 1.7 & 2.3 & 23 & ... & 6.3\times{10}^{21} & 3.2\times {10}^{-2} & 2.3\times{10}^{-12} & 3.1 & 719 & 17, 40\\
c Boo & F5V & 1.5 & 1.2 & 3.3 & 35.8 & 5.8\times{10}^{21} & 3.7\times {10}^{-4} & 3.7\times{10}^{-13} & 3.2 & 562 & 1, 23\\
$\beta$ Cir & A3V & 1.8 & 2.1 & 19 & ... & ... & ... & 2.5\times{10}^{-12} & ... & 661 & 8, 9\\
5 Ser & F8IV & 2.1 & 1.2 & 4.2 & 7.74 & 4.4\times{10}^{20} & 1.3\times {10}^{-4} & 1.3\times{10}^{-12} & 1.4 & 466 & 12, 18, 20\\
g Lup & F3V & 1.3 & 1.3 & 3.3 & 71.7 & 1.1\times{10}^{22} & 2.1\times {10}^{-1} & 2.3\times{10}^{-13} & 4.2 & 627 & 23, 45, 73, 74\\
$\lambda$ Ser & G0IV & 1.3 & 1.3 & 1.9 & 2.56 & 6.8\times{10}^{20} & 2.0\times {10}^{-4} & 1.2\times{10}^{-13} & 2.0 & 608 & 1, 7, 20\\
$\beta$ TrA & F1V & 1.8 & 1.6 & 9.4 & 164 & 1.6\times{10}^{21} & 2.3\times {10}^{-2} & 1.7\times{10}^{-12} & 2.1 & 570 & 7, 8, 9\\
$\chi$ Her & G0V & 1.7 & 1.5 & 3.1 & 5.76 & 8.9\times{10}^{19} & 2.7\times{10}^{-3} & 3.1\times{10}^{-13} & 1.0 & 571 & 1, 20, 23, 75 \\
$\gamma$ Ser & F6IV & 1.4 & 1.2 & 2.9 & 12.2  & 3.5\times{10}^{20} & 5.5\times{10}^{-4} & 3.3\times{10}^{-13} & 1.6 & 553 & 1, 7, 16, 61\\
12 Oph & K1V & 0.79 & 0.85 & 0.44 & 1.91 & 2.0\times{10}^{21} & 6.0\times{10}^{-5} & 9.9\times{10}^{-15} & 3.4 & 641 & 1, 12, 13, 20\\
HD 152391 & G8.5V & 0.83 & 0.92 & 0.56 & 3.80 & 5.0\times{10}^{21} & 8.5\times{10}^{-4} & 1.3\times{10}^{-14} & 4.2 & 651 & 1, 12, 13, 20 \\
$\lambda$ Ara & F4V & 1.6 & 1.3 & 4.8 & 46.3 & 4.4\times{10}^{21} & 7.9\times {10}^{-2} & 7.1\times{10}^{-13} & 2.8 & 553 & 8, 9, 23\\
58 Oph & F5V & 1.4 & 1.2 & 2.9 & \overline{20.9} & \overline{2.5}\times{10}^{21} & \overline{2.9}\times {10}^{-4} & 3.0\times{10}^{-13} & \overline{2.6} & 566 & 15, 25\\
$\mu$ Her & G5IV & 1.7 & 1.3 & 2.6 & 2.17 & 5.1\times{10}^{20} & 5.0\times{10}^{-5} & 2.9\times{10}^{-13} & 1.6 & 548 & 1, 7, 13, 16\\
$\gamma$ Oph\tablenotemark{$\flat$} & A1V & 1.8 & 2.2 & 22 & 274 & 2.7\times{10}^{21} & 1.8\times{10}^{-2} & 2.6\times{10}^{-12} & 2.4 & 687 & 29, 40, 61, 76 \\
$\zeta$ Ser & F2V & 1.8 & 1.9 & 6.6 & 92.7 & 7.6\times{10}^{21} & 9.5\times {10}^{-2} & 6.0\times{10}^{-13} & 3.1 & 636 & 8, 9, 23\\
70 Oph A & K0V & 0.91 & 1.1 & 0.59 & 2.30 & 1.6\times{10}^{21} & 4.3\times{10}^{-4} & 1.2\times{10}^{-14} & 3.4 & 679 & 12, 20, 43, 66, 77\\
72 Oph & A5V & 2.2 & 2.0 & 20 & ... & 8.6\times{10}^{20} & 8.1\times {10}^{-3} & 4.6\times{10}^{-12} & 1.6 & 589 & 57\\
36 Dra & F5V & 1.8 & 1.2 & 4.1 & 101 & 1.2\times{10}^{22} & 4.5\times {10}^{-4} & 8.8\times{10}^{-13} & 3.5 & 501 & 17, 18\\
$\alpha$ CrA & A2V & 2.4 & 2.2 & 33 & ... & ... & ... & 9.2\times{10}^{-12} &  ... & 592 & 19\\
$\iota$ Cyg & A5V & 3.0 & 2.2 & 38 & ... & ... & ... & 1.8\times{10}^{-11} & ... & 534 & 19\\
$\sigma$ Dra & G9V & 0.79 & 0.76 & 0.42 & 1.47 & 2.5\times{10}^{20} & 3.1\times{10}^{-4} & 1.2\times{10}^{-14} & 2.0 & 606 & 1, 7, 13, 20\\
$o$ Aql & F8V & 1.4 & 1.7 & 2.6 & 6.98 & 1.1\times{10}^{21} & 2.2\times {10}^{-4} & 1.0\times{10}^{-13} & 2.2 & 684 & 17, 20, 36\\
$\epsilon$ Pav & A0V & 1.9 & 2.3 & 32 & 151 & ... & ... & 4.3\times{10}^{-12} &  ... & 695 & 19, 28, 29\\
HD 190360 & G7IV-V & 1.1 & 0.98 & 1.2 & 1.46 & \overline{2.2}\times{10}^{19} & \overline{7.0}\times{10}^{-5} & 6.7\times{10}^{-14} & \overline{0.89} & 583 & 20, 44, 78\\
$\rho$ Aql & A1V & 2.0 & 2.1 & 22 & 202 & ... & ... & 4.0\times{10}^{-12} & ... & 624 & 19, 29, 79 \\
29 Cyg\tablenotemark{$\flat$} & A2V & 2.2 & 2.1 & 25 & 69.2 & 3.3\times{10}^{21} & 1.9\times {10}^{-2} & 5.3\times{10}^{-12} & 2.3 & 614 & 19, 29, 49, 80\\
$\phi^{1}$ Pav\tablenotemark{$\sharp$} & F0V & 1.7 & 1.5 & 8.1 & 138 & 1.1\times{10}^{21} & 2.0\times{10}^{-2} & 1.2\times{10}^{-12} & 2.0 & 584 & 8, 9, 81, 82\\
$\eta$ Ind & A9IV & 1.6 & 1.6 & 7.6 & ... & ... & ... & 7.9\times{10}^{-13} & ... & 627 & 8, 9\\
$\psi$ Cap & F5V & 1.5 & 1.6 & 4.0 & 48.0 & 9.3\times{10}^{21} & 1.6\times {10}^{-1} & 2.9\times{10}^{-13} & 3.6 & 637 & 7, 8, 9, 83\\
61 Cyg A & K5V & 0.67 & 0.69 & 0.15 & 0.961 & 1.6\times{10}^{20} & 8.9\times{10}^{-4} & 6.4\times{10}^{-15} & 1.9 & 629 & 12, 13, 20, 42, 84\\
61 Cyg B & M0.5V & 0.59 & 0.60 & 0.085 & 0.792 & 1.6\times{10}^{20} & 8.1\times {10}^{-3} & 9.6\times{10}^{-16} & 2.0 & 623 & 12, 20, 84, 85\\
$\tau$ Cyg & F2IV & 2.6 & 1.7 & 13 & 72.2 & 7.5\times{10}^{21} & 2.4\times {10}^{-3} & 4.5\times{10}^{-12} & 2.6 & 504 & 17, 19, 86\\
$\gamma$ Pav & F9V & 1.1 & 0.95 & 1.5 & 3.85 & 6.3\times{10}^{19} & 9.3\times {10}^{-5} & 1.1\times{10}^{-13} & 1.2 & 574 & 7, 25\\
HN Peg & G0V & 1.0 & 1.1 & 1.2 & 10.1 & 1.0\times{10}^{22} & 1.5\times{10}^{-3} & 4.2\times{10}^{-14} & 4.5 & 643 & 1, 12, 13, 20\\
HD 207129 & G2V & 1.1 & 1.1 & 1.2 & 3.71 & 1.2\times{10}^{21} & 3.0\times {10}^{-4} & 5.2\times{10}^{-14} & 2.5& 621 & 1, 21, 22, 23\\
HD 210277 & G8V & 1.1 & 1.3 & 0.95 & 1.31 & 6.7\times{10}^{20} & 9.0\times{10}^{-5} & 2.5\times{10}^{-14} & 2.2 & 678 & 1, 13, 67, 87\\
$\alpha$ Lac\tablenotemark{$\flat$} & A1V & 2.1 & 2.2 & 28 & ... & 2.9\times{10}^{21} & 1.6\times {10}^{-2} & 5.6\times{10}^{-12} & 2.2 & 625 & 18, 61\\
$\upsilon$ Aqr & F7V & 1.5 & 1.3 & 3.6 & 38.3 & 1.0\times{10}^{22} & 4.4\times {10}^{-4} & 3.6\times{10}^{-13} & 3.7 & 581 & 8, 9, 23\\
HD 214953\tablenotemark{$\sharp$} & F9V & 1.1 & 0.81 & 1.4 & 3.59 & 5.1\times{10}^{20} & 2.1\times {10}^{-4} & 1.2\times{10}^{-13} & 2.0 & 541 & 1, 15, 31\\
$\xi$ Peg\tablenotemark{$\sharp$} & F6V & 1.8 & 1.2 & 4.5 & 9.29 & 6.6\times{10}^{20} & 1.5\times{10}^{-4} & 9.3\times{10}^{-13} & 1.6 & 509 & 1, 16\\
$\epsilon$ Gru & A2IV & 3.3 & 1.9 & 54 & ... & 1.0\times{10}^{21} & 2.9\times{10}^{-3} & 5.8\times{10}^{-11} & 1.4 & 466 & 8, 9, 88\\
$\tau^{1}$ Gru & G0V & 1.7 & 1.6 & 3.5 & 8.17 & 6.3\times{10}^{21} & 3.8\times {10}^{-4} & 3.3\times{10}^{-13} & 3.0 & 586 & 1, 15\\
51 Peg & G2IV & 1.2 & 0.72 & 1.3 & 1.59 & 5.6\times{10}^{19} & 6.0\times {10}^{-5} & 1.8\times{10}^{-13} & 1.0 & 487 & 13, 20, 36, 89\\
7 And & F1V & 1.8 & 1.6 & 8.0 & 156 & 1.2\times{10}^{21} & 2.1\times{10}^{-2} & 1.2\times{10}^{-12} & 1.9 & 579 & 19, 60\\
HR 8832 & K3V & 0.78 & 0.76 & 0.26 & 0.931 & 4.0\times{10}^{19} & 1.1\times {10}^{-4} & 5.8\times{10}^{-15} & 1.2 & 612 & 13, 17, 48, 90\\
HD 219482 & F6V & 1.2 & 1.2 & 1.9 & 26.7 & 1.9\times{10}^{22} & 7.1\times {10}^{-4} & 1.2\times{10}^{-13} & 4.9 & 607 & 23, 25\\
$\gamma$ Tuc & F4V & 2.5 & 1.3 & 11 & 54.3 & 3.4\times{10}^{21} & 4.9\times{10}^{-3} & 6.8\times{10}^{-12} & 2.1 & 444 & 8, 9, 23, 65\\
\enddata
\tablecomments{Underlined values and overlined ones indicate that the values are the lower limit and the upper limit of the quantity, respectively.}
\tablenotetext{\flat}{The magnetic field strength $B_\star$ is used to derive the Rossby number $Ro$ from Eq.~(\ref{magnetic-field-strength}) and then the X-ray luminosity $L_\mathrm{X}$ from Eq.~(\ref{rossby-number}).}
\tablenotetext{\sharp}{The rotation velocity $V_\star$ is used to derive the Rossby number $Ro$ from Eqs.~(\ref{rotation-velocity}) and (\ref{turnover-time}), and then the X-ray luminosity $L_\mathrm{X}$ from Eq.~(\ref{rossby-number}).}
\tablerefs{(1) \citet{valenti-fischer2005}; (2) \citet{miller-et-al2015}; (3) \citet{che-et-al2011}; (4) \citet{panzera-et-al1999}; (5) \citet{udovichenko-et-al1994}; (6) \citet{suarezmascareno-et-al2017}; (7) \citet{schmitt1997}; (8) \citet{david-hillenbrand2015}; (9) \citet{anderson-francis2012}; (10) \citet{huensch-et-al1999}; (11) \citet{simpson-et-al2010}; (12) \citet{pizzolato-et-al2003}; (13) \citet{marsden-et-al2014}; (14) \citet{lovis-et-al2011}; (15) \citet{piters-et-al1998}; (16) \citet{wright-et-al2004}; (17) \citet{huensch-et-al1998b}; (18) \citet{boyajian-et-al2013}; (19) \citet{zorec-royer2012}; 
(20) \citet{baliunas-et-al1996}; (21) \citet{delgadomena-et-al2015}; (22) \citet{watson-et-al2011}; (23) \citet{schmitt-liefke2004}; (24) \citet{wright-et-al2011}; (25) \citet{fuhrmann-chini2015}; (26) \citet{johnson-wright1983}; (27) \citet{huensch-et-al1998a}; (28) \citet{booth-et-al2013}; (29) \citet{royer-et-al2007}; (30) \citet{takeda2007}; (31) \citet{saar-osten1997}; (32) \citet{folsom-et-al2018}; (33) \citet{nordstroem-et-al2004}; (34) \citet{mcalister-et-al2005}; (35) \citet{doyle1996}; (36) \citet{ghezzi-et-al2010}; (37) \citet{telleschi-et-al2005}; (38) \citet{donascimento-et-al2016}; (39) \citet{faramaz-et-al2014}; (40) \citet{malagnini-morossi1990}; (41) \citet{difolco-et-al2004}; (42) \citet{wood-et-al2002}; (43) \citet{johnstone-guedel2015}; (44) \citet{maldonado-et-al2013}; (45) \citet{rhee-et-al2007}; (46) \citet{meshkat-et-al2015}; (47) \citet{hillenbrand-et-al2008}; (48) \citet{boyajian-et-al2012}; (49) \citet{bohlender-landstreet1990}; (50) \citet{diaz-et-al2011}; (51) \citet{mennesson-et-al2014}; (52) \citet{schroeder-et-al2009}; (53) \citet{krist-et-al2012} (54) \citet{diaz-et-al2016}; (55) \citet{gaidos-gonzalez2002}; (56) \citet{spina-et-al2016}; (57) \citet{schroeder-schmitt2007}; (58) \citet{landstreet1982}; (59) \citet{dasilva1975}; (60) Preliminary results of Robrade et al. (2018, in preparation); (61) \citet{monin-et-al2002}; (62) \citet{simon-et-al2002}; (63) \citet{jones-et-al2015}; (64) \citet{lebreton-et-al2016}; (65) \citet{hubrig-et-al2006}; (66) \citet{wood-et-al2005}; (67) \citet{barnes2001}; (68) \citet{vidotto-et-al2012}; (69) \citet{shorlin-et-al2002}; (70) \citet{mathys1994}; (71) \citet{bruntt-et-al2008}; (72) \citet{kurtz-et-al1994}; (73) \citet{kalas-et-al2006}; (74) \citet{reiners2006}; (75) \citet{boesgaard1974}; (76) \citet{su-et-al2008}; (77) \citet{bruntt-et-al2010}; (78) \citet{sanzforcada-et-al2011}; (79) \citet{morales-et-al2016}; (80) \citet{casas-et-al2009}; (81) \citet{moor-et-al2015}; (82) \citet{glebocki-gnacinski2005}; (83) \citet{reiners-et-al2001}; (84) \citet{kervella-et-al2008}; (85) \citet{bychkov-et-al2009}; (86) \citet{bychkov-et-al2003}; (87) \citet{cantomartins-et-al2011}; (88) \citet{schroeder-et-al2008}; (89) \citet{poppenhaeger-et-al2009}; (90) \citet{motalebi-et-al2015}.}
\end{deluxetable*}




\begin{thebibliography}{}

\bibitem[Absil et al.(2009)]{absil-et-al2009}
Absil, O., Mennesson, B., Le Bouquin, J.-B., et al.\ 2009, \apj, 704, 150

\bibitem[Absil et al.(2013)]{absil-et-al2013}
Absil, O., Defr\`{e}re, D., Coud\'{e} du Foresto, V., et al.\ 2013, \aap, 555, A104

\bibitem[Allen(2000)]{allen2000}
Allen, C.~W.\ 2000, Allen's Astrophysical Quantities, (4th ed.; New York: Springer)

\bibitem[Anderson \& Francis(2012)]{anderson-francis2012}
Anderson, E., \& Francis, C.\ 2012, Astron. Lett., 38, 331

\bibitem[Andriesse(1979)]{andriesse1979}
Andriesse, C.~D.\ 1979, \apss, 61, 205

\bibitem[Andriesse(2000)]{andriesse2000}
Andriesse, C.~D.\ 2000, \apj, 539, 364

\bibitem[Babel \& Montmerle(1997)]{babel-montmerle1997}
Babel, J., \& Montmerle, T.\ 1997, \aap, 323, 121

\bibitem[Baliunas et al.(1996)]{baliunas-et-al1996}
Baliunas, S., Sokoloff, D., \& Soon, W.\ 1996, \apj, 457, L99

\bibitem[Barnes(2001)]{barnes2001}
Barnes, S.~A.\ 2001, \apj, 561, 1095

\bibitem[Belcher \& MacGregor(1976)]{belcher-macgregor1976}
Belcher, J.~W., \& MacGregor, K.~B.\ 1976, \apj, 210, 498

\bibitem[Belton(1966)]{belton1966}
Belton, M.~J.~S.\ 1966, Science, 151, 35

\bibitem[Bergh\"{o}fer et al.(1996)]{berghoefer-et-al1996}
Bergh\"{o}fer, T.~W., Schmitt, J.~H.~M.~M., \& Cassinelli, J.~P.\ 1996, \aaps, 118, 481

\bibitem[Boesgaard(1974)]{boesgaard1974}
Boesgaard, A.~M.\ 1974, \apj, 188, 567

\bibitem[Bohlender \& Landstreet(1990)]{bohlender-landstreet1990}
Bohlender, D.~A., \& Landstreet, J.~D.\ 1990, \mnras, 247, 606

\bibitem[Bonsor et al.(2013)]{bonsor-et-al2013}
Bonsor, A., Kennedy, G.~M., Crepp, J.~R., et al.\ 2013, \mnras, 431, 3025

\bibitem[Booth et al.(2013)]{booth-et-al2013}
Booth, M., Kennedy, G., Sibthorpe, B., et al.\ 2013, \mnras, 428, 1263

\bibitem[Borra et al.(1984)]{borra-et-al1984}
Borra, E.~F., Edwards, G., \& Mayor, M.\ 1984, \apj, 284, 211

\bibitem[Boyajian et al.(2012)]{boyajian-et-al2012}
Boyajian, T.~S., von Braun, K., van Belle, G., et al.\ 2012, \apj, 757, 112

\bibitem[Boyajian et al.(2013)]{boyajian-et-al2013}
Boyajian, T.~S., von Braun, K., van Belle, G., et al.\ 2013, \apj, 771, 40

\bibitem[Bruhweiler et al.(1991)]{bruhweiler-et-al1991}
Bruhweiler, F.~C., Kondo, Y., \& Grady, C.~A.\ 1991., \apj, 371, L27

\bibitem[Bruining(1954)]{bruining1954}
Bruining, H.\ 1954, Physics and Applications of Secondary Electron Emission. Pergamon Press, London.

\bibitem[Bruntt et al.(2007)]{bruntt-et-al2007}
Bruntt, H., Su\'{a}rez, J.~C., Bedding, T.~R., et al.\ 2007, \aap, 461, 619

\bibitem[Bruntt et al.(2008)]{bruntt-et-al2008}
Bruntt, H., North, J.~R., Cunha, M., et al.\ 2008, \mnras, 386, 2039

\bibitem[Bruntt et al.(2010)]{bruntt-et-al2010}
Bruntt, H., Bedding, T.~R., Quirion, P.~O., et al.\ 2010, \mnras, 405, 1907

\bibitem[Burns et al.(1979)]{burns-et-al1979}
Burns, J.~A., Lamy, P.~L., and Soter, S.\ 1979, Icarus, 40, 1

\bibitem[Brus(1983)]{brus1983}
Brus L. E.\ 1983, \jcp, 79, 5566

\bibitem[Bychkov et al.(2003)]{bychkov-et-al2003}
Bychkov, V.~D., Bychkova, L.~V., \& Madej, J.\ 2003, \aap, 407, 631

\bibitem[Bychkov et al.(2009)]{bychkov-et-al2009}
Bychkov, V.~D., Bychkova, L.~V., \& Madej, J.\ 2009, \mnras, 394, 1338

\bibitem[Canto Martins et al.(2011)]{cantomartins-et-al2011}
Canto Martins, B.~L., das Chagas, M.~L., Alves, S., et al.\ 2011, \aap, 530, A73

\bibitem[Casas et al.(2009)]{casas-et-al2009}
Casas, R., Moya, A., Su\'{a}rez, J.~C., et al.\ 2009, \apj, 697, 522

\bibitem[Castelli \& Kurucz(2003)]{castelli-kurucz2003}
Castelli, F., \& Kurucz, R.~L.\ 2003, In Modelling of Stellar Atmospheres, (ed.) N.~E. Piskunov, W.~W. Weiss, and D.~F. Gray, ASP- S210, Astronomical Society of the Pacific, pp.~A20

\bibitem[Castor et al.(1975)]{castor-et-al1975}
Castor, J.~I., Abbott, D.~C., \& Klein, R.~I.\ 1975, \apj, 195, 157

\bibitem[Che et al.(2011)]{che-et-al2011}
Che, X., Monnier, J.~D., Zhao, M., et al.\ 2011, \apj, 732, 68

\bibitem[Clarke \& Fox(1969)]{clarke-fox1969}
Clarke, J.~T., \& Fox, B.~R.\ 1969, \jcp, 51, 3231

\bibitem[Czechowski \& Kleimann(2017)]{czechowski-kleimann2017}
Czechowski, A., \& Kleimann, J.\ 2017, Ann. Geophys., 35, 1033

\bibitem[Czechowski \& Mann(2010)]{czechowski-mann2010}
Czechowski, A., \& Mann, I.\ 2010, \apj, 714, 89

\bibitem[Da Silva(1975)]{dasilva1975}
da Silva, L.\ 1975, \aap, 41, 287

\bibitem[David \& Hillenbrand(2015)]{david-hillenbrand2015}
David, T.~J., \& Hillenbrand, L.	A.\ 2015, \apj, 804, 146

\bibitem[Delgado Mena et al.(2015)]{delgadomena-et-al2015}
Delgado Mena, E., Bertr\'{a}n de Lis, S., Adibekyan, V.~Z., et al.\ 2015, \aap, 576, A69

\bibitem[D\'{i}az et al.(2011)]{diaz-et-al2011}
D\'{i}az, C.~G., Gonz\'{a}lez, J.~F., Levato, H., \& Grosso, M.\ 2011, \aap, 531, A143

\bibitem[D\'{i}az et al.(2016)]{diaz-et-al2016}
D\'{i}az, R.~F., S\'{e}gransan, D., Udry, S., et al.\ 2016, \aap, 585, A134

\bibitem[Di Folco et al.(2004)]{difolco-et-al2004}
Di Folco, E., Th\'{e}venin, F., Kervella, P., et al.\ 2004, \aap, 426, 601

\bibitem[do Nascimento et al.(2016)]{donascimento-et-al2016}
do Nascimento Jr., J.~D., Vidotto, A.~A., Petit, P., et al.\ 2016, \apjl, 820, L15

\bibitem[Doyle(1996)]{doyle1996}
Doyle, J.~G.\ 1996, \aap, 307, L45

\bibitem[Draine \& Salpeter(1979)]{draine-salpeter1979}
Draine, B.~T., \& Salpeter, E.~E.\ 1979, \apj, 231, 77

\bibitem[Ertel et al.(2014)]{ertel-et-al2014}
Ertel, S., Absil, O., Defr\`{e}re, D.\ 2014, \aap, 570, A128

\bibitem[Ertel et al.(2016)]{ertel-et-al2016}
Ertel, S., Defr\`{e}re, D., Absil, O., et al.\ 2016, \aap, 595, A44

\bibitem[Faramaz et al.(2014)]{faramaz-et-al2014}
Faramaz, V., Beust, H., Th\'{e}bault, P., et al.\ 2014, \aap, 563, A72

\bibitem[Faramaz et al.(2016)]{faramaz-et-al2016}
Faramaz, V., Ertel, S., Booth, M., Cuadra, J., \& Simmonds, C.\ 2016, \mnras, 465, 2352

\bibitem[Ferland et al.(2013)]{ferland-et-al2013}
Ferland, G.~J., Porter, R.~L., van Hoof, P.~A.~M., et al.\ 2013, RMxAA, 49, 137

\bibitem[Feuerbacher \& Fitton(1972)]{feuerbacher-fitton1972}
Feuerbacher, B., \& Fitton, B.\ 1972, J. Appl. Phys., 43, 1563

\bibitem[Feuerbacher et al.(1972)]{feuerbacher-et-al1972}
Feuerbacher, B., Anderegg, M., Fitton, B., et a.\ 1972, \gca~Suppl., 3, 2655

\bibitem[Folsom et al.(2018)]{folsom-et-al2018}
Folsom, C.~P., Bouvier, J., Petit, P., et al.\ 2018, \mnras, 474, 4956

\bibitem[Fuhrmann \& Chini(2015)]{fuhrmann-chini2015}
Fuhrmann, K., \& Chini, R.\ 2015, \apj, 809, 107

\bibitem[Gaidos \& Gonzalez(2002)]{gaidos-gonzalez2002}
Gaidos, E.~J., \& Gonzalez, G.\ 2002, New Astron., 7, 211

\bibitem[Gallet et al.(2017)]{gallet-et-al2017}
Gallet, F., Charbonnel, C., Amard, L., et al.\ 2017, \aap, 597, A14

\bibitem[Ghezzi et al.(2010)]{ghezzi-et-al2010}
Ghezzi, L., Cunha, K., Smith, V.~V., et al.\ 2010, \apj, 720, 1290

\bibitem[G\l\c{e}bocki \& Gnaci\'{n}ski(2005)]{glebocki-gnacinski2005}
G\l\c{e}bocki, R., \& Gnaci\'{n}ski, P.\ 2005, In 13th Cambridge Workshop on Cool Stars, Stellar Systems and the Sun, (ed.) F. Favata, G.A.J. Hussain, and B. Battrick, ESA SP-560, European Space Agency, pp. 571

\bibitem[Grard(1973)]{grard1973}
Grard, R.~J.~L.\ 1973, \jgr, 78, 2885

\bibitem[Greaves et al.(2014)]{greaves-et-al2014}
Greaves, J.~S., Kennedy, G.~M., Thureau, N., et al.\ 2014, \mnras, 438, L31

\bibitem[G\"{u}nther et al.(2012)]{guenther-et-al2012}
G\"{u}nther, H.~M., Wolk, S.~J., Drake, J.~J., et al.\ 2012, \apj, 750, 78

\bibitem[Hadjara et al.(2014)]{hadjara-et-al2014}
Hadjara, M., Domiciano de Souza, A., Vakili, F., et al.\ 2014, \aap, 569, A45

\bibitem[Hempel et al.(2005)]{hempel-et-al2005}
Hempel, M., Robrade, J., Ness, J.-U., \& Schmitt, J.~H.~M.~M.\ 2005, \aap, 440, 727

\bibitem[Hill et al.(2010)]{hill-et-al2010}
Hill, G., Gulliver, A.~F., \& Adelman, S.~J.\ 2010, \apj, 712, 250

\bibitem[Hillenbrand et al.(2008)]{hillenbrand-et-al2008}
Hillenbrand, L.~A., Carpenter, J.~M., Kim, J.~S., et al.\ 2008, \apj, 677, 630

\bibitem[H\"{u}nsch et al.(1998a)]{huensch-et-al1998a}
H\"{u}nsch, M., Schmitt, J.~H.~M.~M., \& Voges, W.\ 1998a, \aaps, 127, 251

\bibitem[H\"{u}nsch et al.(1998b)]{huensch-et-al1998b}
H\"{u}nsch, M., Schmitt, J.~H.~M.~M., \& Voges, W.\ 1998b, \aap~Suppl.~Ser. 132, 155

\bibitem[H\"{u}nsch et al.(1999)]{huensch-et-al1999}
H\"{u}nsch, M., Schmitt, J.~H.~M.~M., Sterzik, M.~F., \& Voges, W.\ 1999,  \aap~Suppl. Ser., 135, 319

\bibitem[Hubrig et al.(2006)]{hubrig-et-al2006}
Hubrig, S., Yudin, R.~V., Sch\"{o}ller, M., \& Pogodin, M.~A.\ 2006, \aap, 446, 1089

\bibitem[Hwang \& Daily(1992)]{hwang-daily1992}
Hwang, J., \& Daily, J.~W.\ 1992, Aerosol Sci. Technol., 16, 113

\bibitem[Ivey(1949)]{ivey1949}
Ivey, H.~F.\ 1949, Phys. Rev., 76, 567

\bibitem[James \& Jeffries(1997)]{james-jeffries1997}
James, D.~J., \& Jeffries, R.~D.\ 1997, \mnras, 292, 252

\bibitem[Jofr\'{e} et al.(2015)]{jofre-et-al2015}
Jofr\'{e}, E., Petrucci, R., Saffe, C., et al.\ 2015, \aap, 574, A50

\bibitem[Jones et al.(2015)]{jones-et-al2015}
Jones, J., White, R.~J., Boyajian, T., et al.\ 2015, \apj, 813, 58

\bibitem[Johnson \& Wright(1983)]{johnson-wright1983}
Johnson, H.~M., \& Wright, C.~D.\ 1983, \apjs, 53, 643

\bibitem[Johnstone \& G\"{u}del(2015)]{johnstone-guedel2015}
Johnstone, C.~P., \ G\"{u}del, M.\ 2015, \aap, 578, A129

\bibitem[Kalas et al.(2006)]{kalas-et-al2006}
Kalas, P., Graham, J.~R., Clampin, M.~C., \& Fitzgerald, M.~P.\ 2006, \apj, 637, L57

\bibitem[Kashyap et al.(2008)]{kashyap-et-al2008}
Kashyap, V.~L., Drake, J.~J., \& Saar, S.~H.\ 2008, \apj, 687, 1339

\bibitem[Kennedy et al.(2015)]{kennedy-et-al2015}
Kennedy, G.M., Matr\`{a}, L., Marmier, M., et al.\ 2015, \mnras, 449, 3121

\bibitem[Kervella et al.(2008)]{kervella-et-al2008}
Kervella, P., Mer\'{a}nd, A., Pichon, B., et al.\ 2008, \aap, 488, 667

\bibitem[Kimura(2016)]{kimura2016}
Kimura, H.\ 2016, \mnras, 459, 2751

\bibitem[Kimura \& Mann(1998a)]{kimura-mann1998a}
Kimura, H., \& Mann, I.\ 1998a, \apj, 499, 454

\bibitem[Kimura \& Mann(1998b)]{kimura-mann1998b}
Kimura, H., \& Mann, I.\ 1998b, Earth Planets Space, 50, 493

\bibitem[Kimura et al.(1997)]{kimura-et-al1997}
Kimura, H., Ishimoto, H., and Mukai, T.\ 1997, \aap, 326, 263

\bibitem[Kimura et al.(2002)]{kimura-et-al2002}
Kimura, H., Mann, I., Biesecker, D.~A., and Jessberger, E.~K.\ 2002, Icarus, 159, 529

\bibitem[Kimura et al.(2014)]{kimura-et-al2014}
Kimura, H., Senshu, H., \& Wada, K.\ 2014, \planss, 100, 64

\bibitem[Kirchschlager et al.(2017)]{kirchschlager-et-al2017}
Kirchschlager, F., Wolf, S., Krivov, A.~V., Mutschke, H., \& Brunngr\"{a}ber, R.\ 2017, \mnras, 467, 1614

\bibitem[K\"{o}hnlein(1996)]{koehnlein1996}
K\"{o}hnlein, W.\ 1996, Solar Phys., 169, 209

\bibitem[Kobayashi et al.(2009)]{kobayashi-et-al2009}
Kobayashi, H., Watanabe, S.-I., Kimura, H., and Yamamoto, T.\ 2009, Icarus 201, 395

\bibitem[Kobayashi et al.(2011)]{kobayashi-et-al2011}
Kobayashi, H., Kimura, H., Watanabe, S.-I., Yamamoto, T., and M\"{u}ller, S.\ 2011, Earth Planets Sp, 63, 1067

\bibitem[Kral et al.(2017)]{kral-et-al2017}
Kral, Q., Krivov, A.~V., Defr\`{e}re, D., et al.\ 2017, Astron. Rev., 13, 69

\bibitem[Krist et al.(2012)]{krist-et-al2012}
Krist, J.~E., Stapelfeldt, K.~R., Bryden, G., \& Plavchan, P.\ 2012, \aj, 144, 45

\bibitem[Krti\v{c}ka \& Kub\'{a}t(2011)]{krticka-kubat2011}
Krti\v{c}ka, J., \& Kub\'{a}t, J.\ 2011, \aap, 534, A97

\bibitem[Krivov et al.(1998)]{krivov-et-al1998}
Krivov, A., Kimura, H., \& Mann, I.\ 1998, Icarus 134, 311

\bibitem[Kurtz et al.(1994)]{kurtz-et-al1994}
Kurtz, D.~W., Sullivan, D.~J., Martinez, P., \& Tripe, P.\ 1994, \mnras, 270, 674

\bibitem[Lamers \& Cassinelli(1999)]{lamers-cassinelli1999}
Lamers, H.~J.~G.~L.~M., \& Cassinelli, J.~P.\ 1999, Introduction to Stellar Winds. Cambridge Univ. Press, Cambridge.

\bibitem[Landstreet(1982)]{landstreet1982}
Landstreet, J.~D.\ 1982, \apj, 258, 639

\bibitem[Laor \& Draine(1993)]{laor-draine1993}
Laor, A., \& Draine, B.~T.\ 1993, \apj, 402, 441

\bibitem[Lebreton et al.(2013)]{lebreton-et-al2013}
Lebreton, J., van Lieshout, R., Augereau, J.-C.\ 2013, \aap, 555, A146

\bibitem[Lebreton et al.(2016)]{lebreton-et-al2016}
Lebreton, J., Beichman, C., Bryden, G., et al.\ 2016, \apj, 817, 165

\bibitem[Lefevre(1975)]{lefevre1975}
Lefevre, J.\ 1975, \aap, 41, 437

\bibitem[Louh et al.(2005)]{louh-et-al2005}
Louh, S.~P., Leu, I.~C., \& Hon, M.~H.\ 2005, Diam. Relat. Mater., 14, 1000

\bibitem[Lovis et al.(2011)]{lovis-et-al2011}
Lovis, C., Dumusque, X., Santos, N.~C., et al.\ 2011, arXiv:1107.5325

\bibitem[Loyd et al.(2016)]{loyd-et-al2016}
Loyd, R.~O.~P., France, K., Youngblood, A., et al.\ 2016, \apj, 824, 102

\bibitem[Malagnini \& Morossi(1990)]{malagnini-morossi1990}
Malagnini, M.~L., \& Morossi, C.\ 1990, \aap~Suppl. Ser., 85, 1015

\bibitem[Maldonado et al.(2013)]{maldonado-et-al2013}
Maldonado, J., Villaver, E., \& Eiroa, C.\ 2013, \aap, 554, A84

\bibitem[Makov et al.(1988)]{makov-et-al1988}
Makov G., Nitzan A., \& Brus L. E.\ 1988, \jcp, 88, 5076

\bibitem[Marsden et al.(2014)]{marsden-et-al2014}
Marsden, S.~C., Petit, P., Jeffers, S.~V., et al.\ 2014, \mnras, 444, 3517

\bibitem[Marshall et al.(2013)]{marshall-et-al2013}
Marshall, J.~P., Krivov, A.~V., del Burgo, C., et al. 2013, \aap, 557, A58

\bibitem[Mathys(1994)]{mathys1994}
Mathys, G.\ 1994, \aap~Suppl. Ser., 108, 547

\bibitem[McAlister et al.(2005)]{mcalister-et-al2005}
McAlister, H.~A., ten Brummelaar, T.~A., Gies, D.~R., et al.\ 2005, \apj, 628, 439

\bibitem[Mennesson et al.(2014)]{mennesson-et-al2014}
Mennesson, B., Millan-Gabet, R., Serabyn, E., et al.\ 2014, \apj, 797, 119

\bibitem[Meshkat et al.(2015)]{meshkat-et-al2015}
Meshkat, T., Kenworthy, M.~A., Reggiani, M., et al.\ 2015, \mnras, 453, 2533

\bibitem[Miller et al.(2015)]{miller-et-al2015}
Miller, B.~P., Gallo, E., Wright, J.~T., \& Pearson, E.~G.\ 2015, \apj, 799, 163

\bibitem[Monin et al.(2002)]{monin-et-al2002}
Monin, D.~N., Fabrika, S.~N., \& Valyavin, G.~G.\ 2002, \aap, 396, 131

\bibitem[Mo\'{o}r et al.(2015)]{moor-et-al2015}
Mo\'{o}r, A., K\'{o}sp\'{a}l, A., \'{A}brah\'{a}m, P., et al.\ 2015, \mnras, 447, 577

\bibitem[Morales et al.(2016)]{morales-et-al2016}
Morales, F.~Y., Bryden, G., Werner, M.~W., \& Stapelfeldt, K.~R.\ 2016, \apj, 831, 97

\bibitem[Motalebi et al.(2015)]{motalebi-et-al2015}
Motalebi, F., Udry, S., Gillon, M., et al.\ 2015, \aap, 584, A72

\bibitem[Mukai(1981)]{mukai1981}
Mukai, T.\ 1981, \aap, 99, 1

\bibitem[Mukai \& Yamamoto(1979)]{mukai-yamamoto1979}
Mukai, T., \& Yamamoto, T.\ 1979, PASJ, 31, 585

\bibitem[Nagahara et al.(1994)]{nagahara-et-al1994}
Nagahara, H., Kushiro, I., \& Mysen, B.~O.\ 1994, \gca, 58, 1951

\bibitem[Nordstr\"{o}m et al.(2004)]{nordstroem-et-al2004}
Nordstr\"{o}m, B., Mayor, M., Andersen, J., et al.\ 2004, \aap, 418, 989

\bibitem[Noyes et al.(1984)]{noyes-et-al1984}
Noyes, R.~W., Hartmann, L.~W., Baliunas, S.~L., Duncan, D.~K., \& Vaughan, A.~H.\ 1984, \apj, 279, 763

\bibitem[Nu\~{n}ez et al.(2017)]{nunez-et-al2017}
Nu\~{n}ez, P.~D., Scott, N.~J., Mennesson, B., et al.\ 2017. \aap, 608, A113

\bibitem[Ohgaito et al.(2002)]{ohgaito-et-al2002}
Ohgaito, R., Mann, I., Kuhn, J.~R., MacQueen, R.~M., \& Kimura, H.\ 2002, \apj, 578, 610

\bibitem[Parker(1958)]{parker1958}
Parker, E.~N.\ 1958, \apj, 128, 664

\bibitem[Panzera et al.(1999)]{panzera-et-al1999}
Panzera, M.~R., Tagliaferri, G., Pasinetti, L., \& Antonello, E.\ 1999, \aap, 348, 161

\bibitem[Parker(1964)]{parker1964}
Parker, E.~N.\ 1964, \apj, 139, 951

\bibitem[Pasinetti Fracassini et al.(2001)]{pasinettifracassini-et-al2001}
Pasinetti Fracassini, L.~E., Pastori, L., Covino, S., \& Pozzi, A.\ 2001, \aap, 367, 521

\bibitem[Pauldrach et al.(1986)]{pauldrach-et-al1986}
Pauldrach, A., Puls, J., \& Kudritzki, R.~P.\ 1986, \aap, 164, 86

\bibitem[Pease et al.(2016)]{pease-et-al2016}
Pease, D.~O., Drake, J.~J., \& Kashyap, V.~L.\ 2006, \apj, 636, 426

\bibitem[Piters et al.(1998)]{piters-et-al1998}
Piters, A.~J.~M., van Paradijs, J., \& Schmitt, J.~H.~M.~M.\ 1998, \aaps, 128, 29

\bibitem[Pizzolato et al.(2003)]{pizzolato-et-al2003}
Pizzolato, N., Maggio, A., Micela, G., Sciortino, S., \& Ventura, P.\ 2003, \aap, 397, 147

\bibitem[Poppenhaeger et al.(2017)]{poppenhaeger-et-al2017}
Poppenhaeger, K., Auchettl, K., \& Wolk, S.~J.\ 2017, \mnras, 468, 4018

\bibitem[Poppenh\"{a}ger et al.(2009)]{poppenhaeger-et-al2009}
Poppenh\"{a}ger, K., Robrade, J., Schmitt, J.~H.~M.~M., \& Hall, J.~C.\ 2009, \aap, 508, 1417

\bibitem[Reiners(2006)]{reiners2006}
Reiners, A.\ 2006, \aap, 446, 267

\bibitem[Reiners et al.(2001)]{reiners-et-al2001}
Reiners, A., Schmitt, J.~H.~M.~M., \& K\"{u}rster, M.\ 2001, \aap, 376, L13

\bibitem[Rhee et al.(2007)]{rhee-et-al2007}
Rhee, J.~H., Song, I., Zuckerman, B., \& McElwain, M.\ 2007, \apj, 660, 1556

\bibitem[Rieke et al.(2016)]{rieke-et-al2016}
Rieke, G.~H., G\'{a}sp\'{a}r, A., \& Ballering, N.~P.\ 2016, \apj, 816, 50

\bibitem[Robrade \& Schmitt(2009)]{robrade-schmitt2009}
Robrade, J., \& Schmitt, J.~H.~M.~M.\ 2009, \aap, 497, 511

\bibitem[Rouleau \& Martin(1991)]{rouleau-martin1991}
Rouleau, F., \& Martin, P.~G.\ 1991, \apj, 377, 526

\bibitem[Royer et al.(2002)]{royer-et-al2002}
Royer, F., Gerbaldi, M., Faraggiana, R., \& G\'{o}mez, A.~E.\ 2002, \aap, 381, 105

\bibitem[Royer et al.(2007)]{royer-et-al2007}
Royer, F., Zorec, J., \& G\'{o}mez, A.~E.\ 2007, \aap, 463, 671

\bibitem[Rusk(1988)]{rusk1988}
Rusk, E.~T.\ 1988, \aj, 96, 1447

\bibitem[Saar \& Osten(1997)]{saar-osten1997}
Saar, S.~H., \& Osten, R.~A.\ 1997, \mnras, 284, 803

\bibitem[Sakurai(1985)]{sakurai1985}
Sakurai, T.\ 1985, \aap, 152, 121

\bibitem[Sanz-Forcada et al.(2011)]{sanzforcada-et-al2011}
Sanz-Forcada, J., Micela, G., Ribas, I., et al.\ 2011, \aap, 532, A6

\bibitem[Schmitt(1997)]{schmitt1997}
Schmitt, J.~H.~M.~M.\ 1997, \aap, 318, 215

\bibitem[Schmitt \& Liefke(2004)]{schmitt-liefke2004}
Schmitt, J.~H.~M.~M., Liefke, C.\ 2004, \aap, 417, 651

\bibitem[Schr\"{o}der \& Schmitt(2007)]{schroeder-schmitt2007}
Schr\"{o}der, C., \& Schmitt, J.~H.~M.~M.\ 2007, \aap, 475, 677

\bibitem[Schr\"{o}der \& Cuntz(2005)]{schroeder-cuntz2005}
Schr\"{o}der, K.-P., \& Cuntz, M.\ 2005, \apj, 630, L73

\bibitem[Schr\"{o}der et al.(2008)]{schroeder-et-al2008}
Schr\"{o}der, C., Hubrig, S., Schmitt, J.~H.~M.~M.\ 2008, \aap, 484, 479

\bibitem[Schr\"{o}der et al.(2009)]{schroeder-et-al2009}
Schr\"{o}der, C., Reiners, A., Schmitt, J.~H.~M.~M.\ 2009, \aap, 493, 1099

\bibitem[Senshu et al.(2015)]{senshu-et-al2015}
Senshu, H., Kimura, H., Yamamoto, T., et al.\ 2015, \planss 116, 18

\bibitem[Shannon et al.(1991)]{shannon-et-al1991}
Shannon, R.~D., Subramanian, M.~A., Hosoya, S., \& Rossman, G.~R.\ 1991, Phys. Chem. Miner., 18, 1

\bibitem[Shorlin et al.(2002)]{shorlin-et-al2002}
Shorlin, S.~L.~S., Wade, G.~A., Donati, J.-F., et al.\ 2002, \aap, 392, 637

\bibitem[Simpson et al.(2010)]{simpson-et-al2010}
Simpson, E.~K., Baliunas, S.~L., Henry, G.~W., Watson, C.~A.\ 2010, \mnras, 408, 1666

\bibitem[Simon et al.(2002)]{simon-et-al2002}
Simon, T., Ayres, T.~R., Redfield, S. \& Linsky, J.~L.\ 2002, \apj, 579, 800

\bibitem[Smith et al.(2001)]{smith-et-al2001}
Smith, R.~K., Brickhouse, N.~S., Liedahl, D.~A., \& Raymond, J.~C.\ 2001, \apj, 556, L91

\bibitem[Smith et al.(2012)]{smith-et-al2012}
Smith, R., Wyatt, M.C., \& Haniff, C.~A.\ 2012, \mnras, 422, 2560

\bibitem[Sodha(1961)]{sodha1961}
Sodha, M.~S.\ 1961, J. Appl. Phys., 32, 2059

\bibitem[Spina et al.(2016)]{spina-et-al2016}
Spina, L., Mel\'{e}ndez, J., \& Ram\'{i}rez, I.\ 2016, \aap, 585, A152

\bibitem[Stock et al.(2010)]{stock-et-al2010}
Stock, N.~D., Su, K.~Y.~L., Liu, W., et al.\ 2010,\apj, 724, 1238

\bibitem[Su et al.(2008)]{su-et-al2008}
Su, K.~Y.~L., Rieke, G.~H., Stapelfeldt, K.~R., et al.\ 2008, \apj, 679, L125

\bibitem[Su et al.(2016)]{su-et-al2016}
Su, K.~Y.~L., Rieke, G.~H., Defr\'{e}re, D., et al.\ 2016, \apj, 818, 45

\bibitem[Su\'{a}rez Mascare\~{n}o et al.(2017)]{suarezmascareno-et-al2017}
Su\'{a}rez Mascare\~{n}o, A., Rebolo, R., Gonz\'{a}lez Hern\'{a}ndez, J. I., \& Esposito, M.\ 2017, \mnras, 468, 4772

\bibitem[Suzuki(2007)]{suzuki2007}
Suzuki, T.~K.\ 2007, \apj, 659, 1592

\bibitem[Takeda(2007)]{takeda2007}
Takeda, Y.\ 2007, \pasj, 59, 335

\bibitem[Telleschi et al.(2005)]{telleschi-et-al2005}
Telleschi, A., G\"{u}del, M., Briggs, K., et al.\ 2005, \apj, 622, 653

\bibitem[Udovichenko et al.(1994)]{udovichenko-et-al1994}
Udovichenko, S.~N., Keir, L.~E., Shtol, V.~G., \& Bychkov, V.~D.\ 1994, Chemically Peculiar \& Magnetic Stars, ed. J. Zverko \& J. \v{Z}i\v{z}\v{n}ovsk\'{y}, Astron. Inst. SAS, 45.

\bibitem[Valenti \& Fischer(2005)]{valenti-fischer2005}
Valenti, J.~A., Fischer, D.~A.\ 2005, \apjs, 159, 141

\bibitem[van Belle et al.(2006)]{vanbelle-et-al2006}
van Belle, G.~T., Ciardi, D.~R., ten Brummelaar, T., et al.\ 2006, \apj, 637, 494

\bibitem[van Lieshout et al.(2014)]{vanlieshout-et-al2014}
van Lieshout, R., Dominik, C., Kama, M., \& Min, M.\ 2014, \aap, 571, A51

\bibitem[Vidotto et al.(2012)]{vidotto-et-al2012}
Vidotto, A.~A., Fares, R., Jardine, M., et al.\ 2012, \mnras, 423, 3285

\bibitem[Watson et al.(2011)]{watson-et-al2011}
Watson, C.~A., Littlefair, S.~P., Diamond, C., et al.\ 2011, \mnras, 413, L71

\bibitem[Weber \& Davis(1967)]{weber-davis1967}
Weber, E.~J., \& Davis Jr., L.\ 1967, \apj, 148, 217

\bibitem[Wong et al.(2003)]{wong-et-al2003}
Wong K., Vongehr S., \& Kresin V. V.\ 2003, \prb, 67, 035406

\bibitem[Wood et al.(2002)]{wood-et-al2002}
Wood, B.~E., M\"{u}ller, H.-R., Zank, G.~P., \& Linsky, J.~L.\ 2002, \apj, 574, 412

\bibitem[Wood et al.(2005)]{wood-et-al2005}
Wood, B.~E., M\"{u}ller, H.-R., Zank, G.~P., Linsky, J.~L., \& Redfield, S.\ 2005, \apj, 628, L143

\bibitem[Wright et al.(2004)]{wright-et-al2004}
Wright, J.~T., Marcy, G.~W., Butler, R.~P., \& Vogt, S.~S.\ 2004, \apjs, 152, 261

\bibitem[Wright et al.(2011)]{wright-et-al2011}
Wright, N.~J., Drake, J.~J., Mamajek, E.~E., \& Henry, G.~W.\ 2011, \apj, 743, 48

\bibitem[Wright \& Drake(2016)]{wright-drake2016}
Wright, N.~J., \& Drake, J.~J.\ 2016, \nat, 535, 526

\bibitem[Zorec \& Royer(2012)]{zorec-royer2012}
Zorec, J., \& Royer, F.\ 2012, \aap, 537, A120

\end{thebibliography}
\end{document}